\documentclass[preprint]{aastex}
\usepackage{color}
\usepackage{graphicx}
\usepackage{lscape}
\usepackage{longtable,lscape}
\usepackage{rotating}
\usepackage{epsfig}
\usepackage{subfig}

\begin{document}

\newcommand{\dg}{$^{\circ}$} 
\newcommand{\sex}{{\it SExtractor}}
\newcommand{\Msol}{M$_{\odot}$}
\newcommand{\gsim}{$\gtrsim$}
\newcommand{\nod}{\nodata}
\newcommand{\lsim}{$\lesssim$}
\newcommand{\txw}{\textwidth}
\newcommand{\snr}{signal-to-noise ratio}
\newcommand{\hypz}{{\tt HyperZ}}
\newcommand{\gfit}{{\tt GalFit}}
\newcommand{\Sersic}{S\'{e}rsic}
\newcommand{\rsrang}{$0.35$\,\,\lsim\,\,$z$\,\,\lsim\,\,$1.5$}
\newcommand{\nuvv}{(NUV--V)$_{rest}$}
\newcommand{\fuvv}{(FUV--V)$_{rest}$}
\newcommand{\gmr}{($g^{\prime}-r^{\prime}$)$_{rest}$}
\renewcommand{\thefootnote}{\arabic{footnote}}

%%%%%%%%%%%% TITLE %%%%%%%%%%%%%%%%%%%%%%%%%%%%%%%%%%%%%%%%%%%%%%%%
% page:title

\title{\bf A Panchromatic Catalog of Early-Type Galaxies at
  Intermediate Redshift in the Hubble Space Telescope Wide Field
  Camera 3 Early Release Science Field}
\author{M. J. Rutkowski\altaffilmark{1}, S. H. Cohen\altaffilmark{1}, S. Kaviraj\altaffilmark{2, 3}, R. W. O'Connell\altaffilmark{4}, N. P. Hathi\altaffilmark{5}, R. A. Windhorst\altaffilmark{1}, R. E. Ryan Jr.\altaffilmark{6}, R. M. Crockett\altaffilmark{2}, H. Yan\altaffilmark{7}, R. A. Kimble\altaffilmark{8}, J. Silk\altaffilmark{2}, P.J. McCarthy\altaffilmark{5}, A. Koekemoer\altaffilmark{6}, B. Balick\altaffilmark{9}, H. E. Bond\altaffilmark{6}, D. Calzetti\altaffilmark{10}, M. J. Disney\altaffilmark{11}, M. A. Dopita\altaffilmark{12,13}, J. A. Frogel\altaffilmark{13,14}, D. N. B.  Hall\altaffilmark{15}, J. A. Holtzman\altaffilmark{16}, F. Paresce\altaffilmark{17}, A. Saha\altaffilmark{18}, J. T. Trauger\altaffilmark{19}, A. R. Walker\altaffilmark{20}, B. C. Whitmore\altaffilmark{6}, E. T. Young\altaffilmark{21}}

\altaffiltext{1}{School of Earth and Space Exploration, Arizona State University, Tempe, AZ 85287-1404, USA}
\altaffiltext{2}{Department of Physics, University of Oxford, Denys Wilkinson Building, Keble Road, Oxford OX1 3RH, UK}
\altaffiltext{3}{Blackett Laboratory, Imperial College London, South Kensington Campus, London SW7 2AZ, UK}
\altaffiltext{4}{Department of Astronomy, University of Virginia, P.O. Box 3818, Charlottesville, VA, 22903, USA}
\altaffiltext{5}{Observatories of the Carnegie Institute of Washington, Pasadena, CA 91101, USA}
\altaffiltext{6}{Space Telescope Science Institute, Baltimore, MD 21218, USA}
\altaffiltext{7}{Center for Cosmology and Astroparticle Physics, Ohio State University, Columbus, OH 43210}
\altaffiltext{8}{NASA Goddard Space Flight Center, Greenbelt, MD 20771, USA }
\altaffiltext{9}{Department of Astronomy, University of Washington, Seattle, WA 98195-1580, USA}
\altaffiltext{10}{Department of Astronomy, University of Massachusetts, Amherst, MA 01003, USA}
\altaffiltext{11}{School of Physics and Astronomy, Cardiff University, Cardiff CF24 3AA, UK}
\altaffiltext{12}{Research School of Physics and Astronomy, The Australian National University, ACT 2611, Australia}
\altaffiltext{13}{Astronomy Department, King Abdulaziz University, P.O. Box 80203, Jeddah, Saudi Arabia}
\altaffiltext{14}{Galaxies Unlimited, 8726 Hickory Bend Trail, Potomac, MD 20854, USA}
\altaffiltext{15}{Institute for Astronomy, University of Hawaii, Honolulu, HI 96822, USA}
\altaffiltext{16}{Department of Astronomy, New Mexico State University, Las Cruces, NM 88003, USA}
\altaffiltext{17}{Istituto di Astrofisica Spaziale e Fisica Cosmica, INAF, Via Gobetti 101, 40129 Bologna, Italy}
\altaffiltext{18}{National Optical Astronomy Observatories, Tucson, AZ 85726-6732, USA}
\altaffiltext{19}{NASA-Jet Propulsion Laboratory, Pasadena, CA 91109, USA}
\altaffiltext{20}{Cerro Tololo Inter-American Observatory, La Serena, Chile}
\altaffiltext{21}{NASA-Ames Research Center, Moffett Field, CA 94035}

%14 =13.89 ; 0.5 = 0.51 
%%%%%%%%%%%% ABSTRACT %%%%%%%%%%%%%%%%%%%%%%%%%%%%%%%%%%%%%%%%%%%%%%%%
%page: abstract
%\newpage
%\section{abstract}
\begin{abstract}

In the first of a series of forthcoming publications, we present a
panchromatic catalog of 102 visually-selected early-type galaxies
(ETGs) from observations in the Early Release Science (ERS) program
with the Wide Field Camera 3 (WFC3) on the {\it Hubble Space
  Telescope} (HST) of the Great Observatories Origins Deep
Survey-South (GOODS-S) field.  Our ETGs span a large redshift range,
\rsrang, with each redshift spectroscopically-confirmed by previous
published surveys of the ERS field.  We combine our measured WFC3 ERS
and ACS GOODS-S photometry to gain continuous sensitivity from the
rest-frame far-UV to near-IR emission for each ETG.  The superior
spatial resolution of the HST over this panchromatic baseline allows
us to classify the ETGs by their small-scale internal structures, as
well as their local environment.  By fitting stellar population
spectral templates to the broad-band photometry of the ETGs, we
determine that the average masses of the ETGs are comparable to the
characteristic stellar mass of massive galaxies,
10$^{11}<\mbox{M}_{\ast}$[\Msol]$<10^{12}$.  %Though this sample is
%representative of a population of moderate to high mass morphological
%ETGs, we show that the selection criteria we use to define the catalog
%implicitly prevents us from defining a {\it complete} catalog of ETGs
%in the ERS field.

By transforming the observed photometry into the GALEX FUV and NUV,
Johnson V, and SDSS $g^{\prime}$ and $r^{\prime}$ bandpasses we
identify a noteworthy diversity in the rest-frame UV-optical colors
and find the mean rest-frame (FUV$-$V) $=3.5$ and (NUV$-$V) $=3.3$,
with 1$\sigma$ standard deviations $\simeq$ 1.0.  The blue rest-frame
UV-optical colors observed for most of the ETGs are evidence for
star-formation during the preceding gigayear, but no systems exhibit
UV-optical photometry consistent with major recent (\,\,\lsim\,\,50
Myr) starbursts. Future publications which address the diversity of
stellar populations likely to be present in these ETGs, and the
potential mechanisms by which recent star-formation episodes are
activated, are discussed.
\end{abstract}

\section{Introduction}\label{sec:intro}
The star-formation histories of early-type galaxies (ellipticals and
S0s, hereafter denoted ETGs) are now known to be considerably more
diverse than had been originally expected. Optical broad-band
photometry initially suggested that ETGs in the local universe
were largely composed of a homogeneous, old ($>$ 10~Gyr), and passively
evolving stellar populations that were formed at a uniformly high
redshift via the ``monolithic collapse'' scenario \citep[e.g.,][]{
  ELS62,T80}.  However, high precision optical spectrophotometry
\citep[e.g.,][]{OC80,R85,W94,T00} shows that a significant fraction of
nearby ETGs experienced prolonged episodes of star-formation, lasting
until a few gigayears ago.  Their inferred luminosity-weighted ages
have recently been found to correlate with velocity dispersion as well
as environment \citep{G09,C09,Sc09}, so the mechanisms driving recent
star-formation activity in ETGs are now coming into better focus.  Cool
interstellar material capable of fueling star-formation is also
frequently present in ETGs \citep[e.g.][and references
  therein]{M06,L07}.  These, and many other lines of evidence,
including fine-structure (e.g., rings, shells, and ripples) in nearby
ETGs \citep{S90,C01,Sa10,K10}, statistics of close pairs \citep{P02},
and the evolution of galaxy morphologies \citep{VD05,VD10}, point
toward a hierarchical, merger-dominated assembly of ETGs over an
extended period \citep[][and references therein]{T72,B01,Ka09,Ka11}.

Ultraviolet (UV) observations of large samples of ETGs, first enabled
by the International Ultraviolet Explorer \citep[IUE, see][]{K87} and
later by the {\it Hubble Space Telescope} (HST) and the {\it Galaxy
  Evolution Explorer} \citep[GALEX,][]{M97}, confirmed the presence of
late star-formation in many ETGs.  The 1200-3000~\AA\,UV continuum is
highly sensitive to small amounts of star-formation during the last
$\sim$1 Gyr \citep[see,][]{FS00,Ka09}.  With GALEX, \cite{Yi05} found
residual star-formation to be readily detectable in ETGs at low
redshifts.  Subsequently, a study of the UV-optical properties of
$\sim$2100 ETGs by \cite{K07} revealed that {\it at least}~30\% of
low-redshift ($ z<0.11$) field ETGs have UV-optical photometry
consistent with active star-formation during the previous $\sim$1~Gyr.
It is therefore of considerable interest to follow the incidence of
rest-frame UV signatures of star-formation in ETGs to redshifts of $z
\sim$1--2 at the HST diffraction limit.

The UV provides a valuable window on older, hot stellar populations as
well.  A UV upturn (UVX)---characterized by a sharp rise in the far-UV
spectrum shortward of $\sim$2000~\AA---has been detected in many
low-redshift ETGs \citep[e.g.,][and references
  therein]{B88,Do07,Je09}, but cannot be attributed to recent
star-formation.  The UVX is believed to arise predominantly from a
small population of highly-evolved, hot, low-mass stars, especially
extreme horizontal branch (EHB) stars (for a review, see
O'Connell\,1999).  These stars have lost most of their hydrogen
envelopes, thus exposing their hot (T\,\,\gsim\,\,20,000 K),
helium-burning cores \citep[$M<0.52$ \Msol,][]{D93}.  Various
mechanisms are capable of reducing the envelopes, including giant
branch mass-loss in metal-rich stars \citep{GR90,D95,Y95,Y98}, binary
interactions \citep{H07}, or extreme aging in a metal-poor population
\citep{PL97}.  Most evidence favors a metal-rich UVX interpretation,
but a much better understanding of the underlying mechanisms could be
obtained, {\it if} we could follow the evolution of the UVX with
look-back time over the past ~5--8 Gyr.  A number of studies have
attempted to determine lookback dependence up to $z\sim$0.5
\citep{B00,B03,Y04,Le05,R07,A09}, but these were inconclusive, either
because of small samples, or because of low \snr.

The high spatial resolution and significant UV sensitivity of the HST
WFC3 are very well suited to the study of low-level star-formation
\citep[see, e.g., ][]{Cr10} and the UVX in intermediate redshift
ETGs. In this paper, we describe the selection and photometric
properties of a sample of intermediate redshift (\rsrang) ETGs
obtained from observations of the GOODS-S field \citep{G04}.  This
paper is the first in a series that will investigate the stellar
population(s) extant in intermediate redshift ETGs in the ERS survey
field.

This paper is organized in the following manner.  In
\S\ref{sec:thedataset}, we briefly describe the ERS program, technical
issues associated with WFC3 UV imaging relevant to this work, and the
observations.  In \S\ref{sec:selectcrit} we present the selection
criteria used to produce the catalog, and in \S\ref{sec:catalogueprod}
we present and describe the photometric catalog.  In
\S\ref{sec:absmag} we discuss the fitting of model stellar populations
defined by a single burst of star-formation to the broad-band spectral
energy distribution (SED) of the ETGs, the results from which we used
to measure the absolute photometry of the ETGs. In \S\ref{sec:GALFIT}
and \S\ref{sec:AGN}, we discuss the multi-wavelength morphological
properties of the ETGs. In \S\ref{sec:completeness}, we discuss the
impact of the ETG selection criteria on catalog completeness. In
\S\ref{sec:analysis} and \S\ref{sec:discusscolors}, we present the
rest-frame photometry transformation and discuss the rest-frame
UV-optical photometry of the ETGs, respectively.

Throughout this paper we assume a $\Lambda$CDM cosmology with
$\Omega_{m}$=0.27, $\Omega_{\Lambda}$=0.73, and H$_{0}$=70 km
s$^{−-1}$ Mpc$^{−-1}$ \citep{Ko11}. We use the following designations
$\colon$ F225W, F275W, F336W, F435W, F606W, F775W, F850LP, F098M,
F125W, and F160W represent the HST filters throughout; $g^{\prime}$
and $r^{\prime}$ represent the Sloan Digital Sky Survey (SDSS) filters
\citep{F96}; FUV and NUV represent the GALEX 150 \& 250 nm filters,
respectively \citep{M05}.  Throughout, we quote all fluxes on the
AB-magnitude system \citep{O83}.

\section{Observations}\label{sec:thedataset}

Our sample of ETGs is drawn from the HST imaging with Advanced Camera
for Surveys (ACS) and WFC3, which was obtained as part of the ERS
program.  Near-UV and near-IR observations were acquired as part of
the WFC3 ERS program (HST Program ID \#11359, PI$\colon$
R. W. O'Connell), a 104 orbit medium-depth survey using the HST UVIS
and IR cameras.  A general introduction to the performance and
calibration of the WFC3 is provided in \cite{W11}.

The ERS program observed approximately 50 square arcminutes in the
GOODS-S field with the HST WFC3 UVIS in three filters$\colon$ F225W
and F275W for 2 orbits, and F336W for 1 orbit, per pointing,
respectively. The program observed approximately 40 square arcminutes
in the same field with the WFC3 IR in three filters$\colon$ F098M,
F125W, and F160W, each for 2 orbits per pointing. The 5$\sigma$ 50\%
point-source completeness limits are$\colon$ F225W=26.3, F275W=26.4,
F336W=26.1, F098M=27.2, F125W=27.5, and F160W=27.2 mag
\citep[see][]{W11}.  The analysis presented here was completed using
mosaicked images produced for each of the UVIS and IR band tilings, and
each image mosaic was drizzled to a pixel scale equal to
0.090\arcsec\ pixel$^{-1}$. The UVIS filters have a small known
red-leak (i.e., contamination by unwanted long-wavelength photons),
which contributes no more than 3.0\% of the total flux, even for ETGs
at moderate redshift (see Appendix A).

%A field-of-view approximately equal to 50 square arcminutes was
%observed with the WFC3 UVIS broad-band filters F225W, F275W, and F336W
%in 2 orbits per pointing for F225W and F275W, and 1 orbit in F336W,
%which reaches 5$\sigma$ 50\% point-source completeness limits of
%F225W=26.3, F275W=26.4, and F336W=26.1 mag,
%respectively.  The UVIS filters have a small known red-leak (i.e.,
%contamination by unwanted long-wavelength photons), which contributes
%no more than 3.0\% of the total flux, even for ETGs at moderate
%redshift (see Appendix A).
%
%A field-of-view approximately equal to 40 square arcminutes was
%observed with the WFC3 IR broad-band filters F098M, F125W, and F160W in
%2 orbits per pointing for each of these filters, which reaches
%5$\sigma$ 50\% point-source completeness limits of F098M=27.2,
%F125W=27.5, and F160W=27.2 mag, respectively
%\citep[see][]{W11}.  The analysis presented here was completed using
%mosaiced images produced for each of the UVIS and IR band tilings, and
%each image mosaic was drizzled to a pixel scale equal to
%0.090\arcsec\ pixel$^{-1}$.

The WFC3 mosaics roughly cover the northern one-third of the GOODS-S
field \citep{G04}, and we incorporate the pre-existing ACS dataset
(F435W, F606W, F775W, and F850LP) with the WFC3 observations.  We
produced mosaicked images of the GOODS-S ACS data, which were binned to
match the pixel scale of the WFC3 UVIS/IR mosaics.

\section{A UV-optical-IR Photometric Catalog of Early-Type Galaxies}\label{sec:catalog}

\subsection{Selection Criteria}\label{sec:selectcrit}

We require our galaxies to have$\colon$ (1) been imaged in all UV and
IR bands, to uniform depth; (2) a spectroscopically-confirmed redshift
in the range \rsrang; and (3) an ETG morphology.

There are many techniques for identifying ETGs at intermediate
redshift.  We are particularly motivated to include in our sample ETGs
that encompass all possible star-formation histories, thus we do {\it
  not} select ETGs using traditional optical color-based methods,
since these may be biased toward specific star-formation histories.  For
example, photometric selection techniques \citep[e.g., optical color
  selection, see][]{B04}--- which assume a quiescent template
SED---will exclude ETGs with on-going or recent star-formation.  The
quantitative morphological classification of galaxies is an
alternative method of identifying a sample of ETGs
\citep[e.g.,][]{C03, A03, L04}.  However, the robustness of
each of these classifiers can be dramatically affected by a variety of
systematics, such as the image \snr~ \citep{C03,L08} and the bandpass
in which the technique is applied \citep{V07,C08}.  In lieu of these
techniques, we select our sample by visual classification.  This
technique is subjective, and as such can introduce new biases, but it
has been successfully applied to the identification of both low
redshift \citep[$z\sim$0.1;][]{S07} and intermediate redshift
\citep[$z$\,\,\lsim\,\,1.3;][]{P05, F09} ETGs.  We will demonstrate in
\S \ref{sec:completeness} that the spectroscopic redshift requirement,
and not the morphological selection technique, is the most significant
source of bias.

To identify our sample, ETGs were identified (M.J.R.) and then
independently confirmed by co-authors (S.K., R.M.C.) by visual
inspection of the GOODS ACS F606W, F775W, and F850LP and ERS WFC3 IR
F098M, F125W, and F160W image mosaics.  The galaxies included in this
sample exhibited the morphological characteristics of early-type
galaxies---i.e. these galaxies exhibited a centrally peaked light
profile, which declines sharply with radius, a high degree of
rotational symmetry, and a lack of visible internal structure.

UV imaging can provide unique insight into the star-formation history
of ETGs. Thus, we require our sample ETGs to be observed in each of
the UV filter mosaics. To ensure that all galaxies were observed to a
similar depth, we also require each ETG in the sample to be observed
in the UV and IR image mosaics for at least the mean exposure time
measured for each filter as given by \cite{W11}.  Since we are
interested in the star-formation histories of ETGs, and the WFC3 UVIS
channel is only sensitive to UV emission at $\lambda \sim1500$~\AA~for
objects at redshift z\,\,\gsim\,\,0.35, we define this redshift as
low-redshift cutoff of the sample.  The high-redshift cutoff was
selected to ensure that the visual inspection and classification of
the ETG --- in the filter set outlined above -- considers the
rest-frame V-band morphology. We are sensitive to at least the
UV-optical SED of every ETG in our catalog.

The spectroscopic redshifts for these ETGs were derived from the
analyses of spectra obtained with the Very Large Telescope
\citep{P09,LeF04,S04,Mi05,R207,V08,P09} and Keck Telescopes
\citep{St04,D05} and the HST ACS Grism (G800L) \citep{D05,P06,F09}.

We find 102 ETGs that satisfy these selection criteria.

\subsection{Photometry}\label{sec:catalogueprod}
We measured object fluxes using \sex~in dual-image mode \citep{B96},
with the WFC3 F160W image as the detection band.  For source
detection, we required sources to be detected in minimally four
connected pixels, each at $\geq$ 0.75$\sigma$ above the local computed
sky-background. For deblending, we adopted a contrast parameter of
10$^{-3}$ with 32 sub-thresholds.  Object photometry was determined
with MAG\_AUTO parameters Kron factor equal to 2.5 and minimum radius
equal to 3.5 pixels.

We adopted gains for each filter using the mean exposure time
calculated for each mosaic as follows$\colon$ F225W and F275W equal to
5688 sec; F336W equal to 2778 sec and F098M, F125W, and F160W equal to
5017 sec \citep[see][]{W11}.  From \cite{K09a,K09b} we assumed
zeropoints for the filter set F225W, F275W, F336W, F098M, F125W, F160W
equal to 24.06, 24.14, 24.64, 25.68, 26.25, 25.96 mag, respectively.
We assumed zeropoints for the filter set F435W, F606W, F775W, and
F850LP equal to 25.673, 26.486, 25.654, and 24.862 mag,
respectively\footnote{For more details, see
  http://archive.stsci.edu/pub/hlsp/goods/v2/h\_goods\_v2.0\_rdm.html}.

In Table \ref{tab:fluxtab} we present the measured photometry for the
ETGs. \sex~non-detections are designated ``\nodata'' (23 galaxies) and
ETG fluxes with detections fainter than the recovery limits (discussed
below) are designated ``---'' (52 galaxies), as explained in the
footnotes of Table \ref{tab:fluxtab}.

The combination of the stable WFC3 UV-optical-IR PSF and high spatial
resolution allows many compact or low surface brightness (SB) ETG
candidates to be detected and measured.  These candidates may meet the
morphological selection criteria in the ``detection'' image, but in
dual-image mode \sex~returns flux measurements for these ETGs which
are significantly below the formal completeness limits in the
``measurement'' image.  Their formal flux uncertainties are larger
than $\sim$1 mag (implying a \snr\,\,\lsim\,\,1).  To ascertain the
reliability of these faint flux measurements in the UV bandpasses, we
inserted simulated galaxies into the images, and performed an object
recovery test to measure the flux level where the signal-to-noise
typically approaches $\sim$1.  To derive 90\% confidence limits, we
inserted $\sim$60,000 simulated galaxy images representing a range of
total magnitudes ($24 \mbox{ mag}<\mbox{m}< 30 \mbox{ mag}$) and
half-light radii ($0.8''\,<\,\mbox{r}_{hl}\,<\,2.25''$) into each of
the UVIS mosaics, and measured the fraction of simulated galaxies
which were recovered by \sex, using the same \sex~configuration as
discussed above. The simulated galaxies were defined with an $r^{1/4}$
(``bulge'') or exponential SB profile (``disk''). From these
simulations, we estimated the 90\% recovery limits for simulated bulge
profiles with half-light radius of 1$\farcs0$ equal to F225W=26.5,
F275W=26.6, F336W=26.4, and F435W=26.7 mag, respectively. We
interpret ETGs with magnitudes fainter than these recovery limits as
1-$\sigma$ upper limits.

%I used bytscl in IDL in the ``scaling' below
In Figure 1, we provide ten-band postage stamp images of the
ETGs. These images are converted to flux units (nJy), and displayed
with the same linear gray-scale.  Each postage stamp measures
11$\farcs$2 on a side. In Table \ref{tab:fluxtab}, the typical
measured photometric uncertainties are small, and the typical
uncertainties associated with an m=25 mag galaxy in the ERS and
GOODS-S object catalog are$\colon$ 0.26 (F225W), 0.24 (F275W), 0.34
(F336W), 0.06 (F435W), 0.05 (F606W), 0.07 (F775W), 0.07 (F850LP), 0.11
(F098M), 0.07 (F125W), and 0.12 mag (F160W), respectively.  On
average, the measured photometric uncertainties are larger for the
UVIS bandpasses for this catalog. This can be largely attributed to
the lower telescope throughput, the lower intrinsic ETG flux, and the
shorter effective exposure time per pixel in each UVIS bandpass,
compared to the ACS and WFC3/IR instruments and image mosaics
\citep[see Fig. 1 in][]{W11}.

In Figure \ref{fig:qualities_p1}, the distribution of these galaxies
is plotted as a solid histogram, and the distribution of all available
spectroscopic redshifts in the CDF-S field is shown as a dashed
histogram.  The redshift peaks in this distribution at $z\,\approx$
0.53, 0.67, 0.73, 1.03, 1.09, 1.22, and 1.3 correspond to known
large-scale structures in the Chandra Deep Field-South (CDF-S)
\citep{G03, P09}.

%\subsection{SAbsolute Photometry}\label{sec:photometry} 
\subsection{Stellar Population Modeling}\label{sec:absmag} 
 
To measure absolute photometric properties of the ETGs, we first fit
the population synthesis models of \cite{BC03} (hereafter, BC03) to
the broad-band observed Optical-IR (F435W, F606W, F775W, F850LP,
F098M, F125W, F160W) SED of each ETG, applying the standard techniques
outlined in \cite{P01}.  The template library of models we used in
this fitting routine was generated for BC03 single burst stellar
templates defined by a Salpeter IMF, solar metallicity, no extinction
from dust, and with the star-formation history of the single burst
defined by an exponentially declining function, weighted by time
constant, $\tau$, i.e.,$\colon$
\begin{equation}\label{eqn:burstmod}
\psi(t)\!\propto\!e^{-t/\tau}
\end{equation}
These models were defined for a grid of time constants\footnote{We
  calculate models for N=15 values of $\tau$ defined with a stepsize
  of $\frac{max(\mbox{\tiny log} (\tau) )-min(\mbox{\tiny log} (\tau)
    )}{(N-1)}$=0.28.} ($-2.0 < \mbox{log}(\tau \mbox{[Gyr]}) <2.0$) and ages
(1$\times10^{8} < t \mbox{(yr)} <13.7\times10^{9}$).

We minimize the goodness-of-fit $\chi^2$ statistic between this
library of synthetic and observed fluxes to determine the optimal
model\footnote{We assume 7 degrees of freedom when determining the
  reduced $\chi^2$ statistic.}.  For each galaxy, we required the
best-fitting age parameter to not be greater than the age of the
Universe at the redshift of the ETG.  From this best-fitting template
the appropriate k-correction was then calculated, yielding an absolute
magnitude for each ETG in the $r^{\prime}$, Johnson V, and F606W
bandpasses (see Figure \ref{fig:absmaglim}).

We fit the observed SEDs in this limited filter selection to ensure
that the rest-frame optical and near-IR emission, which provides the
best indication to the majority (old) stellar populations extant in
the ETGs, is included in determining the best-fit spectral template.
Fitting single burst models to the limited SED also ensures that
rest-frame UV emission is largely {\it excluded} from the fitting.
Emission at UV wavelengths that arises from multiply-aged young ($< 1$
Gyr) {\it and} old ($\gg1$ Gyr) stellar populations or minority
UV-bright old stellar populations (see \S \ref{sec:intro}) may not be
well-fit with these single burst models.  A detailed modeling of these
complex stellar populations is beyond the intended scope of this work
and we will present a more detailed analysis of the stellar
populations extant in the ETGs in future work (Rutkowski et al., in
prep.).

Typical reduced $\chi^2$ determined from the SED-fitting were small
($\left<\chi^2_{\nu}\right>\!=\!1.1$) for ETGs at redshift
$z$\,\,\lsim\,\,0.6 (22 ETGs\footnote{we excluded a single
  poorly-fitted faint (M$_{V}=-$17), compact ETG (J033244.97-274309.1)
  from this set when calculating these averages}).  For this subset of
ETGs, the mean mass, age and log($\tau$ [Gyr]) were derived from the
broad-band Optical-IR SED fitting, and measured to be equal to
1.1$\times10^{10}$\Msol, 2.8$\times10^{9}$yr, and $-$0.3,
respectively.  At redshifts $z$\,\,\gsim\,\,$0.6$, the optical GOODS
filter set is sensitive to significant rest-frame near-UV emission,
the stellar source of which is not inherently well-described by the
models in the single burst library used in this analysis. Nonetheless,
the majority of ETGs at $z > 0.6$ are well-fitted by the single burst
models. Only 13 ETGs were ``failures'' (which we define as ETGs with
minimum $\chi_{\nu}^2 > 5$); 11 of these ETGs had spectroscopic
redshifts greater than 1.  At this high redshift, the F435W ACS is
sensitive to UV emission ($\lambda >$ 1800~\AA) exclusively. Excluding
these ``failures'', the mean mass, age and log($\tau$[Gyr]) of this
high redshift subset of the catalog are measured to be equal
to$\colon$ 9.2$\times10^{9}$\Msol, 2.8$\times10^9$yr, and $-$0.3,
respectively.  This SED analysis demonstrates that we have identified
a population of galaxies that are generally$\colon$ 1) `peaky'' (i.e.,
low $\tau$) in their star-formation history; 2) old (i.e., bulk
stellar population formed $>$1 Gyr ago); and 3) have stellar masses
comparable to the characteristic stellar mass of red galaxies
\citep[$\sim$10$^{11}$\Msol, see][cf.  Figure \ref{fig:absmaglim}
  below]{M09} at these redshifts. At low redshift, these
characteristics of the bulk stellar populations of galaxies are found
more often to be associated with early-type galaxies \citep[see,
  e.g.,][]{B04}.  Thus, we initially conclude that we have selected
galaxies representative of the class of intermediate to high mass
ETGs.

\subsection{Source Classification}\label{sec:GALFIT}

In this section, we discuss the morphological properties and
classification definitions for our ETGs.  Although optical colors were
not used to select or exclude ETGs, the color of the ETGs and/or
neighboring galaxies may aid in understanding the star-formation
history of the ETGs \cite[see][]{P10}.  In the following comments, the
definition of the ETG ``companion(s)'' is made strictly based on the
close proximity---in projection---of any two or more galaxies.
Furthermore, the classifications below are not mutually exclusive.
When galaxies meet the qualifications for multiple classifications, we
provide only the unique classifications and/or the most general
classification.  To qualitatively assess the primary ETG, its local
environment, and any possible companions, we inspect the GOODS
three-color, four panel 7\farcs0 x 7\farcs0 cutouts prepared for four
permutations\footnote{The cutouts are available online at$\colon$
  http$\colon$//archive.stsci.edu/eidol\_v2.php.  Specifically, the
  color cutouts are generated for $BVi'$, $BVz'$, $Bi'z'$, and $Vi'z'$
  colors; where $BVi'z'$ refer to the ACS F435W, F606W, F775W and
  F850LP filters, respectively} of the GOODS ACS F435W, F606W, F775W
and F850LP images.  In Figure \ref{fig:ofinterest}, we provide the
GOODS color cutouts of an ETG representative of each of the following
classes.

\begin{itemize}
\item Comment ``Comp.'' --- {\it ETG identified with companions}. We
  note cases where the colors of galaxies in the color cutouts are
  similar to our ETG.  This similarity could suggest that the
  galaxies are at a similar redshift, which would indicate that our
  ETG is a member of a small group.  We define two sub-classes $\colon$
\begin{itemize}
\item Comment ``LSB-Comp.'' --- {\it Low surface brightness companions}.  ETGs
  with low SB companions are candidates for future work to
  study the role of minor mergers in moderating star-formation in
  intermediate redshift ETGs. 

\item Comment ``b-Comp'' --- {\it ETG has blue companions(s)}.  We
  note objects which have projected companions that are bluer than the
  primary ETG in all color cutouts.  We speculate the enhanced
  emission in the F435W and/or F606W bands suggest that these possible
  companions have higher star-formation rates than the primary ETG.
\end{itemize}

 \item Comment ``d'' --- {\it ETG exhibits dust lane}.  The existence of
  a dust lane in an ETG has implications for the
  merger and star-formation history of the ETG. 

\item Comment ``c'' --- {\it Compact profile}.  These ETGs are notably
  more compact than the typical ETG in our sample, but were not
  identified as stars in \cite{R207} or \cite{W11}.

\item Comment ``DC'' --- {\it ETG has double core}.  This designation
  applies to a single ETG (J033210.76-274234.6), which
  appears to be an ongoing major merger of two spheroidal galaxies
  both of which have prominent central cores.

\item Comment ``S0'' --- {\it S0 candidate}. These ETGs show evidence
  for a bright core-bulge component, continuous light distribution,
  and an extended disk-like profile.

\item Comment ``VGM'' --- {\it Visual group member}.  These ETGs exist
  in a region of probable local overdensity of both early- and
  late-type galaxies, as well as low SB companions.

\end{itemize}
In Table \ref{tab:tab3} (Column 4), we comment on the morphology,
light-profile, and the environment of our ETGs. 

\subsection{Active Galactic Nuclei}\label{sec:AGN}

While {\it weak} active galactic nuclei (AGN) do not dominate the
optical SED of their host ETG \citep{K85}, these may contribute
emission in the UV spectrum of galaxies \citep[][and references
  therein]{vdB01}. Therefore, to understand the stellar sources of UV
flux in our ETGs, we must identify and account for weak AGN
contamination.  AGN were flagged in our catalog by matching the
positions of our ETGs to the X-ray \citep{G02,L10} and radio
\citep{M08} source catalogs.  In Table \ref{tab:tab3}, we denote X-ray
and/or radio sources as ``X*'' ``R*'', respectively (or ``XR*'' if the
ETG was identified in both catalogs). We give the AGN classifications
(Table \ref{tab:tab3}, Column 3) from \cite{S04}, which are based on
the X-ray luminosity, hardness ratios, and optical line-widths.  Nine
ETGs in the catalog were matched with sources in the X-ray and/or
radio.

\section{Catalog Completeness}\label{sec:completeness}

While our morphological selection criteria ensure our galaxies are
generally representative of the class of ETGs, the high spatial
resolution HST ACS and WFC3 imaging allows us to identify
sub-structures (e.g., dust lanes, which would be unresolved in
ground-based imaging) which ensures the catalog better captures the
morphological diversity of ETGs.  We also conclude that the ETG masses
are approximately equal to the characteristic stellar mass parameter,
$10^{10}\leq$M$_{\ast}$[\Msol]$\leq 10^{12}$ (see \S \ref{sec:absmag}
and Figure \ref{fig:absmaglim}).  Thus, our catalog is representative
of the class of intermediate to high mass ETGs.  Yet, our selection
criteria must necessarily imply that the catalog is an {\it
  incomplete} assessment of the ETGs in the ERS survey volume.  In
this section, we discuss the extent to which selection criteria
affects catalog completeness.

To quantify the number of ETGs we exclude from the catalog by
enforcing the selection criteria, we inspected a randomly-selected
region in the F160W mosaic with an area equal to $\sim$10\% of the
total area of the ERS field.  Therein, we identified $\sim$180
galaxies which have sufficiently high \snr, surface brightness, and
spatial extent to be morphologically-classified.  We visually
classified 45 of these galaxies as ETGs. Approximately 35\% (17 ETGs)
of these galaxies were included in the catalog. If we extrapolate this
observed fraction to the full ERS field, we estimate that there are
$\sim$1800 visually-classifiable galaxies in the field, of which
$\sim$280 galaxies {\it could} have been morphologically classified as
ETGs, but were not included in the catalog because they lacked
spectroscopically-confirmed redshifts\footnote{This sample likely
  includes morphological ETGs at $z$\,\,\gsim\,\,1.5, but we can
  reasonably assume \citep[cf.][]{B09,R10} that the number of ETGs at
  high redshift is a small fraction of the lower redshift ($z<1.5$)
  ETG population}. Thus, we can assume that as a result of the
spectroscopic redshift incompleteness, we are likely excluding a
population of ETGs approximately 2--3 times larger than the catalog in
\S\ref{sec:catalogueprod}.  We conclude that the requirement that each
ETG have a spectroscopically-confirmed redshift most strongly prevents
our definition of a complete sample of ETGs in the ERS field.

At low to intermediate redshift ($z$\,\,\lsim\,\,0.6), this
incompleteness disproportionately affects fainter galaxies. Large
(greater than a few square degrees) spectroscopic surveys of thousands
of galaxies have noted the paucity of low-luminosity (M$_{B}>-18$) red
galaxies \citep{Wie05,Wi06}.  This paucity can be partly attributed to
the difficulty associated with the measurement of spectroscopic
redshifts for galaxies with largely featureless spectra, with few (or
no) weak lines, that are common to quiescent faint galaxies. At high
redshift (z\,\,\gsim\,\,1), the measurement of spectroscopic redshifts
is increasingly more difficult because ground-based optical-near IR
spectrometers can not adequately constrain the 3648~\AA\, Balmer
break. Furthermore, color-based candidate galaxy selection at optical
wavelengths \citep[e.g., F775W$-$ F850LP$>0.6$ mag;][]{V08} will
intrinsically select high redshift ($z>1$) ETGs with bluer rest-frame
UV-optical colors.  As a result, these technical limitations and color
selections promote spectroscopic redshift incompleteness in surveys of
red galaxies at high redshift across the mass spectrum.

We can not rule out the effect of cosmic variance in the ERS field as
an additional source of incompleteness in the catalog.  \cite{Wi06}
measured the best-fit Schechter luminosity function parameters from
11,000 galaxies at z\,\,\lsim\,\,1 in the DEEP2 Survey \citep{D07},
and provide these results for two sub-populations, ``red'' and
``blue'' galaxies\footnote{``Red'' and ``blue'' galaxies are
  distinguished using the color criterion {\it U}$-${\it
    B}=$-$0.032(M$_B + 21.52$)+0.204}.  Assuming the best-fit
Schechter parameters for the ``red'' sample measured at $z=0.5$, we
estimate that the ERS survey volume defined for $0.4<z<0.7$ contains
only $\sim$1 luminous (M$_V < -22$ mag) ETG.

\section{Conversion to Rest-Frame UV-Optical Photometry}\label{sec:analysis}

Our measured rest-frame FUV to optical photometry provides a uniform
basis for studying the star-formation histories of our ETGs.  Here we
describe and apply an interpolation method to transform the observed
photometry to a ``standard'' set of FUV, NUV, $g^{\prime}$,
$r^{\prime}$, and Johnson V bandpasses. We select this filter set for
our analysis because there are now extensive references in the
literature which use the same set in the study of nearby and
low-redshift ETGs \citep[e.g.][and references
  therein]{K07,S07,R07,K10}.

First, we generated a library of hybrid spectral templates defined by
two instantaneous bursts of star-formation. The first burst occurred
at a fixed redshift ($z=3$) with a fixed solar metallicity
($Z_1=Z_{\odot}$). The second burst was modeled assuming a variable
stellar mass fraction ($f_2$), age of burst ($t_2$), dust content
characterized by $\mbox{E (B}-\mbox{V)}$, and metallicity ($Z_2$). The full parameter
space represented in the library of hybrid spectral templates is
provided in Table \ref{tab:params}. Next, we identified a set (Table
\ref{tab:proxy}) of WFC3 and ACS ``proxy'' filters that most closely
trace the bandpasses corresponding to the desired filters (FUV, NUV,
Johnson V, $g^{\prime}$, and $r^{\prime}$) at the relevant redshift.
Finally, we folded the library of spectral templates with this filter
set to determine the proxy and desired rest-frame colors.  To define a
general transformation function for each redshift, we fit a
second-order polynomial to the desired colors as a function of proxy
color.  These transformations can be considered as a generalized
k-correction.

The BC03 models are known to be an incomplete representation of the UV
spectrum of ETGs with ages $>$3~Gyr \citep[see][]{K07,K08} due to
their treatment of the UV upturn.  The UV energy distribution in the
BC03 models does not include the effects of extreme HB stars which are
expected to dominate this region of the spectrum of old stellar
populations. Therefore, we use a set of templates which are a hybrid
of BC03 models and \citet{Y99,Y03} for stellar populations of ages
\gsim\,3~Gyr. This hybrid library has been demonstrated
\citep{K07,K08} to fit observed ETGs across a large redshift range
($0<z<1$) with both young and old UV-bright stellar populations.

The rest-frame UV-optical colors are given for our sample in Table
\ref{tab:plotcolors}.  Following the convention of Table
\ref{tab:fluxtab}, we designate \sex~non-detections in the blue proxy
band as ``\nodata''.  ETGs not detected at or above the 1$\sigma$
completeness limits (\S \ref{sec:catalogueprod}) in the bandpasses
used to determine the rest-frame UV--optical colors are designated
``---''.

The \gmr\,and rest-frame Johnson V and $r^{\prime}$ apparent
magnitudes are also provided in Table \ref{tab:plotcolors}. The
\gmr\,colors were calculated using a method similar to the one outlined
above for converting the observed photometry to rest-frame UV-optical
colors, though the \gmr\,transformation function was calculated for a
different proxy filter set (see Table \ref{tab:proxy}). To calculate
the Johnson V and $r^{\prime}$ apparent magnitudes presented in Table
\ref{tab:plotcolors}, the F606W filter was fixed as the proxy filter
and a linear transformation function was fit to the proxy and desired
apparent magnitudes measured from the hybrid template library.
Typically, we measure the difference for any proxy-desired bandpass
pair to be small (less than 0.1 mag), but at higher redshifts, the
redshifting of the Balmer break in the spectrum through the bandpass
can produce larger offsets.  Particularly, between the F606W and
Johnson V bandpasses, these offsets can be as large as $\sim$1.1 mag

\section{Discussion of Rest-Frame Panchromatic Photometry}\label{sec:discusscolors}

In the upper panel of Figures \ref{fig:nuvcolors} and
\ref{fig:fuvcolors}, the apparent colors and associated photometric
uncertainties bars are plotted for reference.  We calculate these
colors by simply differencing the apparent magnitudes in the proxy
bandpasses for each redshift bin (see Table \ref{tab:proxy}). We show
in the lower panel of Figures \ref{fig:nuvcolors} and
\ref{fig:fuvcolors} the \nuvv\, and \fuvv\, colors, which are
calculated using the best-fit transformation function from \S
\ref{sec:analysis}. Each ETG is plotted with its measured photometric
and systematic (i.e., associated with the transformation)
uncertainties. An asterisk indicates that the ETG was identified by
the radio or X-ray surveys of the CDF-S (see \S \ref{sec:AGN}).  Since
many nearby ellipticals have strong internal UV-optical color
gradients \citep{O98,Je09}, we show the integrated \nuvv\, and \fuvv\,
colors from the GALEX UV Atlas of Nearby Galaxies \citep{G07} for NGC
221 (M32), 1399, and 1404 (triangles).  We select these specific ETGs,
since they well-represent the evolved red sequence of ETGs in the
local Universe.

We also show the rest-frame colors of three model galaxies, generated
using the BC03 single burst templates (see \S\ref{sec:absmag}) for
three star-formation histories defined by Equation \ref{eqn:burstmod}
for log($\tau$[Gyr])$\simeq$ 1.1 (blue),$-$0.3 (green), and $-$2.0
(red). For each model, we assume solar metallicity, a Salpeter IMF, no
dust, and formation redshift $z_f=4.0$.  The time since $z_f$ is
plotted as the upper abcissa in each Figure.  This formation redshift
can be considered to represent the {\it effective} start of
star-formation in ETGs, because it is approximately halfway in cosmic
time between the start of cosmic star-formation at $z \simeq 10$
\citep{Ko11} and the (broad) peak of the cosmic star-formation history
at $z\simeq 2$ \citep{M98}.

%We show the distribution of \gmr\, colors in Figure \ref{fig:optcmd}
%with respect to the absolute $r^{\prime}$ magnitudes calculated for
%our sample. The \gmr\, colors of ETGs in Figure \ref{fig:optcmd} span
%\,\,\lsim\,\,1 mag.  The observed and transformed rest-frame optical
%photometry is unaffected by large amplitude systematic or photometric
%uncertainties, since this color distribution of ETGs is bounded by
%reasonable population synthesis models (see Figures
%\ref{fig:nuvcolors} and \ref{fig:fuvcolors}).  Therefore, the observed
%bimodality in the \gmr\, colors, which distinguishes luminous red ETGs
%from lower luminosity blue ETGs, seen in our sample is not an
%artifact.  However, the optical colors are a poor discriminator of
%recent star-formation history of ETGs, and we limit the discussion of
%the star-formation history of our ETGs to the rest-frame UV-optical
%colors, which span a range of $\sim$4--6 mag (Figures
%\ref{fig:nuvcolors}, \ref{fig:fuvcolors}, and \ref{fig:UVOPTcolors}).

Over the surveyed redshift range, Figures \ref{fig:nuvcolors} and
\ref{fig:fuvcolors} show that the majority of ETGs have UV--optical
colors {\it no bluer} than the log($\tau$[Gyr]) $\simeq 1.1$
single burst model, suggesting that these ETGs have {\it not}
undergone a significant, recent star-formation event which would be
identified by \nuvv\,\,\lsim\,\,$-$1.0 mag.  Secondly, only a minority
of ETGs can be well-described by a quiescent, instantaneous
star-formation history that assumes a high formation redshift
($z_f=4.0$).  Finally, we note that the ``red envelope'' of the
\fuvv~and \nuvv~colors, the latter is most sensitive to recent
star-formation, remains constant across the intermediate redshift
$z<0.5$.

Furthermore, few (1-2) ETGs at intermediate redshift
($z$\,\,\lsim\,\,0.6) have measured rest-frame colors as {\it red} as
those observed for the {\it strongest} UV upturn galaxy in the local
Universe, NGC 1399.  In \S\ref{sec:absmag} and
\S\ref{sec:completeness} we showed that our selection criteria
(unavoidably) defined a catalog that is deficient in bright (M$<-22$
mag) ETGs.  If we assume a stellar mass-to-light ratio of these bright
ETGs approximately equal to unity, then the masses of these ETGs are
greater than $\sim10^{11}$ \Msol, e.g., early-type brightest cluster
galaxies (BCGs, with stellar masses $10^{10.5}<$M$_{\ast}\,$[\Msol]$ <
10^{11.5}$; see von den Linden 2007) and cD-type galaxies
(M$_{\ast}$[\Msol]\,\,\gsim\,\,$10^{12}$), which is consistent with
the results presented with Figure \ref{fig:absmaglim}.  From theory
and observations of the UVX in low-redshift ETGs, we expect that an
optimal sample for the study of the UVX at intermediate redshift would
include the oldest \citep[\,\,\gsim\,\,6 Gyr,][]{T96} and brightest
ETGs.  The latter is due to the observation that the strength of the
UVX is positively correlated with host galaxy luminosity \citep{B88}.
Thus, the analysis in \S\ref{sec:absmag} and \S\ref{sec:completeness}
suggested, and Figures \ref{fig:nuvcolors} and \ref{fig:fuvcolors}
confirm, that our catalog is deficient in ETGs of the variety
best-suited for the analysis of the UVX across cosmic time.

Some ETGs are likely to contain UVX stellar populations, but these
ETGs are likely to be dominated in the UV by emission from young, not
old, stellar populations.  Any future work that seeks to model the UVX
evolution over cosmic time using our catalog must do so with caution,
and take care to include multiple stellar populations in the SED
analysis.

At higher redshift ($z$\,\,\gsim\,\,$0.5$), the rest-frame UV-optical
colors are uniquely sensitive to recent star-formation, because the
older evolved stellar populations do not contribute significantly to
the UV SED of the host ETGs \citep{FS00,Yi05,Ka07,Ka09}.  If the
measured rest-frame colors of the ETGs are compared with the results
from \cite{Yi05} and \cite{K07}, these colors indicate a wide range of
star-formation histories ranging from continuous star-formation
(log($\tau$[Gyr])$=1.1$) to nearly-quiescent
(log($\tau$[Gyr])$=-0.3$), assuming a uniform formation redshift of
the majority stellar population.  %Conversely, at lower redshift
%($0.35$\,\,\lsim\,\,$z$\,\,\lsim\,\,$0.5$) UVX stellar populations may
%influence the rest-frame UV--optical color of ETGs. 

In Figure \ref{fig:UVOPTcolors} we show the rest-frame UV-optical
color-color diagram for the ETGs that are {\it brighter} than the
simulated 1$\sigma$ 90\% recovery limits (see \S
\ref{sec:catalogueprod}), with photometric and systematic
uncertainties included. Furthermore, we color-code the data to
correspond with the redshift of the ETG; the color scheme is defined
in Figure \ref{fig:UVOPTcolors}.  In Figure \ref{fig:UVOPTcolors}, the
\gmr\, colors of ETGs span \,\,\lsim\,\,1 mag. The \gmr\, colors of
the ETGs are also well-distributed as a function of redshift and
color, which indicates that UV-optical transformation function defined
in \S \ref{sec:analysis} is not affected by any large systematic
uncertainties. In Figure \ref{fig:optcmd}, we show the \gmr\, colors
of the ETGs with respect to the absolute $r^{\prime}$
magnitudes. Since the color distribution is bounded by reasonable
population synthesis models (Figures \ref{fig:nuvcolors} and
\ref{fig:fuvcolors}) and the the rest-frame optical photometry is
unaffected by large amplitude systematic or photometric uncertainties
(Figure \ref{fig:UVOPTcolors}), the bimodality in the \gmr\, colors
which distinguishes luminous red ETGs from lower luminosity blue ETGs
present in the Figure is not an artifact.  Though the optical colors
of ETGs are a poor discriminator of recent star-formation history of
ETGs, the distribution of rest-frame optical colors supports the
previous conclusion that there exists a diversity in the
star-formation histories of these ETGs.

Finally, in Figures \ref{fig:nuvcolors} and \ref{fig:fuvcolors}, we
note a transition from peaky star-formation histories for the highest
redshift ($z$\,\,\gsim\,\,$1$) ETGs to a more gradual and sustained
star-formation in ETGs at low to intermediate redshifts
($z$\,\,\lsim\,\,$1$) across the entire surveyed redshift range.
Specifically, at high redshift ($z$\,\,\gsim\,\,1), many ETGs appear
to cluster near to the log($\tau$[Gyr])$=-0.3$ curve, whereas no low
to intermediate redshift ($z$\,\,\lsim\,\,1) ETGs exist on this curve
and few have (FUV$-$V)$_{rest}$\,\,\lsim\,\,6.  But we do not make an
interpretation of this trend as it may not represent a physical
transition. In \S\ref{sec:completeness} we outlined a number of biases
implicit in optical-near IR spectroscopic redshift surveys that
specifically select {\it against} red ETGs, both at intermediate and
high redshift.  The paucity of red ETGs at low to intermediate
redshift may be partially attributed to the spectroscopic redshift
incompleteness, and as a result, this transition would not indicate a
physical evolution in the star-formation histories of these ETGs.  To
determine the significance of this apparent transition we require a
catalog of ETGs selected in such a way that the biases introduced by
spectroscopic redshift incompleteness are minimized. To produce this
catalog, future selection and spectroscopic observations of
intermediate and high redshift ETGs in the ERS field must be made in
the near-IR.

\section{Summary}\label{sec:discussion}

HST WFC3 provides novel insight into more than $\sim$50\% of the epoch
of cosmic star-formation history at redshift $z$\,\,\gsim\,\,$0.35$,
particularly due to its unique panchromatic coverage and sensitivity
to rest-frame UV emission.  In this first publication in a series, we
present a ten-band (from HST ACS and WFC3, covering wavelengths
between 2200~\AA\,--1.6$\mu$m) catalog of visually-selected ETGs with
spectroscopically-confirmed redshifts (\rsrang). Results from this
work extends conclusions drawn from studies of lower redshifts
($z$\,\,\lsim\,\,$0.35$) ETG populations with GALEX and SDSS
\citep{Y04,K07}.  In particular, we have found significant diversity
in the UV--optical colors of this population of ETGs, which is likely
the result of recent star-formation.

Though the GOODS-S field contains 1000s of spectroscopic redshifts
acquired from multiple, extensive ground-based surveys, spectroscopic
redshifts have been measured for only a fraction of the
morphologically-classified ETGs in the ERS field. We demonstrate that
the catalog is {\it representative} of the class of ETGs, but is not a
{\it complete} catalog of ETGs in the field.  A panchromatic analysis
of a mass-complete sample of intermediate to high
($z$\,\,\gsim\,\,0.6) redshift ETGs would be tremendously beneficial
for understanding the star-formation histories of ETGs at an epoch
during which the Universal star-formation rate is declining from the
peak rate at $z\simeq$2.  To produce such a sample for the ERS
catalog, for which HST WFC3 imaging has provided unprecedented access
to the rest-frame UV-optical SEDs of the ETGs, a ground-based
spectroscopic campaign specifically designed to measure redshifts for
the remainder of ETGs selected on the basis of this HST near-IR
photometry is necessary.

In a sequel paper, we will present a complete analysis of the stellar
population(s) extant in the ETGs using the panchromatic
(UV-optical-near IR) broad-band photometry, which will extend the
discussion in \S\ref{sec:absmag} to include a consideration of the
local environment parameters (\S \ref{sec:GALFIT}) and new analysis of
the ETG light profiles \citep[i.e., measured S\'{e}rsic index,
  blue/red-Core ETGs, see e.g.,][]{S10}.  This analysis will be
specifically focused on investigating the role that small galaxy group
dynamics and minor mergers may play in modifying the star-formation
history of the ETGs.  Until a successor observatory to HST is
available, with comparable UV sensitivity and spatial resolution,
these public ERS, and similar HST UVIS and IR, data represent the best
opportunity for this panchromatic study of the evolution of ETGs and
their stellar populations at intermediate redshift.

%Do we want to include this? 10/17
%Increasing the catalog size by either a) relaxing the
%morphological criteria or b) acquiring more spectroscopic redshifts
%for galaxies in the ERS field in an attempt to include more massive
%galaxies would not improve the the number of viable massive ETGs
%candidates required for a study of the intermediate redshift evolution
%of the UVX.  END KEEP

%\section{Conclusion}

\section{Acknowledgments}
This paper is based on Early Release Science observations made by the
WFC3 Scientific Oversight Committee.  We thank the Director of the
Space Telescope Science Institute for awarding Director's
Discretionary time for this program.  We thank an anonymous referee
for comments and suggestions that have improved the scientific outline
of this manuscript.  Finally, we are deeply indebted to the crew of
STS-125 for refurbishing and repairing HST. Support for program
\#11359 was provided by NASA through a grant from the Space Telescope
Science Institute, which is operated by the Association of
Universities for Research Inc., under NASA contract NAS 5-26555.

%\clearpage
%\newpage

\appendix 
\section{Red-Leak}\label{sec:appendix}

Ultraviolet observations of objects with weak UV emission and red SEDs
may be prone to significant red-leaks, where long-wavelength photons
can be incorrectly counted as UV photons.  Despite significant efforts
by the WFC3 instrument team to minimize red-leaks, it is important to
understand this effect on the photometry of a typical ETG.

We measure the red-leak associated with each of the WFC3 UVIS filter
response curves (see Figure \ref{fig:redleakcurve}) for model SEDs
defined over a range of redshift \rsrang by measuring the ratio of
flux at $\lambda > 4000\mbox{\AA}$ to the total$\colon$
\begin{equation}
{\cal R}=\frac{\left<F_{\lambda>4000~\mbox{\AA}}\right>}{\left<F_{\lambda}\right>}=\frac{\int_0^{\nu_0}F_{\nu}T_{\nu}d\nu/\nu}{\int_{0}^{\infty} F_{\nu}T_{\nu}d\nu/\nu}
\end{equation}
where $\nu_{0}=c/4000~\mbox{\AA}$, $F_{\nu}$ represents the flux per
unit frequency associated with the model spectrum, and T$_{\nu}$ is
the filter response\footnote{The response curves are provided by the
  {\it syn}thetic {\it phot}ometry IRAF package {\it synphot}, which
  was prepared by STSCI for the HST instrument suite; more details are
  available online at
  www.stsci.edu/resources/software\_hardware/stsdas/synphot}.

Because the UV emission profile of an homogeneously old ETG model can
vary significantly with the models of the UVX stellar populations (see
\S \ref{sec:completeness}), we measured the effect of filter red-leak
for two template spectra.  We used the \cite{CWW80} Elliptical and a
BC03 exponentially-declining star-formation template with
log($\tau$[Gyr])$=-2.0$ and an absolute age of $\sim$12 Gyr (even when
current cosmology dictates that such an old model is infeasible) to
define our model SEDs.  We consider the grid of model spectra for the
redshift range, \rsrang, and provide the maximum red-leak measured for
this grid in Table \ref{tab:redleaktab}.  We conclude that the filter
red-leak in this redshift range is never larger than 3.5\%, even for
the bluest F225W filter.

\begingroup
\tiny
\begin{landscape}
\begin{longtable}{cccccccccccccc}
\caption{Early-Type Galaxies Catalog, Measured Photometry}\\
\hline \hline
\multicolumn{1}{c}{GOODS ID} &
\multicolumn{1}{c}{R.A.} &
\multicolumn{1}{c}{Dec.} &
\multicolumn{1}{c}{F225W} &
\multicolumn{1}{c}{F275W} &
\multicolumn{1}{c}{F336W} &
\multicolumn{1}{c}{F435W} &
\multicolumn{1}{c}{F606W} &
\multicolumn{1}{c}{F775W} &
\multicolumn{1}{c}{F850LP} & 
\multicolumn{1}{c}{F098M} &
\multicolumn{1}{c}{F125W} &
\multicolumn{1}{c}{F160W} &
\multicolumn{1}{c}{Redshift} \\
\multicolumn{1}{c}{} &
\multicolumn{1}{c}{} &
\multicolumn{1}{c}{} &
\multicolumn{1}{c}{$\Delta$m} &
\multicolumn{1}{c}{$\Delta$m} &
\multicolumn{1}{c}{$\Delta$m} &
\multicolumn{1}{c}{$\Delta$m} &
\multicolumn{1}{c}{$\Delta$m} &
\multicolumn{1}{c}{$\Delta$m} &
\multicolumn{1}{c}{$\Delta$m} &
\multicolumn{1}{c}{$\Delta$m} &
\multicolumn{1}{c}{$\Delta$m} &
\multicolumn{1}{c}{$\Delta$m} &
\multicolumn{1}{c}{} \\ \hline
\endfirsthead

\multicolumn{14}{c}{Table 1$\colon$ETG Catalog, Measured Phot. (Continued)}\\%[0.7ex]
\hline \hline
\multicolumn{1}{c}{GOODS ID} &
\multicolumn{1}{c}{R.A.} &
\multicolumn{1}{c}{Dec.} &
\multicolumn{1}{c}{F225W} &
\multicolumn{1}{c}{F275W} &
\multicolumn{1}{c}{F336W} &
\multicolumn{1}{c}{F435W} &
\multicolumn{1}{c}{F606W} &
\multicolumn{1}{c}{F775W} &
\multicolumn{1}{c}{F850LP} & 
\multicolumn{1}{c}{F098M} &
\multicolumn{1}{c}{F125W} &
\multicolumn{1}{c}{F160W} &
\multicolumn{1}{c}{Redshift} \\
\multicolumn{1}{c}{} &
\multicolumn{1}{c}{} &
\multicolumn{1}{c}{} &
\multicolumn{1}{c}{$\Delta$m} &
\multicolumn{1}{c}{$\Delta$m} &
\multicolumn{1}{c}{$\Delta$m} &
\multicolumn{1}{c}{$\Delta$m} &
\multicolumn{1}{c}{$\Delta$m} &
\multicolumn{1}{c}{$\Delta$m} &
\multicolumn{1}{c}{$\Delta$m} &
\multicolumn{1}{c}{$\Delta$m} &
\multicolumn{1}{c}{$\Delta$m} &
\multicolumn{1}{c}{$\Delta$m} &
\multicolumn{1}{c}{} \\ \hline
\endhead

%\multicolumn{3}{l}{{Continued on Next Page\ldots}}
%\endfoot
\hline \hline
\multicolumn{14}{l}{\textbf{Notes-} Objects detected in the ERS F160W mosaic but not measured by \sex~ in the ERS or GOODS  mosaics are designated ``\nodata''. 1$\sigma$ 90\% recovery limits}\\
\multicolumn{14}{l}{~~~~~~~~~~~~~were calculated in Section \ref{sec:catalogueprod} to be equal to F225W=26.5, F275W=26.6, F336W=26.4, \& F435W=26.7 mag.  Detections fainter }\\
\multicolumn{14}{l}{~~~~~~~~~~~~~than these recovery limits are designated ``---''.   Measured photometric uncertainties, $\Delta$m, are provided for each ETG.}
\endlastfoot
%%\nodata = aaaa;---=bbbb

J033202.71-274310.8&03$\colon$32$\colon$02.71&-27$\colon$43$\colon$10.87&23.07&23.30&21.62&20.24&18.82&18.28&18.01&17.91&17.69&17.50&0.493\\[-1.0pt]&&&0.17&0.20&0.06&0.00&0.00&0.00&0.00&0.00&0.00&0.00&\\[-1.0pt]
J033203.29-274511.4&03$\colon$32$\colon$03.29&-27$\colon$45$\colon$11.47&26.00&25.89&25.50&25.13&24.42&23.78&23.61&23.44&23.37&23.18&0.542\\[-1.0pt]&&&0.35&0.29&0.27&0.06&0.02&0.02&0.02&0.02&0.02&0.02&\\[-1.0pt]
J033205.09-274514.0&03$\colon$32$\colon$05.09&-27$\colon$45$\colon$14.03&24.88&24.92&24.80&24.51&23.94&23.23&22.99&22.98&22.74&22.59&0.763\\[-1.0pt]&&&0.18&0.17&0.21&0.03&0.02&0.02&0.01&0.02&0.01&0.01&\\[-1.0pt]
J033205.13-274351.0&03$\colon$32$\colon$05.13&-27$\colon$43$\colon$51.05&24.28&24.04&24.06&23.84&23.36&22.63&22.45&22.41&22.26&22.18&0.806\\[-1.0pt]&&&0.09&0.06&0.09&0.02&0.01&0.01&0.01&0.01&0.01&0.01&\\[-1.0pt]
J033206.27-274536.7&03$\colon$32$\colon$06.27&-27$\colon$45$\colon$36.68&---&---&25.62&25.67&23.00&21.54&21.04&20.85&20.44&20.06&0.669\\[-1.0pt]&&&---&---&0.72&0.20&0.01&0.01&0.01&0.00&0.00&0.00&\\[-1.0pt]
J033206.48-274403.6&03$\colon$32$\colon$06.48&-27$\colon$44$\colon$03.68&---&25.78&26.07&---&24.43&23.03&22.11&21.85&21.41&21.03&0.958\\[-1.0pt]&&&---&0.46&0.83&---&0.04&0.02&0.01&0.01&0.00&0.00&\\[-1.0pt]
J033206.81-274524.3&03$\colon$32$\colon$06.81&-27$\colon$45$\colon$24.37&25.61&---&26.37&26.12&25.42&23.91&23.18&22.75&22.03&21.65&1.373\\[-1.0pt]&&&0.38&---&0.94&0.20&0.09&0.04&0.02&0.02&0.01&0.01&\\[-1.0pt]
J033207.55-274356.6&03$\colon$32$\colon$07.55&-27$\colon$43$\colon$56.68&---&---&---&---&25.13&23.76&22.81&22.40&21.88&21.50&1.370\\[-1.0pt]&&&---&---&---&---&0.14&0.06&0.03&0.02&0.01&0.01&\\[-1.0pt]
J033207.95-274212.1&03$\colon$32$\colon$07.95&-27$\colon$42$\colon$12.18&26.47&---&---&26.46&24.96&23.64&23.17&23.01&22.68&22.39&0.740\\[-1.0pt]&&&0.66&---&---&0.23&0.05&0.02&0.02&0.01&0.01&0.01&\\[-1.0pt]
J033208.41-274231.3&03$\colon$32$\colon$08.41&-27$\colon$42$\colon$31.37&26.31&\nodata&25.99&24.83&22.87&21.74&21.34&21.19&20.85&20.53&0.540\\[-1.0pt]&&&0.94&\nodata&0.85&0.09&0.01&0.01&0.01&0.00&0.00&0.00&\\[-1.0pt]
J033208.45-274145.9&03$\colon$32$\colon$08.44&-27$\colon$41$\colon$45.95&25.06&25.20&24.57&25.15&23.55&22.00&21.44&21.22&20.81&20.42&0.730\\[-1.0pt]&&&0.41&0.43&0.25&0.14&0.03&0.01&0.01&0.01&0.00&0.00&\\[-1.0pt]
J033208.53-274217.7&03$\colon$32$\colon$08.53&-27$\colon$42$\colon$17.78&24.10&25.02&24.60&24.35&22.70&21.27&20.76&20.57&20.16&19.80&0.730\\[-1.0pt]&&&0.22&0.46&0.43&0.10&0.02&0.01&0.01&0.00&0.00&0.00&\\[-1.0pt]
J033208.55-274231.1&03$\colon$32$\colon$08.55&-27$\colon$42$\colon$31.14&26.31&26.23&25.93&26.55&25.10&23.79&23.52&23.31&23.08&22.83&0.509\\[-1.0pt]&&&0.76&0.65&0.67&0.34&0.07&0.04&0.04&0.02&0.01&0.01&\\[-1.0pt]
J033208.65-274501.8&03$\colon$32$\colon$08.65&-27$\colon$45$\colon$01.84&---&---&26.32&25.31&23.11&21.62&20.98&20.84&20.50&20.20&0.873\\[-1.0pt]&&&---&---&1.01&0.10&0.01&0.01&0.00&0.00&0.00&0.00&\\[-1.0pt]
J033208.90-274344.3&03$\colon$32$\colon$08.90&-27$\colon$43$\colon$44.36&25.39&25.51&25.23&24.60&23.35&22.77&22.59&22.52&22.38&22.23&0.580\\[-1.0pt]&&&0.29&0.30&0.32&0.05&0.01&0.01&0.01&0.01&0.01&0.01&\\[-1.0pt]
J033209.09-274510.8&03$\colon$32$\colon$09.09&-27$\colon$45$\colon$10.85&25.83&25.57&25.27&25.38&24.54&24.24&23.97&24.00&24.01&23.92&0.401\\[-1.0pt]&&&0.27&0.20&0.21&0.06&0.02&0.03&0.03&0.03&0.02&0.02&\\[-1.0pt]
J033209.19-274225.6&03$\colon$32$\colon$09.19&-27$\colon$42$\colon$25.66&---&---&---&25.80&23.57&22.10&21.61&21.38&21.00&20.64&0.720\\[-1.0pt]&&&---&---&---&0.22&0.02&0.01&0.01&0.01&0.00&0.00&\\[-1.0pt]
J033210.04-274333.1&03$\colon$32$\colon$10.04&-27$\colon$43$\colon$33.15&26.06&25.23&25.55&25.30&23.74&22.15&21.14&20.87&20.34&19.95&1.009\\[-1.0pt]&&&1.11&0.48&0.86&0.19&0.03&0.01&0.01&0.00&0.00&0.00&\\[-1.0pt]
J033210.12-274333.3&03$\colon$32$\colon$10.12&-27$\colon$43$\colon$33.37&---&\nodata&\nodata&26.46&24.69&23.20&22.23&21.91&21.44&21.06&1.009\\[-1.0pt]&&&---&\nodata&\nodata&0.26&0.04&0.02&0.01&0.01&0.00&0.00&\\[-1.0pt]
J033210.16-274334.3&03$\colon$32$\colon$10.16&-27$\colon$43$\colon$34.38&\nodata&\nodata&\nodata&25.84&24.25&22.61&21.65&21.45&20.91&20.53&0.990\\[-1.0pt]&&&\nodata&\nodata&\nodata&0.34&0.06&0.02&0.01&0.01&0.00&0.00&\\[-1.0pt]
J033210.76-274234.6&03$\colon$32$\colon$10.76&-27$\colon$42$\colon$34.65&23.46&23.33&23.05&21.73&19.89&19.00&18.64&18.50&18.17&17.85&0.419\\[-1.0pt]&&&0.15&0.12&0.13&0.01&0.00&0.00&0.00&0.00&0.00&0.00&\\[-1.0pt]
J033210.86-274441.2&03$\colon$32$\colon$10.86&-27$\colon$44$\colon$41.24&26.45&26.17&---&---&24.69&23.43&22.98&22.88&22.55&22.23&0.676\\[-1.0pt]&&&0.66&0.47&---&---&0.03&0.02&0.01&0.01&0.01&0.01&\\[-1.0pt]
J033211.21-274533.4&03$\colon$32$\colon$11.21&-27$\colon$45$\colon$33.44&26.31&---&---&26.07&24.62&23.18&22.16&21.79&21.32&20.99&1.215\\[-1.0pt]&&&0.72&---&---&0.20&0.04&0.02&0.01&0.01&0.00&0.00&\\[-1.0pt]
J033211.61-274554.1&03$\colon$32$\colon$11.61&-27$\colon$45$\colon$54.13&---&25.50&25.71&25.80&24.15&22.73&21.75&21.38&20.93&20.55&1.039\\[-1.0pt]&&&---&0.41&0.67&0.21&0.03&0.01&0.01&0.01&0.00&0.00&\\[-1.0pt]
J033212.20-274530.1&03$\colon$32$\colon$12.19&-27$\colon$45$\colon$30.04&25.04&24.93&24.44&24.28&22.32&21.06&20.64&20.50&20.17&19.86&0.676\\[-1.0pt]&&&0.48&0.41&0.34&0.09&0.01&0.01&0.00&0.00&0.00&0.00&\\[-1.0pt]
J033212.31-274527.4&03$\colon$32$\colon$12.31&-27$\colon$45$\colon$27.43&\nodata&25.53&25.03&25.57&23.46&22.17&21.77&21.61&21.29&21.01&0.680\\[-1.0pt]&&&\nodata&0.45&0.40&0.18&0.02&0.01&0.01&0.01&0.00&0.00&\\[-1.0pt]
J033212.47-274224.2&03$\colon$32$\colon$12.47&-27$\colon$42$\colon$24.24&---&---&25.50&24.87&23.04&22.19&21.92&21.78&21.55&21.30&0.417\\[-1.0pt]&&&---&---&0.41&0.06&0.01&0.01&0.01&0.01&0.00&0.00&\\[-1.0pt]
J033214.26-274254.2&03$\colon$32$\colon$14.26&-27$\colon$42$\colon$54.28&---&\nodata&26.19&26.38&24.96&23.90&23.48&23.34&22.94&22.68&0.814\\[-1.0pt]&&&---&\nodata&0.62&0.18&0.05&0.03&0.03&0.02&0.01&0.01&\\[-1.0pt]
J033214.45-274456.6&03$\colon$32$\colon$14.45&-27$\colon$44$\colon$56.58&---&---&25.39&---&24.81&23.37&22.95&22.80&22.43&22.14&0.737\\[-1.0pt]&&&---&---&0.37&---&0.05&0.02&0.02&0.02&0.01&0.01&\\[-1.0pt]
J033214.65-274136.6&03$\colon$32$\colon$14.65&-27$\colon$41$\colon$36.56&25.75&25.67&---&26.12&25.42&23.84&23.00&22.51&21.77&21.33&1.338\\[-1.0pt]&&&0.63&0.55&---&0.29&0.13&0.05&0.03&0.02&0.01&0.01&\\[-1.0pt]
J033214.68-274337.1&03$\colon$32$\colon$14.69&-27$\colon$43$\colon$37.10&26.49&25.42&25.28&25.08&24.31&23.44&22.95&22.88&22.58&22.42&0.910\\[-1.0pt]&&&0.70&0.24&0.28&0.05&0.02&0.02&0.01&0.02&0.01&0.01&\\[-1.0pt]
J033214.73-274153.3&03$\colon$32$\colon$14.73&-27$\colon$41$\colon$53.32&\nodata&26.07&---&25.07&23.40&22.56&22.22&22.09&21.85&21.60&0.490\\[-1.0pt]&&&\nodata&0.51&---&0.07&0.01&0.01&0.01&0.01&0.01&0.01&\\[-1.0pt]
J033214.78-274433.1&03$\colon$32$\colon$14.78&-27$\colon$44$\colon$33.11&---&---&---&---&24.52&23.11&22.63&22.41&21.93&21.57&0.736\\[-1.0pt]&&&---&---&---&---&0.04&0.02&0.01&0.01&0.01&0.00&\\[-1.0pt]
J033214.83-274157.1&03$\colon$32$\colon$14.83&-27$\colon$41$\colon$57.13&---&---&---&25.18&23.54&22.34&21.96&21.84&21.53&21.25&0.680\\[-1.0pt]&&&---&---&---&0.09&0.02&0.01&0.01&0.01&0.00&0.00&\\[-1.0pt]
J033215.98-274422.9&03$\colon$32$\colon$15.99&-27$\colon$44$\colon$22.96&25.60&25.80&25.63&24.48&22.96&21.78&21.41&21.28&21.00&20.75&0.735\\[-1.0pt]&&&0.68&0.75&0.84&0.06&0.01&0.01&0.01&0.01&0.00&0.00&\\[-1.0pt]
J033216.19-274423.1&03$\colon$32$\colon$16.20&-27$\colon$44$\colon$23.14&25.48&25.86&25.63&24.82&23.25&22.45&22.15&22.08&21.81&21.63&0.419\\[-1.0pt]&&&0.50&0.65&0.70&0.06&0.01&0.01&0.01&0.01&0.01&0.01&\\\hline\\
J033217.11-274220.9&03$\colon$32$\colon$17.11&-27$\colon$42$\colon$20.90&26.31&26.17&24.69&24.99&25.15&25.11&25.26&25.09&24.32&25.25&1.240\\[-1.0pt]&&&0.37&0.32&0.11&0.04&0.04&0.06&0.09&0.06&0.02&0.06&\\[-1.0pt]
J033217.12-274407.7&03$\colon$32$\colon$17.12&-27$\colon$44$\colon$07.73&\nodata&---&---&26.51&24.59&23.14&22.66&22.53&22.15&21.83&0.730\\[-1.0pt]&&&\nodata&---&---&0.25&0.03&0.01&0.01&0.01&0.01&0.01&\\[-1.0pt]
J033217.14-274303.3&03$\colon$32$\colon$17.14&-27$\colon$43$\colon$03.30&24.16&23.92&23.37&23.07&21.69&20.81&20.53&20.37&20.15&19.81&0.556\\[-1.0pt]&&&0.13&0.10&0.08&0.02&0.00&0.00&0.00&0.00&0.00&0.00&\\[-1.0pt]
J033217.49-274436.7&03$\colon$32$\colon$17.49&-27$\colon$44$\colon$36.73&25.43&25.08&24.78&24.68&23.01&21.80&21.36&21.23&20.89&20.60&0.734\\[-1.0pt]&&&0.67&0.45&0.45&0.11&0.02&0.01&0.01&0.01&0.00&0.00&\\[-1.0pt]
J033217.91-274122.7&03$\colon$32$\colon$17.91&-27$\colon$41$\colon$22.70&25.77&26.44&\nodata&26.65&24.48&22.96&22.04&21.73&21.24&20.87&1.039\\[-1.0pt]&&&0.62&1.08&\nodata&0.48&0.05&0.02&0.01&0.01&0.00&0.00&\\[-1.0pt]
J033218.31-274233.5&03$\colon$32$\colon$18.31&-27$\colon$42$\colon$33.52&23.96&23.80&24.72&23.51&21.44&20.37&19.99&19.88&19.54&19.24&0.519\\[-1.0pt]&&&0.20&0.16&0.47&0.04&0.00&0.00&0.00&0.00&0.00&0.00&\\[-1.0pt]
J033218.64-274144.4&03$\colon$32$\colon$18.64&-27$\colon$41$\colon$44.43&---&26.23&\nodata&---&27.29&25.56&24.66&24.11&23.38&23.01&1.325\\[-1.0pt]&&&---&0.46&\nodata&---&0.35&0.12&0.07&0.03&0.01&0.01&\\[-1.0pt]
J033218.74-274415.8&03$\colon$32$\colon$18.73&-27$\colon$44$\colon$15.90&25.15&25.03&24.90&24.20&22.28&21.21&20.86&20.76&20.45&20.14&0.509\\[-1.0pt]&&&0.41&0.34&0.40&0.06&0.01&0.01&0.00&0.00&0.00&0.00&\\[-1.0pt]
J033219.02-274242.7&03$\colon$32$\colon$19.02&-27$\colon$42$\colon$42.73&26.25&---&26.33&25.70&24.75&23.38&22.61&22.15&21.72&21.41&1.019\\[-1.0pt]&&&0.95&---&1.25&0.19&0.07&0.03&0.02&0.01&0.01&0.00&\\[-1.0pt]
J033219.48-274216.8&03$\colon$32$\colon$19.48&-27$\colon$42$\colon$16.81&24.37&25.08&24.08&23.01&21.31&20.49&20.19&20.05&19.76&19.50&0.382\\[-1.0pt]&&&0.28&0.50&0.29&0.03&0.00&0.00&0.00&0.00&0.00&0.00&\\[-1.0pt]
J033219.59-274303.8&03$\colon$32$\colon$19.59&-27$\colon$43$\colon$03.80&---&24.98&25.02&24.62&22.79&21.42&21.02&20.86&20.58&20.27&0.735\\[-1.0pt]&&&---&0.33&0.44&0.09&0.01&0.01&0.01&0.00&0.00&0.00&\\[-1.0pt]
J033219.77-274204.0&03$\colon$32$\colon$19.77&-27$\colon$42$\colon$04.00&---&\nodata&---&26.70&25.70&24.17&23.29&23.02&22.59&22.32&1.044\\[-1.0pt]&&&---&\nodata&---&0.42&0.10&0.04&0.02&0.02&0.01&0.01&\\[-1.0pt]
J033220.02-274104.2&03$\colon$32$\colon$20.02&-27$\colon$41$\colon$04.25&25.34&25.63&25.06&25.59&23.41&21.89&21.43&21.23&20.81&20.46&0.681\\[-1.0pt]&&&0.50&0.60&0.41&0.19&0.02&0.01&0.01&0.01&0.00&0.00&\\[-1.0pt]
J033220.09-274106.7&03$\colon$32$\colon$20.09&-27$\colon$41$\colon$06.75&---&26.54&25.57&---&25.19&23.20&22.25&21.74&20.97&20.56&1.309\\[-1.0pt]&&&---&1.41&0.67&---&0.11&0.03&0.02&0.01&0.00&0.00&\\[-1.0pt]
J033220.67-274446.4&03$\colon$32$\colon$20.67&-27$\colon$44$\colon$46.42&24.54&24.82&26.26&25.21&23.28&21.95&21.47&21.26&20.82&20.45&0.726\\[-1.0pt]&&&0.25&0.31&1.45&0.14&0.02&0.01&0.01&0.01&0.00&0.00&\\[-1.0pt]
J033221.28-274435.6&03$\colon$32$\colon$21.28&-27$\colon$44$\colon$35.60&25.62&25.34&24.82&23.76&21.55&20.34&19.89&19.70&19.31&18.96&0.620\\[-1.0pt]&&&0.55&0.40&0.32&0.03&0.00&0.00&0.00&0.00&0.00&0.00&\\[-1.0pt]
J033222.33-274226.5&03$\colon$32$\colon$22.33&-27$\colon$42$\colon$26.54&\nodata&\nodata&\nodata&\nodata&25.31&23.55&22.63&22.34&21.82&21.41&1.018\\[-1.0pt]&&&\nodata&\nodata&\nodata&\nodata&0.09&0.03&0.02&0.01&0.01&0.01&\\[-1.0pt]
J033222.58-274141.2&03$\colon$32$\colon$22.58&-27$\colon$41$\colon$41.18&---&\nodata&25.19&24.31&22.36&21.32&20.96&20.85&20.53&20.24&0.509\\[-1.0pt]&&&---&\nodata&0.48&0.06&0.01&0.00&0.00&0.00&0.00&0.00&\\[-1.0pt]
J033222.58-274152.1&03$\colon$32$\colon$22.58&-27$\colon$41$\colon$52.04&---&---&---&26.38&25.28&24.71&24.55&24.78&24.65&24.59&0.529\\[-1.0pt]&&&---&---&---&0.15&0.04&0.04&0.04&0.06&0.03&0.04&\\[-1.0pt]
J033223.01-274331.5&03$\colon$32$\colon$23.02&-27$\colon$43$\colon$31.49&---&---&26.27&26.49&23.89&22.45&21.97&21.78&21.43&21.10&0.740\\[-1.0pt]&&&---&---&0.87&0.35&0.03&0.01&0.01&0.01&0.00&0.00&\\[-1.0pt]
J033224.36-274315.2&03$\colon$32$\colon$24.37&-27$\colon$43$\colon$15.18&---&26.26&25.62&24.44&24.47&24.64&24.60&24.46&24.07&24.75&1.271\\[-1.0pt]&&&---&0.29&0.22&0.02&0.02&0.04&0.04&0.04&0.02&0.05&\\[-1.0pt]
J033224.98-274101.5&03$\colon$32$\colon$24.98&-27$\colon$41$\colon$01.52&24.79&24.20&23.56&23.45&22.43&21.38&20.99&20.85&20.48&20.20&0.569\\[-1.0pt]&&&0.29&0.15&0.11&0.03&0.01&0.01&0.01&0.00&0.00&0.00&\\[-1.0pt]
J033225.11-274425.6&03$\colon$32$\colon$25.11&-27$\colon$44$\colon$25.59&25.60&25.04&25.48&25.38&25.20&24.82&24.35&24.32&24.12&23.88&1.220\\[-1.0pt]&&&0.29&0.16&0.29&0.07&0.05&0.05&0.04&0.04&0.02&0.03&\\[-1.0pt]
J033225.29-274224.2&03$\colon$32$\colon$25.29&-27$\colon$42$\colon$24.20&---&---&26.05&24.59&23.09&22.41&22.21&22.14&22.05&21.92&0.612\\[-1.0pt]&&&---&---&0.43&0.03&0.01&0.01&0.01&0.01&0.00&0.00&\\[-1.0pt]
J033225.47-274327.6&03$\colon$32$\colon$25.47&-27$\colon$43$\colon$27.55&\nodata&25.35&23.87&24.55&21.98&20.52&20.04&19.87&19.47&19.10&0.690\\[-1.0pt]&&&\nodata&0.70&0.26&0.11&0.01&0.00&0.00&0.00&0.00&0.00&\\[-1.0pt]
J033225.85-274246.1&03$\colon$32$\colon$25.85&-27$\colon$42$\colon$46.12&25.74&25.31&---&26.11&25.20&23.94&23.12&23.02&22.40&22.05&1.182\\[-1.0pt]&&&0.60&0.38&---&0.30&0.11&0.06&0.03&0.03&0.01&0.01&\\[-1.0pt]
J033225.97-274312.5&03$\colon$32$\colon$25.97&-27$\colon$43$\colon$12.56&---&---&---&---&26.46&24.80&24.00&23.84&23.27&22.87&0.972\\[-1.0pt]&&&---&---&---&---&0.17&0.06&0.03&0.03&0.01&0.01&\\[-1.0pt]
J033225.98-274318.9&03$\colon$32$\colon$25.98&-27$\colon$43$\colon$18.93&26.31&---&\nodata&26.65&25.42&23.89&22.87&22.52&22.02&21.66&1.215\\[-1.0pt]&&&0.67&---&\nodata&0.32&0.08&0.03&0.02&0.01&0.01&0.01&\\[-1.0pt]
J033226.05-274236.5&03$\colon$32$\colon$26.05&-27$\colon$42$\colon$36.54&\nodata&---&---&---&27.09&24.93&23.92&23.31&22.16&21.65&1.125\\[-1.0pt]&&&\nodata&---&---&---&0.41&0.10&0.05&0.03&0.01&0.01&\\[-1.0pt]
J033226.71-274340.2&03$\colon$32$\colon$26.71&-27$\colon$43$\colon$40.15&26.05&\nodata&---&25.00&23.10&21.91&21.51&21.41&21.07&20.77&0.550\\[-1.0pt]&&&0.71&\nodata&---&0.10&0.01&0.01&0.01&0.01&0.00&0.00&\\[-1.0pt]
J033227.18-274416.5&03$\colon$32$\colon$27.18&-27$\colon$44$\colon$16.46&24.49&23.73&23.76&22.51&20.57&19.63&19.28&19.15&18.82&18.51&0.610\\[-1.0pt]&&&0.40&0.18&0.25&0.02&0.00&0.00&0.00&0.00&0.00&0.00&\\[-1.0pt]
J033227.62-274144.9&03$\colon$32$\colon$27.62&-27$\colon$41$\colon$44.91&24.39&25.25&24.85&24.02&22.74&21.59&21.27&21.20&20.83&20.49&0.667\\[-1.0pt]&&&0.14&0.28&0.26&0.03&0.01&0.00&0.00&0.00&0.00&0.00&\\[-1.0pt]
J033227.70-274043.7&03$\colon$32$\colon$27.70&-27$\colon$40$\colon$43.69&---&\nodata&---&25.94&23.90&22.43&21.56&21.32&20.90&20.57&0.967\\[-1.0pt]&&&---&\nodata&---&0.21&0.03&0.01&0.01&0.01&0.00&0.00&\\[-1.0pt]
J033227.84-274136.8&03$\colon$32$\colon$27.84&-27$\colon$41$\colon$36.82&25.95&24.91&---&25.47&24.06&22.72&21.89&21.54&21.07&20.71&1.042\\[-1.0pt]&&&1.22&0.45&---&0.23&0.06&0.03&0.02&0.01&0.01&0.00&\\[-1.0pt]
J033227.86-274313.6&03$\colon$32$\colon$27.86&-27$\colon$43$\colon$13.58&\nodata&\nodata&---&25.97&25.73&25.00&24.36&24.01&23.13&22.80&1.338\\[-1.0pt]&&&\nodata&\nodata&---&0.11&0.08&0.07&0.05&0.03&0.01&0.01&\\[-1.0pt]
J033228.88-274129.3&03$\colon$32$\colon$28.87&-27$\colon$41$\colon$29.32&25.70&25.96&24.92&24.32&22.61&21.07&20.58&20.38&19.97&19.61&0.732\\[-1.0pt]&&&0.92&1.08&0.56&0.09&0.02&0.01&0.01&0.00&0.00&0.00&\\\hline\\ \\
J033229.04-274432.2&03$\colon$32$\colon$29.04&-27$\colon$44$\colon$32.21&\nodata&\nodata&---&---&27.36&25.16&24.45&23.93&22.86&22.38&1.202\\[-1.0pt]&&&\nodata&\nodata&---&---&0.45&0.10&0.06&0.04&0.01&0.01&\\[-1.0pt]
J033229.30-274244.8&03$\colon$32$\colon$29.30&-27$\colon$42$\colon$44.85&---&25.89&\nodata&25.72&25.00&23.91&23.28&23.03&22.50&22.25&0.880\\[-1.0pt]&&&---&0.40&\nodata&0.11&0.05&0.03&0.02&0.02&0.01&0.01&\\[-1.0pt]
J033229.64-274030.3&03$\colon$32$\colon$29.64&-27$\colon$40$\colon$30.25&26.31&26.24&25.34&25.72&25.02&24.01&23.31&23.17&22.79&22.46&1.136\\[-1.0pt]&&&0.64&0.55&0.32&0.12&0.06&0.04&0.03&0.02&0.01&0.01&\\[-1.0pt]
J033230.56-274145.7&03$\colon$32$\colon$30.56&-27$\colon$41$\colon$45.69&24.94&24.73&24.14&24.29&23.55&22.64&22.32&22.22&21.86&21.66&0.837\\[-1.0pt]&&&0.22&0.16&0.13&0.04&0.02&0.01&0.01&0.01&0.01&0.01&\\[-1.0pt]
J033231.84-274329.4&03$\colon$32$\colon$31.84&-27$\colon$43$\colon$29.41&\nodata&---&25.64&---&25.35&24.14&23.25&22.92&22.43&22.04&1.024\\[-1.0pt]&&&\nodata&---&0.65&---&0.11&0.06&0.03&0.03&0.01&0.01&\\[-1.0pt]
J033232.34-274345.8&03$\colon$32$\colon$32.33&-27$\colon$43$\colon$45.83&26.12&26.09&25.89&25.58&25.10&24.37&23.94&23.82&23.54&23.38&1.026\\[-1.0pt]&&&0.40&0.36&0.40&0.08&0.04&0.03&0.03&0.03&0.02&0.02&\\[-1.0pt]
J033232.57-274133.8&03$\colon$32$\colon$32.57&-27$\colon$41$\colon$33.79&26.12&---&23.88&26.37&25.95&25.45&25.26&25.43&25.21&25.33&0.736\\[-1.0pt]&&&0.35&---&0.05&0.12&0.07&0.08&0.08&0.09&0.05&0.07&\\[-1.0pt]
J033232.96-274106.8&03$\colon$32$\colon$32.96&-27$\colon$41$\colon$06.77&23.88&23.76&23.63&23.27&22.40&21.90&21.67&21.59&21.51&21.31&0.472\\[-1.0pt]&&&0.07&0.06&0.07&0.01&0.01&0.01&0.01&0.01&0.00&0.00&\\[-1.0pt]
J033233.28-274236.0&03$\colon$32$\colon$33.29&-27$\colon$42$\colon$35.97&---&---&---&---&27.60&25.74&24.67&24.32&23.62&23.11&1.215\\[-1.0pt]&&&---&---&---&---&0.44&0.14&0.06&0.05&0.02&0.01&\\[-1.0pt]
J033233.40-274138.9&03$\colon$32$\colon$33.40&-27$\colon$41$\colon$38.92&26.01&24.97&24.66&24.52&23.59&22.39&21.62&21.43&21.02&20.74&1.045\\[-1.0pt]&&&0.76&0.27&0.27&0.06&0.02&0.01&0.01&0.01&0.00&0.00&\\[-1.0pt]
J033233.87-274357.6&03$\colon$32$\colon$33.87&-27$\colon$43$\colon$57.55&---&25.85&26.19&26.42&25.00&23.46&22.64&22.41&21.98&21.64&0.978\\[-1.0pt]&&&---&0.41&0.77&0.25&0.06&0.02&0.01&0.01&0.01&0.01&\\[-1.0pt]
J033234.34-274350.1&03$\colon$32$\colon$34.35&-27$\colon$43$\colon$50.10&24.18&24.11&23.37&24.24&22.47&21.24&20.86&20.71&20.39&20.09&0.660\\[-1.0pt]&&&0.29&0.25&0.17&0.09&0.01&0.01&0.01&0.01&0.00&0.00&\\[-1.0pt]
J033235.10-274410.7&03$\colon$32$\colon$35.10&-27$\colon$44$\colon$10.61&24.80&25.27&24.92&24.44&23.91&23.36&23.03&22.86&22.12&21.71&0.838\\[-1.0pt]&&&0.23&0.33&0.31&0.06&0.03&0.03&0.03&0.03&0.01&0.01&\\[-1.0pt]
J033235.63-274310.2&03$\colon$32$\colon$35.63&-27$\colon$43$\colon$10.03&25.21&25.82&25.88&25.27&24.54&22.96&21.93&21.54&20.97&20.59&1.190\\[-1.0pt]&&&0.51&0.81&1.18&0.18&0.08&0.03&0.01&0.01&0.00&0.00&\\[-1.0pt]
J033236.72-274406.4&03$\colon$32$\colon$36.72&-27$\colon$44$\colon$06.41&24.56&24.88&24.74&24.37&23.12&21.99&21.58&21.43&21.04&20.73&0.665\\[-1.0pt]&&&0.24&0.30&0.35&0.07&0.02&0.01&0.01&0.01&0.01&0.01&\\[-1.0pt]
J033237.32-274334.3&03$\colon$32$\colon$37.32&-27$\colon$43$\colon$34.30&25.79&---&24.14&24.56&23.09&21.79&21.37&21.17&20.81&20.49&0.660\\[-1.0pt]&&&0.99&---&0.25&0.11&0.02&0.01&0.01&0.01&0.00&0.00&\\[-1.0pt]
J033237.38-274126.2&03$\colon$32$\colon$37.38&-27$\colon$41$\colon$26.21&25.63&24.66&24.26&23.78&21.35&19.93&19.47&19.29&18.91&18.54&0.671\\[-1.0pt]&&&0.82&0.31&0.29&0.05&0.01&0.00&0.00&0.00&0.00&0.00&\\[-1.0pt]
%J033238.03-274404.6&03$\colon$32$\colon$38.03&-27$\colon$44$\colon$04.57&25.32&25.91&---&---&26.34&25.19&24.62&24.36&24.05&23.63&1.401\\[-1.0pt]&&&0.27&0.44&---&---&0.16&0.09&0.07&0.07&0.04&0.03&\\[-1.0pt]
J033238.06-274128.4&03$\colon$32$\colon$38.05&-27$\colon$41$\colon$28.35&25.99&\nodata&24.43&25.39&22.84&21.36&20.87&20.69&20.28&19.93&0.665\\[-1.0pt]&&&1.06&\nodata&0.32&0.21&0.02&0.01&0.01&0.01&0.00&0.00&\\[-1.0pt]
J033238.36-274128.4&03$\colon$32$\colon$38.36&-27$\colon$41$\colon$28.38&\nodata&26.36&25.92&25.85&23.85&22.60&22.17&22.00&21.67&21.36&0.869\\[-1.0pt]&&&\nodata&0.80&0.71&0.18&0.03&0.02&0.01&0.01&0.01&0.00&\\[-1.0pt]
J033238.44-274019.6&03$\colon$32$\colon$38.44&-27$\colon$40$\colon$19.55&25.76&---&\nodata&26.08&24.45&23.00&22.09&21.82&21.30&20.90&1.033\\[-1.0pt]&&&0.78&---&\nodata&0.36&0.06&0.03&0.02&0.01&0.01&0.00&\\[-1.0pt]
J033238.48-274313.8&03$\colon$32$\colon$38.48&-27$\colon$43$\colon$13.76&25.22&24.71&24.28&23.24&22.20&21.79&21.65&21.58&21.48&21.36&0.430\\[-1.0pt]&&&0.60&0.35&0.31&0.03&0.01&0.01&0.01&0.01&0.01&0.01&\\[-1.0pt]
J033239.17-274026.5&03$\colon$32$\colon$39.16&-27$\colon$40$\colon$26.54&25.76&24.73&24.60&24.52&22.94&21.57&21.16&21.04&20.69&20.42&0.768\\[-1.0pt]&&&0.72&0.26&0.31&0.08&0.02&0.01&0.01&0.01&0.00&0.00&\\[-1.0pt]
J033239.17-274257.7&03$\colon$32$\colon$39.17&-27$\colon$42$\colon$57.75&24.59&25.39&23.65&22.22&20.35&19.47&19.16&19.06&18.75&18.45&0.419\\[-1.0pt]&&&0.49&0.94&0.26&0.02&0.00&0.00&0.00&0.00&0.00&0.00&\\[-1.0pt]
J033239.18-274329.0&03$\colon$32$\colon$39.18&-27$\colon$43$\colon$29.00&---&26.53&25.99&---&25.75&24.46&23.45&23.09&22.49&22.05&1.178\\[-1.0pt]&&&---&1.01&0.84&---&0.15&0.08&0.04&0.03&0.01&0.01&\\[-1.0pt]
J033239.52-274117.4&03$\colon$32$\colon$39.52&-27$\colon$41$\colon$17.42&26.46&---&\nodata&26.60&24.53&23.06&22.10&21.82&21.34&20.98&1.039\\[-1.0pt]&&&0.97&---&\nodata&0.38&0.06&0.03&0.01&0.01&0.00&0.00&\\[-1.0pt]
J033240.38-274338.3&03$\colon$32$\colon$40.38&-27$\colon$43$\colon$38.27&25.14&24.85&24.93&24.72&24.29&23.26&22.28&21.94&21.46&21.07&1.179\\[-1.0pt]&&&0.51&0.36&0.52&0.12&0.07&0.05&0.02&0.02&0.01&0.01&\\[-1.0pt]
J033241.63-274151.5&03$\colon$32$\colon$41.63&-27$\colon$41$\colon$51.41&25.77&---&---&25.28&24.44&23.32&22.60&22.07&21.34&20.97&1.427\\[-1.0pt]&&&0.62&---&---&0.14&0.07&0.04&0.03&0.01&0.00&0.00&\\[-1.0pt]
J033242.36-274238.0&03$\colon$32$\colon$42.35&-27$\colon$42$\colon$37.96&25.27&---&25.26&23.82&21.56&20.34&19.94&19.81&19.47&19.12&0.566\\[-1.0pt]&&&0.64&---&0.78&0.06&0.01&0.00&0.00&0.00&0.00&0.00&\\[-1.0pt]
J033243.93-274232.4&03$\colon$32$\colon$43.93&-27$\colon$42$\colon$32.32&---&26.09&25.75&26.01&25.14&23.56&22.62&22.03&21.12&20.63&1.193\\[-1.0pt]&&&---&0.88&0.87&0.32&0.12&0.05&0.02&0.01&0.00&0.00&\\[-1.0pt]
J033244.97-274309.1&03$\colon$32$\colon$44.97&-27$\colon$43$\colon$09.02&---&\nodata&---&26.68&24.87&24.32&24.01&23.69&22.59&21.78&0.444\\[-1.0pt]&&&---&\nodata&---&0.34&0.05&0.05&0.05&0.04&0.01&0.01
\label{tab:fluxtab}
\end{longtable}
\end{landscape}
\endgroup
  % Table
\newpage
\clearpage

\begingroup
%\begin{subfig}
\captionsetup[subfigure]{labelformat=empty,format=hang,justification=justified,singlelinecheck=false,margin=0pt}
\begin{sidewaysfigure}
\centering
\subfloat[Figure 1a$\colon$ Ten-band thumbnails of the first ten ETGs listed in Table 1 ordered, from left to
  right, by increasing wavelength with the GOODS Object ID. Each image
  has been converted into flux units (nJy), and all are displayed with
  the same scale.  All postage stamps are 11.2 arcseconds (128 pixels)
  on a
  side.  Please contact the author---mjrutkow@asu.edu---for a high resolution FITS/JPG of each galaxy]{\includegraphics[angle=0,scale=1,width=1.0\txw]{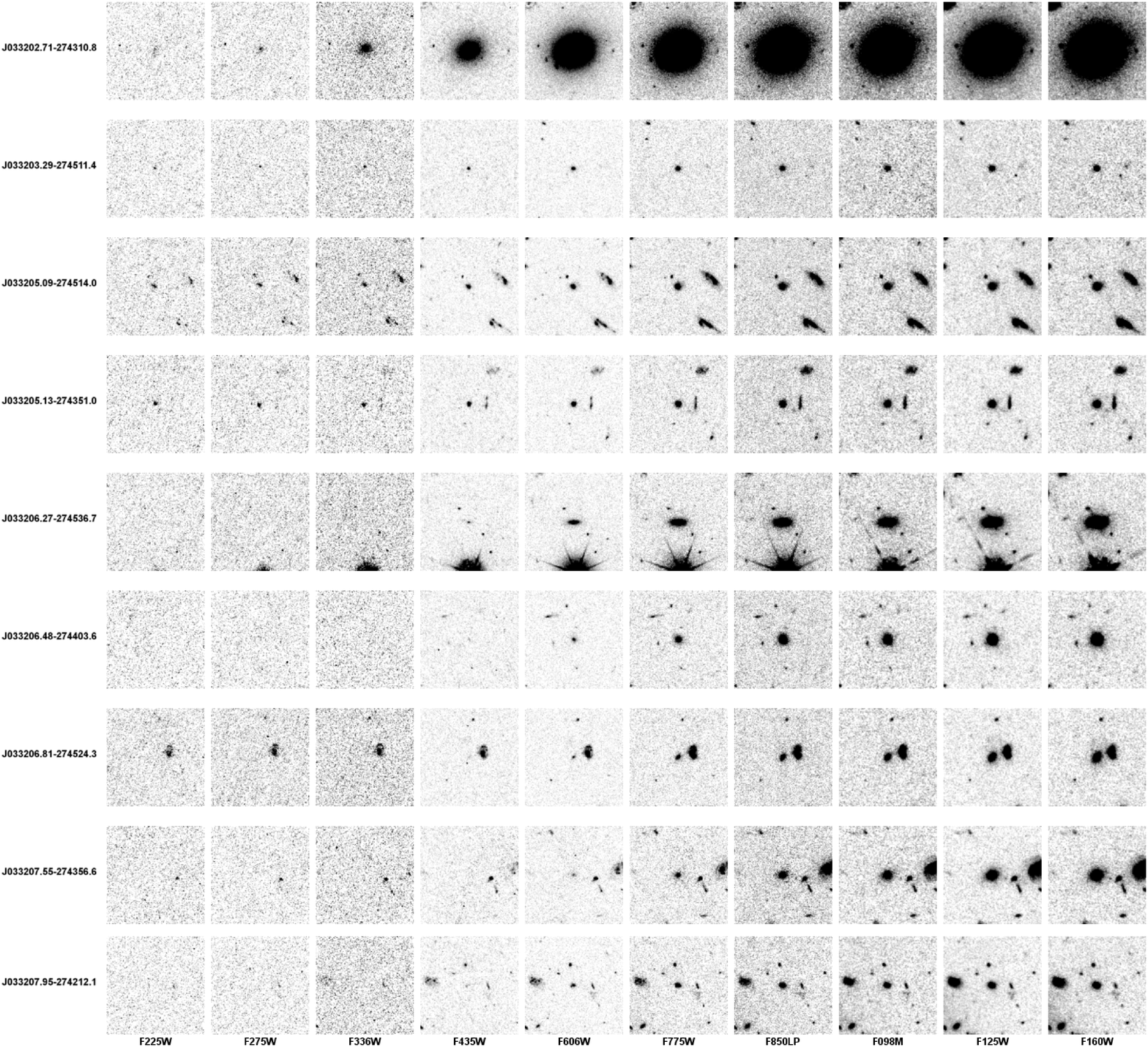}
  }
\end{sidewaysfigure}

\clearpage
\newpage

\endgroup

 % FIGURE cp of WORK/*-subf.tex Figure 
\clearpage

\begin{figure}[htbc]
\addtocounter{figure}{12}
\centering
\epsfig{file=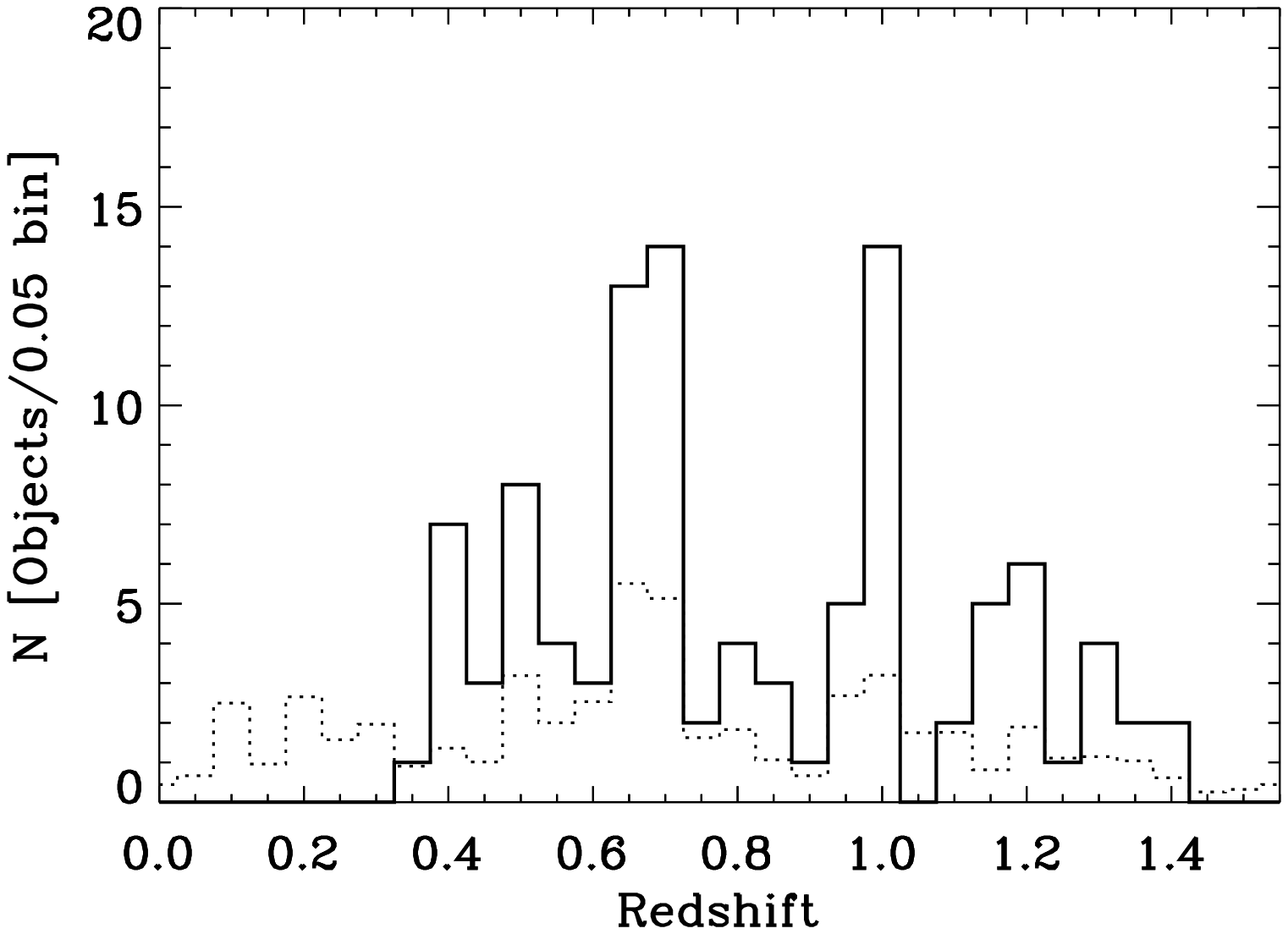, width=0.6\txw,clip=} 
\caption{\small The redshift distribution of ETGs is plotted as a
  solid histogram.  The redshift distribution of objects in the CDF-S
  is plotted as a dot-dashed histogram.  The number of objects for
  which redshifts have been determined in the CDF-S has been scaled
  downwards by a factor of 75, so that both redshift distributions
  could be plotted on the same axis for comparison.  The peaks in this
  distribution indicate large-scale structure in the CDF-S.  Our
  selection of ETGs amplifies these peaks because ETGs are known to be
  more strongly clustered than field galaxies.}
\label{fig:qualities_p1}
\end{figure}
 % Figure
\clearpage
%\LTchunksize=40
\begingroup
\tiny
\begin{center}
\begin{longtable}{lccc}
\caption{Early-Type Galaxies Catalog, Additional Parameters}\\
\hline \hline 
\multicolumn{1}{c}{GOODS ID}&
\multicolumn{1}{c}{X-ray/Radio}&
\multicolumn{1}{c}{AGN Note}&
\multicolumn{1}{c}{Comments}\\
\multicolumn{1}{c}{}&
\multicolumn{1}{c}{Source?}&
\multicolumn{1}{c}{}&
\multicolumn{1}{c}{} \\ \hline
\endfirsthead

\multicolumn{4}{c}{Table 2$\colon$ETG Catalog, Additional Parameters (Continued)}\\%[0.7ex]
\hline \hline
\multicolumn{1}{c}{GOODS ID}&
\multicolumn{1}{c}{X-ray/Radio}&
\multicolumn{1}{c}{AGN Note}&
\multicolumn{1}{c}{Comments}\\
\multicolumn{1}{c}{}&
\multicolumn{1}{c}{Source?}&
\multicolumn{1}{c}{}&
\multicolumn{1}{c}{}\\ \hline
\endhead

%% footnote here
\hline \hline
 \multicolumn{4}{l}{\textbf{Notes-}}\\
 \multicolumn{4}{l}{\textit{Col. 1} $\colon$ GOODS Identifier String}\\
 \multicolumn{4}{l}{\textit{Col. 2} $\colon$ Galaxies identified in X-ray, Radio, or both surveys are denoted}\\
 \multicolumn{4}{l}{~~~~~~~~~~here by ``X*'', ``R*'' or ``XR*'', respectively.}\\
 \multicolumn{4}{l}{\textit{Col. 3} $\colon$ X-ray and optical spectral classification of ETGs are from}\\
 \multicolumn{4}{l}{~~~~~~~~~\cite{S04}. For X-ray classifications, objects are}\\
 \multicolumn{4}{l}{~~~~~~~~~primarily distinguished by the hardness ratio (HR) of the X-ray }\\
 \multicolumn{4}{l}{~~~~~~~~~spectrum$\colon$ $\leq$ 0.2 for AGN-1 ($>$-0.2 for AGN-2).  For Optical}\\
 \multicolumn{4}{l}{~~~~~~~~~classification, ``BLAGN'' denotes a broad-line AGN source;}\\
 \multicolumn{4}{l}{~~~~~~~~~``HEX'' (``LEX'') indicates ``high'' (``low'') degree of excitation;}\\
 \multicolumn{4}{l}{~~~~~~~~~ ``ABS'' denotes a typical galaxy absorption line system; for more}\\
 \multicolumn{4}{l}{~~~~~~~~~ details on these designations see \cite{S04}.}\\
 \multicolumn{4}{l}{\textit{Col. 4} $\colon$ Comments flags$\colon$ Comp --potential satellites or companion;}\\
 \multicolumn{4}{l}{~~~~~~~~~b-Comp. -- blue companions; LSB-Comp. -- low surface brightness}\\
 \multicolumn{4}{l}{~~~~~~~~~companions; c -- compact; DC -- Double Core; d -- potential dust}\\
 \multicolumn{4}{l}{~~~~~~~~ lane; S0 -- S0 candidate; VGM -- visual group member. For details}\\
\multicolumn{4}{l}{~~~~~~~~~~regarding each of these designations, see \S \ref{sec:GALFIT}.}
\endlastfoot

J033202.71-274310.8&--&--&LSB-Comp\\[-1.0pt]
J033203.29-274511.4&--&--&--\\[-1.0pt]
J033205.09-274514.0&--&--&Comp\\[-1.0pt]
J033205.13-274351.0&--&--&Comp\\[-1.0pt]
J033206.27-274536.7&X*&ABS&S0\\[-1.0pt]
J033206.48-274403.6&--&--&LSB-Comp\\[-1.0pt]
J033206.81-274524.3&--&--&--\\[-1.0pt]
J033207.55-274356.6&--&--&--\\[-1.0pt]
J033207.95-274212.1&--&--&--\\[-1.0pt]
J033208.41-274231.3&--&--&Comp\\[-1.0pt]
J033208.45-274145.9&--&--&Comp\\[-1.0pt]
J033208.53-274217.7&X*&ABS&b-Comp\\[-1.0pt]
J033208.55-274231.1&--&--&Comp\\[-1.0pt]
J033208.65-274501.8&--&--&b-Comp\\[-1.0pt]
J033208.90-274344.3&--&--&--\\[-1.0pt]
J033209.09-274510.8&--&--&--\\[-1.0pt]
J033209.19-274225.6&X*&ABS&S0\\[-1.0pt]
J033210.04-274333.1&--&--&VGM\\[-1.0pt]
J033210.12-274333.3&--&--&VGM\\[-1.0pt]
J033210.16-274334.3&--&--&VGM\\[-1.0pt]
J033210.76-274234.6&--&--&DC\\[-1.0pt]
J033210.86-274441.2&--&--&--\\[-1.0pt]%
J033211.21-274533.4&--&--&LSB-Comp\\[-1.0pt]
J033211.61-274554.1&--&--&S0\\[-1.0pt]
J033212.20-274530.1&XR*&AGN-2,LEX&VGM\\[-1.0pt]
J033212.31-274527.4&--&--&VGM\\[-1.0pt]
J033212.47-274224.2&--&--&--\\[-1.0pt]%
J033214.26-274254.2&--&--&Comp\\[-1.0pt]%
J033214.45-274456.6&--&--&Comp\\[-1.0pt]
J033214.65-274136.6&--&--&--\\[-1.0pt]
J033214.68-274337.1&--&--&S0,Comp\\[-1.0pt]
J033214.73-274153.3&--&--&--\\[-1.0pt]
J033214.78-274433.1&--&--&--\\[-1.0pt]
J033214.83-274157.1&--&--&m\\[-1.0pt]
J033215.98-274422.9&--&--&Comp\\[-1.0pt]
J033216.19-274423.1&--&--&Comp\\[-1.0pt]
J033217.11-274220.9&--&--&c,b-Comp\\[-1.0pt]
J033217.12-274407.7&--&--&--\\[-1.0pt]
J033217.14-274303.3&XR*&AGN-1,BLAGN&LSB-Comp\\[-1.0pt]
J033217.49-274436.7&--&--&--\\[-1.0pt]
J033217.91-274122.7&--&--&--\\[-1.0pt]
J033218.31-274233.5&--&--&S0,VGM\\[-1.0pt]
J033218.64-274144.4&--&--&c\\[-1.0pt]
J033218.74-274415.8&--&--&VGM\\[-1.0pt]%
J033219.02-274242.7&--&--&VGM\\[-1.0pt]%
J033219.48-274216.8&--&--&--\\[-1.0pt]
J033219.59-274303.8&--&--&VGM\\[-1.0pt]
J033219.77-274204.0&--&--&c,LSB-Comp\\[-1.0pt]
J033220.02-274104.2&--&--&LSB-Comp\\[-1.0pt]
J033220.09-274106.7&--&--&Comp.\\[-1.0pt]
J033220.67-274446.4&--&--&S0,Comp\\[-1.0pt]
J033221.28-274435.6&XR*&--&m,VGM\\[-1.0pt]
J033222.33-274226.5&--&--&S0\\[-1.0pt]
J033222.58-274141.2&--&--&Comp\\[-1.0pt]
J033222.58-274152.1&--&--&c\\[-1.0pt]
J033223.01-274331.5&--&--&Comp\\[-1.0pt]
J033224.36-274315.2&--&--&c\\[-1.0pt]
J033224.98-274101.5&X*&AGN-2,LEX&Comp\\[-1.0pt]
J033225.11-274425.6&--&--&c\\[-1.0pt]
J033225.29-274224.2&--&--&LSB-Comp\\[-1.0pt]
J033225.47-274327.6&--&--&Comp\\[-1.0pt]
J033225.85-274246.1&--&--&c,VGM\\[-1.0pt]
J033225.97-274312.5&--&--&c\\[-1.0pt]
J033225.98-274318.9&--&--&S0,VGM\\[-1.0pt]%
J033226.05-274236.5&--&--&b-Comp\\[-1.0pt]%
J033226.71-274340.2&--&--&Comp\\[-1.0pt]%
J033227.18-274416.5&--&--&S0,m\\[-1.0pt]%
J033227.62-274144.9&X*&AGN-2,HEX&S0\\[-1.0pt]
J033227.70-274043.7&--&--&S0\\[-1.0pt]
J033227.84-274136.8&--&--&Comp\\[-1.0pt]
J033227.86-274313.6&--&--&c\\[-1.0pt]
J033228.88-274129.3&--&--&d,Comp\\[-1.0pt]
J033229.04-274432.2&--&--&c\\[-1.0pt]
J033229.30-274244.8&--&--&--\\[-1.0pt]%
J033229.64-274030.3&--&--&--\\[-1.0pt]%
J033230.56-274145.7&--&--&m,b-Comp\\[-1.0pt]
J033231.84-274329.4&--&--&c\\[-1.0pt]%
J033232.34-274345.8&--&--&c\\[-1.0pt]%
J033232.57-274133.8&--&--&c\\[-1.0pt]
J033232.96-274106.8&--&--&LSB-Comp\\[-1.0pt]
J033233.28-274236.0&--&--&c\\[-1.0pt]
J033233.40-274138.9&--&--&--\\[-1.0pt]%
J033233.87-274357.6&--&--&--\\[-1.0pt]%
J033234.34-274350.1&X*&AGN-2,LEX&b-Comp\\[-1.0pt]
J033235.10-274410.7&--&--&c,VGM\\[-1.0pt]
J033235.63-274310.2&--&--&S0,Comp\\[-1.0pt]
J033236.72-274406.4&--&--&S0\\[-1.0pt]
J033237.32-274334.3&--&--&LSB-Comp\\[-1.0pt]
J033237.38-274126.2&--&--&Comp.\\[-1.0pt]
J033238.06-274128.4&--&--&b-Comp.\\[-1.0pt]
J033238.36-274128.4&--&--&LSB-Comp\\[-1.0pt]
J033238.44-274019.6&--&--&--\\[-1.0pt]%
J033238.48-274313.8&--&--&--\\[-1.0pt]%
J033239.17-274026.5&--&--&m\\[-1.0pt]
J033239.17-274257.7&--&--&--\\[-1.0pt]\hline\\%
J033239.18-274329.0&--&--&--\\[-1.0pt]
J033239.52-274117.4&--&--&--\\[-1.0pt]%
J033240.38-274338.3&--&--&--\\[-1.0pt]%
J033241.63-274151.5&--&--&--\\[-1.0pt]
J033242.36-274238.0&--&--&Comp\\[-1.0pt]
J033243.93-274232.4&--&--&c\\[-1.0pt]%
J033244.97-274309.1&--&--&c\\[-1.0pt]%
\label{tab:tab3}
\end{longtable}
\end{center}
\endgroup
%\end{minipage}
  % Table

\begin{figure}[htbc]
\centering
\epsfig{file=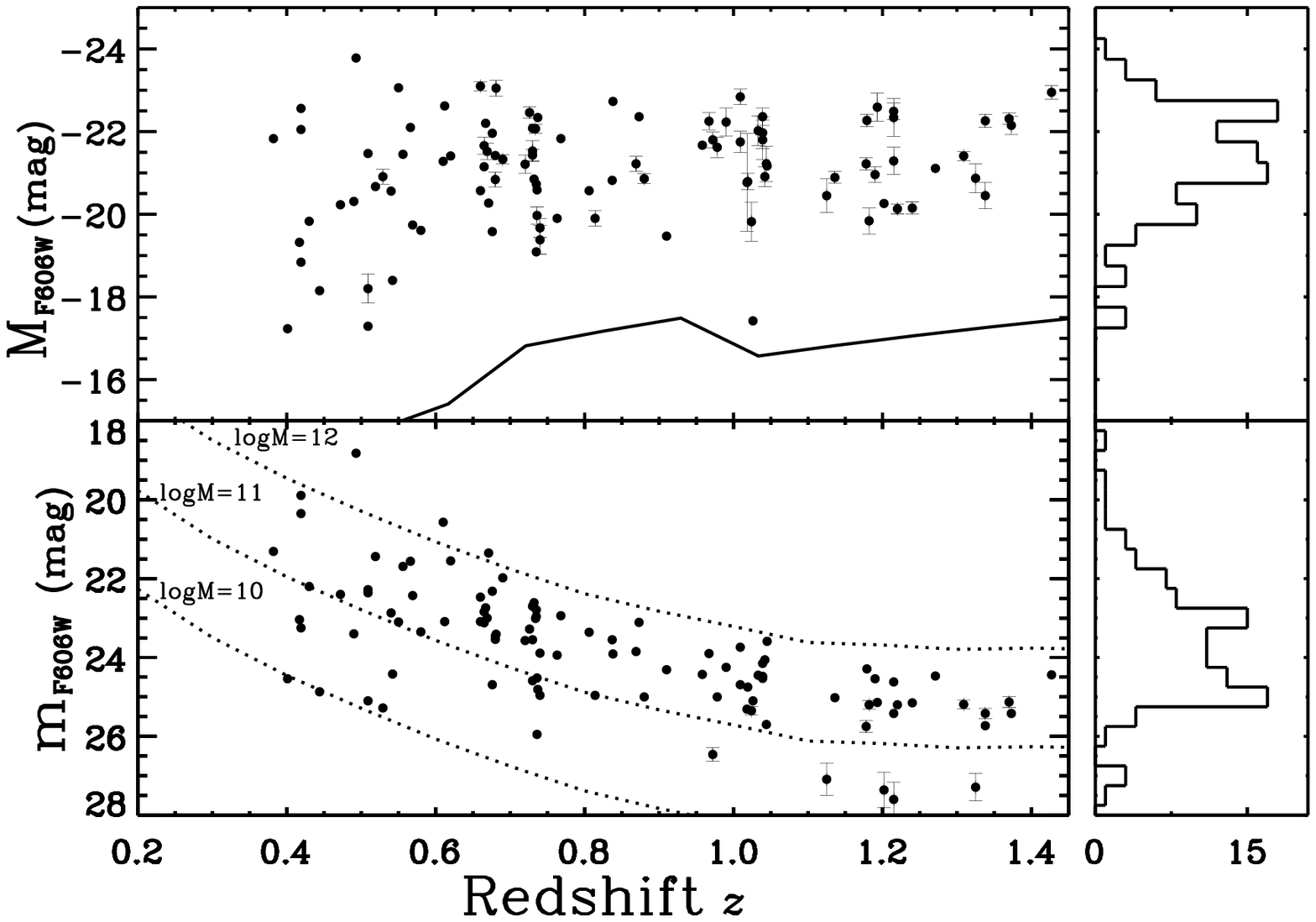, width=1.0\txw,clip=} 
\caption{\small Absolute and apparent magnitudes of catalog ETGs are
  plotted versus redshift.  For clarity, photometric uncertainties are
  only overplotted if the uncertainty is greater than 0.1 mag. {\bf
    Upper Panel}$\colon$ The absolute F606W magnitudes were measured
  for the ETGs using the best-fit single burst stellar population
  model to the SED of each ETG as outlined in \S\ref{sec:absmag}. We
  overplot the photometric completeness limits (solid curve), which we
  derived from the recovery limits (see \S
  \ref{sec:catalogueprod}). {\bf Lower Panel} $\colon$ In addition to
  the apparent F606W magnitudes measured for the ETGs, we overplot the
  apparent F606W magnitudes of a maximally old BC03 model galaxy with
  a star-formation history defined by Equation \ref{eqn:burstmod},
  with log($\tau$[Gyr])=$-$0.3 and $z_f$=4.0. For each model, we
  assume no dust, solar metallicity and a Salpeter IMF.  The only free
  parameter was the stellar mass of the template galaxy, which we
  overplot for each curve.  The majority of ETGs are bounded by the
  10$<$log(M [\Msol])$<$12 curves; in comparison to published mass
  functions of massive galaxies \citep[e.g.,][]{M09} this suggests that
  these ETGs are near or above the characteristic stellar mass.  We
  provide for both panels, at right, a number histogram, corresponding
  to the plotted absolute (apparent) magnitudes.}
\label{fig:absmaglim}
\end{figure}
 % Figure
\begin{deluxetable}{cc}
\tabletypesize{\scriptsize}
\tablecolumns{2} 
\tablewidth{0pc} 
\tablecaption{Model Galaxy Template Parameters}
\tablehead{ 
\colhead{Parameter} & 
\colhead{Range}}
\startdata
%\multicolumn{6}{c}{F$_{bkg,s} =$ 2.167 (-14), F$_{bkg,h}=$ 2.433 (-13) ||| min $\chi^2=$ 616.22 } \\ 
$t_2$ & 0.001 - 13 Gyr \\
$f_2$ & 0.001 - 1 \\
$Z_2$ & 0.1 - 2.5 Z$_{\odot}$ \\
E(B--V) & 0 - 0.5 \\
%\cline{1-12}
\enddata 
\label{tab:params}
\tablecomments{The parameter space represented in the grid of spectral model templates used to determine the (NUV--V), (FUV--V), (g$^{\prime}$--r$^{\prime}$) colors is provided here.  The variable parameters outlined here are as follows $\colon$ $t_2$ = time of second star formation burst; $f_2$ = fraction of stars generated in second burst; $Z_2$ = stellar metallicity of second burst; E(B--V) = dust extinction parameter. For complete details of the model templates and their star formation histories, see Section \ref{sec:analysis}}
\end{deluxetable}

 % Table
\begin{deluxetable}{ccccccc}
%\label{tab:proxy}
\tabletypesize{\scriptsize} \tablecaption{Proxy Filter List for (UV--V) Rest-Frame Color Conversions}
\tablewidth{0pt} 
%\tablehead{\colhead{Filter} & \colhead{t$_{eff}$\tablenotemark{a}} & \colhead{Zeropoint (AB)} & \colhead{Number of Detections (3$\sigma$)}&
\tablehead{\colhead{Redshift} & \colhead{GALEX FUV Proxy} & \colhead{GALEX NUV Proxy} & \colhead{Sloan g$^{\prime}$ Proxy} & \colhead{Sloan r$^{\prime}$ Proxy} & \colhead{Johnson V Proxy}}
\startdata 
%   0.10     & F225W   & F275W  &    F435W   &   F606W     & F606W \\
%   0.15     & F225W   & F275W  &    F606W   &   F775W     & F606W \\ 
%   0.20     & F225W   & F275W  &    F606W   &   F775W     & F606W \\
%   0.25     & F225W   & F275W  &    F606W   &   F775W     & F606W \\ 
   0.30     & F225W   & F275W  &    F606W   &   F775W     & F775W \\
   0.35     & F225W   & F336W  &    F606W   &  F850LP     & F775W \\ 
   0.40     & F225W   & F336W  &    F606W   &  F850LP     & F775W \\
   0.45     & F225W   & F336W  &    F606W   &  F850LP     & F775W \\
   0.50     & F225W   & F336W  &    F775W   &   F098M    & F850LP \\
   0.55     & F225W   & F336W  &    F775W   &   F098M    & F850LP \\
   0.60     & F225W   & F336W  &    F775W   &   F098M    & F850LP \\
   0.65     & F225W   & F336W  &    F775W   &   F098M    & F850LP \\
   0.70     & F275W   & F435W  &    F775W   &   F098M     & F098M \\
   0.75     & F275W   & F435W  &   F850LP   &   F098M     & F098M \\
   0.80     & F275W   & F435W  &   F850LP   &   F098M     & F098M \\
   0.85     & F275W   & F435W  &   F850LP   &   F098M     & F098M \\
   0.90     & F275W   & F435W  &   F850LP   &   F098M     & F098M \\
   1.00     & F275W   & F435W  &    F098M   &   F125W     & F098M \\
   1.10     & F336W   & F435W  &    F098M   &   F125W     & F125W \\
   1.20     & F336W   & F435W  &    F098M   &   F125W     & F125W \\
   1.30     & F336W   & F606W  &    F098M   &   F125W     & F125W \\
   1.40     & F336W   & F606W  &    F098M   &   F125W     & F125W \\
\enddata
%\caption{Effective Exposure Time per filter in the Mosaiced Image}
\label{tab:proxy}
\comment{}
\end{deluxetable}
    
 % Table
\begin{figure}[htbc]
%\begin{minipage}{7.11in}
%\begin{center}
\centering
\begin{tabular}{ccc}
\includegraphics[width=5in,angle=0,scale=0.6]{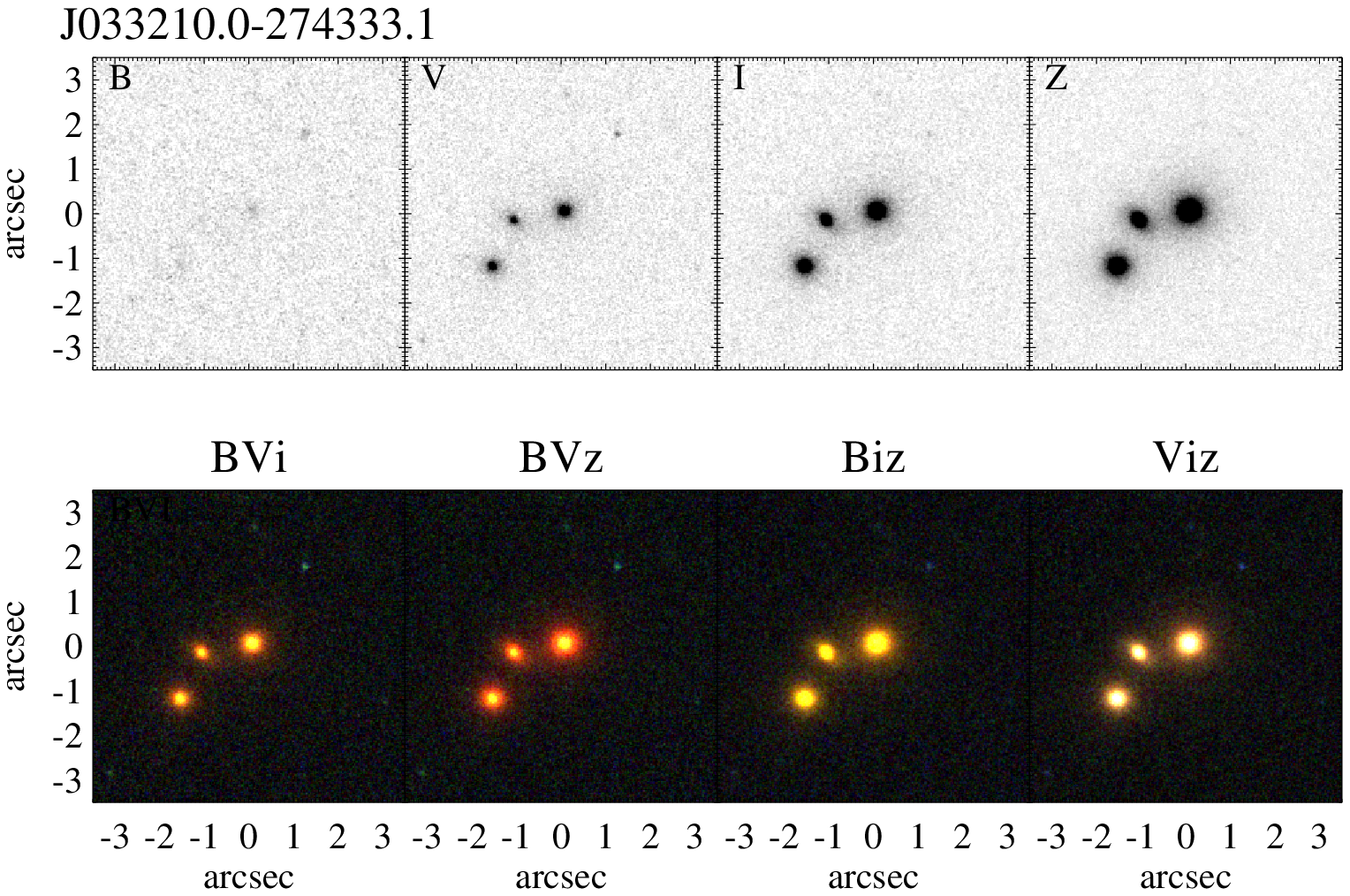} & 
\includegraphics[width=5in,angle=0,scale=0.6]{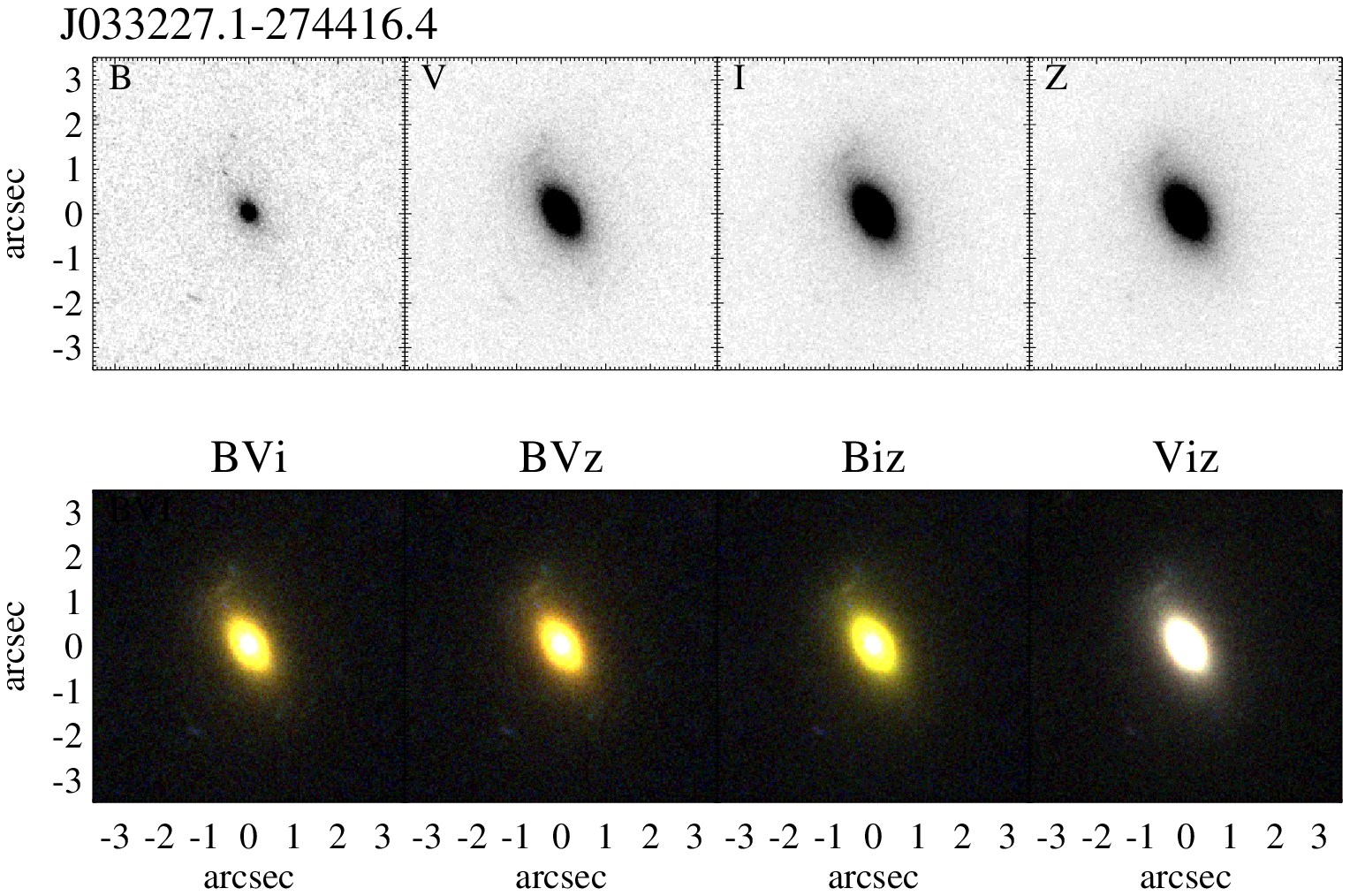} \\
\includegraphics[width=5in,angle=0,scale=0.6]{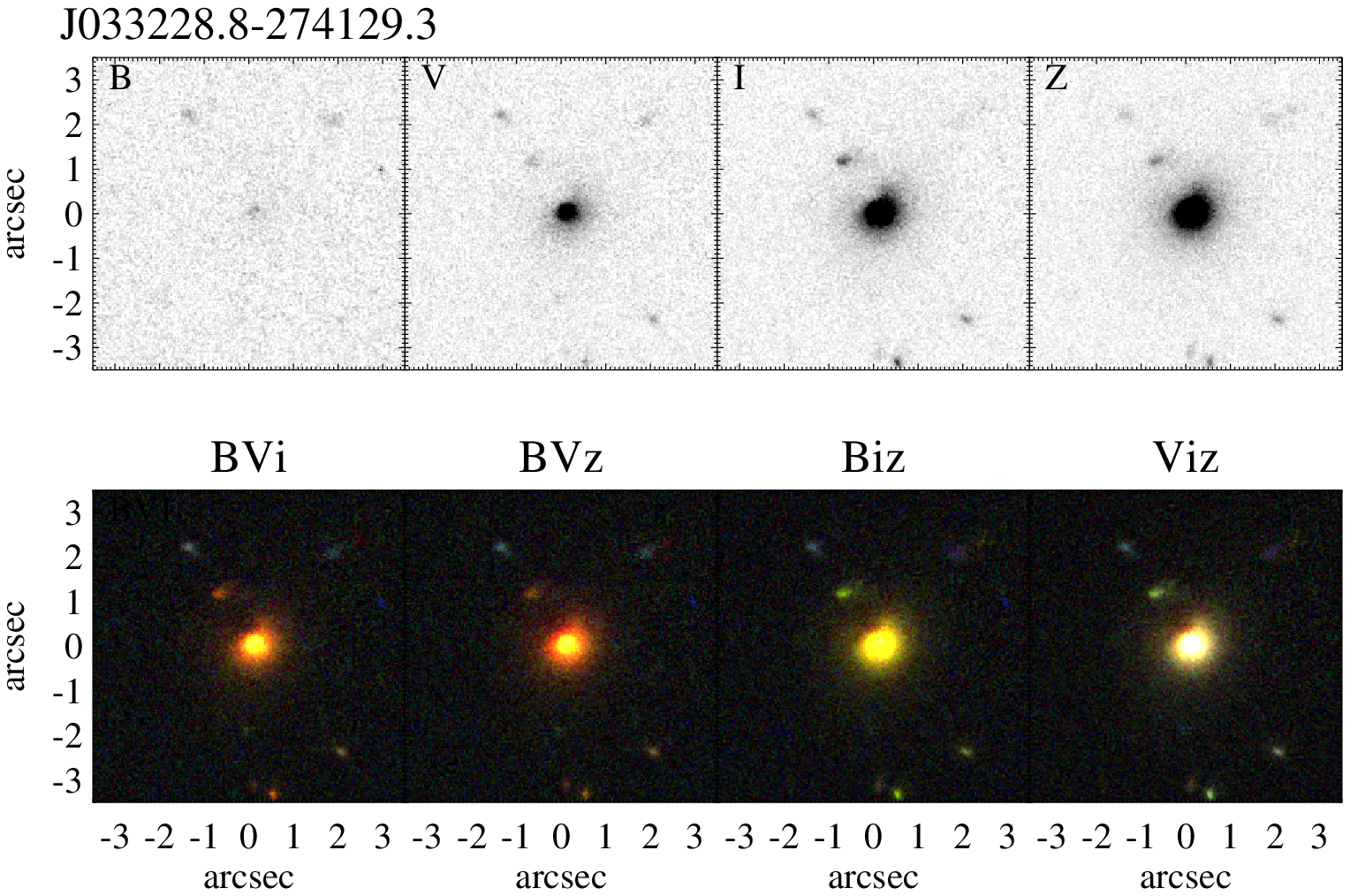} & 
\includegraphics[width=5in,angle=0,scale=0.6]{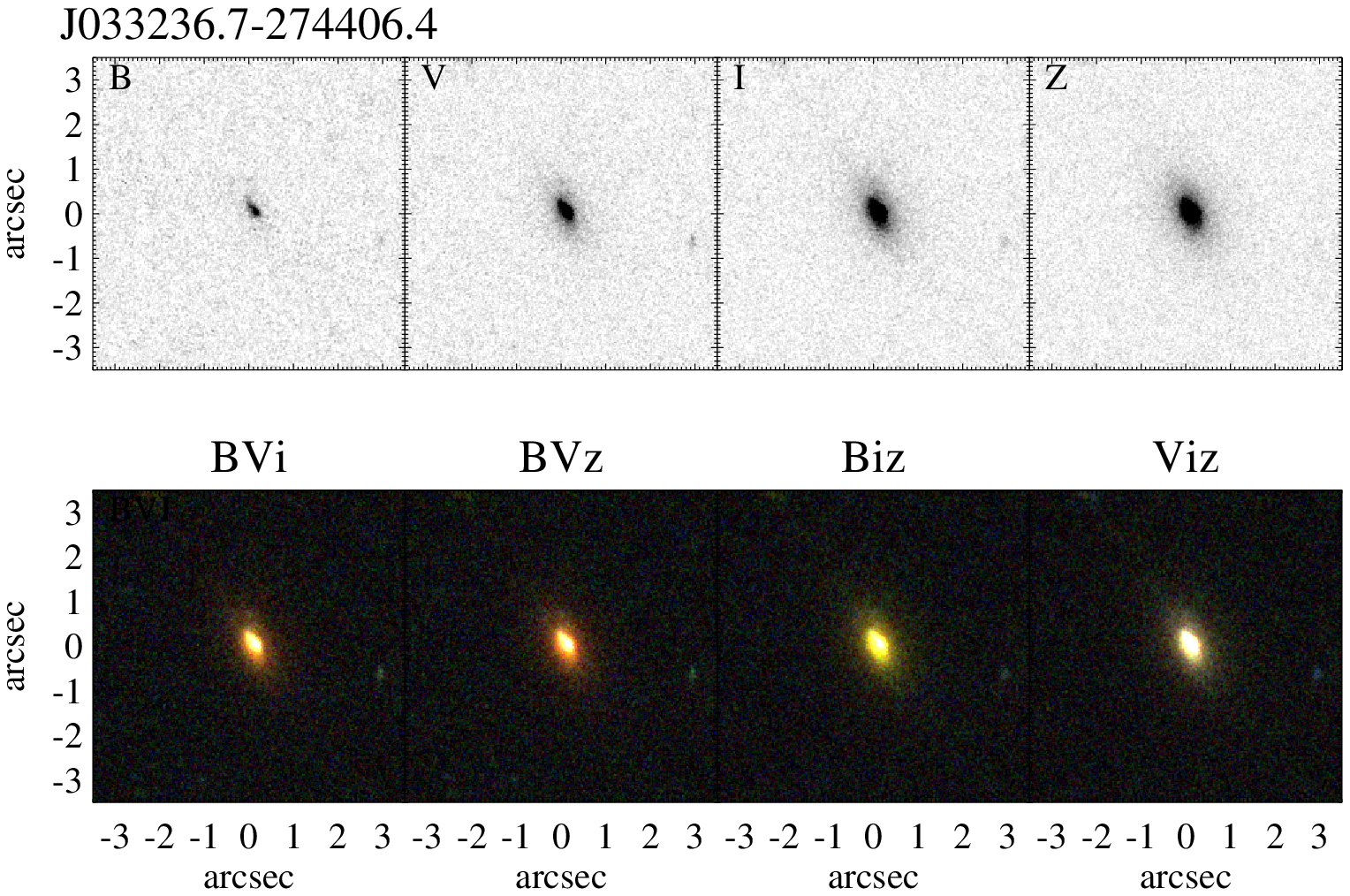} \\
\includegraphics[width=5in,angle=0,scale=0.6]{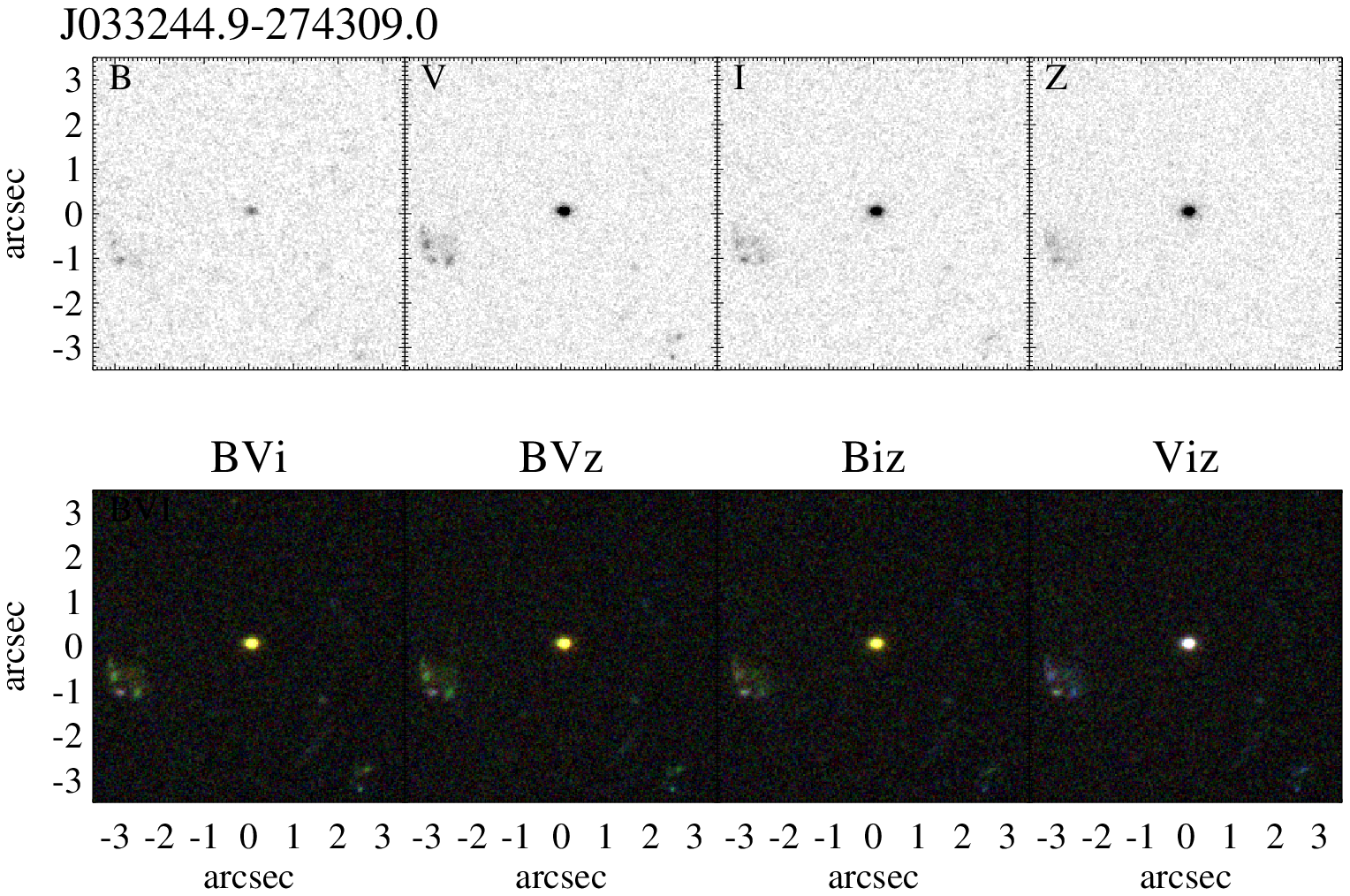} 
\end{tabular}
\caption{Cutouts of six ETGs selected to represent one of each of
  the comment classes defined in Section \ref{sec:GALFIT}. The galaxy,
  and the comment class it represents is defined as follows $\colon$
  J033210.0-274333.1 --- Visual Group Member; J033227.1-274416.4 ---
  Low Surface Brightness Companion (North-east, roughly parallel to minor axis); J033228.8-274129.3 --- dust; J033236.7-274406.4 --- S0;
  J033244.9-274309.0 --- compact.  These images were generated using the GOODS ACS Cutout Tool, available online at http$\colon$\mbox{//}archive.stsci.edu/eidol\_v2.php}
\label{fig:ofinterest}
\end{figure}
 % Figure
\begingroup
\tiny
\begin{landscape}
\begin{longtable}{lccccccccccc}
\caption{Early-Type Galaxies Catalog, Converted Photometry}\\
\hline \hline
\multicolumn{1}{c}{GOODS ID} &
\multicolumn{1}{c}{(FUV--V)$_p$} &
\multicolumn{1}{c}{(FUV--V)$_r$} &
\multicolumn{1}{c}{(NUV--V)$_p$} &
\multicolumn{1}{c}{(NUV--V)$_r$} &
\multicolumn{1}{c}{(g$^{\prime}$--r$^{\prime}$)$_p$}&
\multicolumn{1}{c}{(g$^{\prime}$--r$^{\prime}$)$_r$}&
\multicolumn{1}{c}{M$_{F606W}$} &
\multicolumn{1}{c}{M$_{V}$} &
\multicolumn{1}{c}{M$_{r^{\prime}}$} & 
\multicolumn{1}{c}{m$_{V}$} & 
\multicolumn{1}{c}{m$_{r^{\prime}}$}\\
%\multicolumn{1}{c}{Redshift}\\
\multicolumn{1}{c}{} &
\multicolumn{1}{c}{$\Delta$m} &
\multicolumn{1}{c}{} &
\multicolumn{1}{c}{$\Delta$m} &
\multicolumn{1}{c}{} &
\multicolumn{1}{c}{$\Delta$m} &
\multicolumn{1}{c}{} &
\multicolumn{1}{c}{} &
\multicolumn{1}{c}{} &
\multicolumn{1}{c}{} &
\multicolumn{1}{c}{} &
%\multicolumn{1}{c}{} &
\multicolumn{1}{c}{} \\ \hline
\endfirsthead

\multicolumn{12}{c}{Table 5$\colon$ ETG Catalog, Converted Phot. (Continued)}\\%[0.7ex]
\hline \hline
\multicolumn{1}{c}{GOODS ID} &
\multicolumn{1}{c}{(FUV--V)$_p$} &
\multicolumn{1}{c}{(FUV--V)$_r$} &
\multicolumn{1}{c}{(NUV--V)$_p$} &
\multicolumn{1}{c}{(NUV--V)$_r$} &
\multicolumn{1}{c}{(g$^{\prime}$--r$^{\prime}$)$_p$}&
\multicolumn{1}{c}{(g$^{\prime}$--r$^{\prime}$)$_r$}&
\multicolumn{1}{c}{M$_{F606W}$} &
\multicolumn{1}{c}{M$_{V}$} &
\multicolumn{1}{c}{M$_{r^{\prime}}$} & 
\multicolumn{1}{c}{m$_{V}$} &
\multicolumn{1}{c}{m$_{r^{\prime}}$} \\
%\multicolumn{1}{c}{Redshift}\\
\multicolumn{1}{c}{} &
\multicolumn{1}{c}{(Error)} &
\multicolumn{1}{c}{} &
\multicolumn{1}{c}{(Error)} &
\multicolumn{1}{c}{} &
\multicolumn{1}{c}{(Error)} &
\multicolumn{1}{c}{} &
\multicolumn{1}{c}{} &
\multicolumn{1}{c}{} &
\multicolumn{1}{c}{} &
\multicolumn{1}{c}{} &
%\multicolumn{1}{c}{} &
\multicolumn{1}{c}{} \\ \hline
\endhead

\hline
%\multicolumn{3}{l}{{Continued on Next Page\ldots}}
%\endfoot
\hline \hline 
\\
\multicolumn{12}{l}{\textbf{Notes}$\colon$ Subscripts on column headings designate whether the colors are observed (``p''-- proxy) or rest-frame (``r'').  Galaxies that were \sex~detections}\\
\multicolumn{12}{l}{~~~~~~~~~~but fell below the 90\%1-$\sigma$ completeness limits (see \S \ref{sec:catalogueprod}) in one or more filters used in the transformation are denoted ``---''. ETGs which were}\\
\multicolumn{12}{l}{~~~~~~~~~~\sex~non-detections in the blue proxy band are denoted ``\nodata''(see \S \ref{sec:thedataset}).  The uncertainties, $\Delta$m, reported for rest-frame quantities include }\\
\multicolumn{12}{l}{~~~~~~~~~~measured photometeric and systematic uncertainties (see \S \ref{sec:catalogueprod} and \ref{sec:analysis}).}\\

\endlastfoot
%% footnote here
%%  \multicolumn{14}{l}{    }\\[-10.5pt]
%\multicolumn{14}{l}{\textbf{Notes.}}\\[-0.5pt]
%\endlastfoot
%
J033202.71-274310.8&4.78&5.08&3.34&3.38&0.81&0.43&-23.87&-23.78&-23.93&19.39&18.57\\[-1.0pt]&&0.17&&0.05&&0.00&0.00&\\[-1.0pt]
J033203.29-274511.4&2.38&2.29&1.88&1.76&0.34&0.41&-18.47&-18.40&-18.53&25.04&24.28\\[-1.0pt]&&0.33&&0.24&&0.03&0.07&\\[-1.0pt]
J033205.09-274514.0&1.94&1.90&1.53&1.50&0.01&0.01&-19.96&-19.90&-20.01&24.42&23.85\\[-1.0pt]&&0.16&&0.03&&0.06&0.04&\\[-1.0pt]
J033205.13-274351.0&1.63&1.61&1.43&1.44&0.04&0.14&-20.59&-20.57&-20.62&23.85&23.10\\[-1.0pt]&&0.06&&0.02&&0.04&0.03&\\[-1.0pt]
J033206.27-274536.7&---&---&4.77&4.16&0.69&0.75&-21.68&-21.52&-21.80&23.54&22.85\\[-1.0pt]&&---&&0.58&&0.01&0.20&\\[-1.0pt]
J033206.48-274403.6&3.93&3.90&---&---&0.25&0.81&-21.81&-21.67&-21.92&25.13&24.18\\[-1.0pt]&&0.45&&---&&0.03&0.05&\\[-1.0pt]
J033206.81-274524.3&4.34&3.85&3.38&3.36&0.72&0.44&-22.26&-22.15&-22.35&26.56&25.41\\[-1.0pt]&&0.86&&0.10&&0.01&0.22&\\[-1.0pt]
J033207.55-274356.6&---&---&3.25&3.20&0.51&0.32&-22.38&-22.31&-22.45&26.27&25.12\\[-1.0pt]&&---&&0.16&&0.01&2.23&\\[-1.0pt]
J033207.95-274212.1&---&---&3.44&3.48&0.63&0.65&-19.80&-19.67&-19.90&25.44&24.87\\[-1.0pt]&&---&&0.25&&0.02&0.23&\\[-1.0pt]
J033208.41-274231.3&4.96&4.80&4.64&4.23&0.55&0.68&-20.71&-20.56&-20.82&23.49&22.73\\[-1.0pt]&&0.92&&0.74&&0.01&0.09&\\[-1.0pt]
J033208.45-274145.9&3.98&3.92&3.93&4.01&0.77&0.80&-21.59&-21.43&-21.71&24.03&23.46\\[-1.0pt]&&0.42&&0.15&&0.01&0.14&\\[-1.0pt]
J033208.53-274217.7&4.45&4.39&3.78&3.84&0.69&0.72&-22.23&-22.08&-22.35&23.18&22.61\\[-1.0pt]&&0.45&&0.11&&0.00&0.10&\\[-1.0pt]
J033208.55-274231.1&2.78&2.67&2.40&2.23&0.47&0.59&-18.33&-18.20&-18.44&25.72&24.96\\[-1.0pt]&&0.73&&0.60&&0.05&0.35&\\[-1.0pt]
J033208.65-274501.8&---&---&4.46&4.48&0.14&0.49&-22.48&-22.36&-22.57&23.60&22.85\\[-1.0pt]&&---&&0.09&&0.00&0.10&\\[-1.0pt]
J033208.90-274344.3&2.79&2.67&2.63&2.49&0.25&0.29&-19.67&-19.61&-19.71&23.97&23.21\\[-1.0pt]&&0.28&&0.29&&0.01&0.05&\\[-1.0pt]
J033209.09-274510.8&1.59&1.72&1.03&1.03&0.56&0.31&-17.26&-17.23&-17.28&25.11&24.29\\[-1.0pt]&&0.29&&0.21&&0.02&0.08&\\[-1.0pt]
J033209.19-274225.6&---&---&4.42&4.55&0.72&0.74&-21.36&-21.21&-21.47&24.05&23.48\\[-1.0pt]&&---&&0.24&&0.01&0.22&\\[-1.0pt]
J033210.04-274333.1&4.35&4.30&4.42&4.48&0.52&0.36&-23.00&-22.84&-23.12&24.65&23.49\\[-1.0pt]&&0.47&&0.16&&0.00&0.19&\\[-1.0pt]
J033210.12-274333.3&\nodata&\nodata&4.54&4.58&0.47&0.31&-21.91&-21.75&-22.02&25.60&24.44\\[-1.0pt]&&\nodata&&0.23&&0.00&0.26&\\[-1.0pt]
J033210.16-274334.3&\nodata&\nodata&4.38&4.42&0.20&0.65&-22.38&-22.23&-22.50&24.95&24.00\\[-1.0pt]&&\nodata&&0.31&&0.03&0.34&\\[-1.0pt]
J033210.76-274234.6&4.45&4.74&4.04&4.01&1.25&0.74&-22.71&-22.56&-22.82&20.46&19.64\\[-1.0pt]&&0.15&&0.12&&0.00&0.01&\\[-1.0pt]
J033210.86-274441.2&3.47&3.33&---&---&0.55&0.60&-19.71&-19.58&-19.80&25.23&24.54\\[-1.0pt]&&0.64&&---&&0.02&0.04&\\[-1.0pt]
J033211.21-274533.4&---&---&4.75&3.80&0.47&0.31&-22.60&-22.49&-22.68&25.81&24.47\\[-1.0pt]&&---&&0.15&&0.00&0.20&\\[-1.0pt]
J033211.61-274554.1&4.12&4.07&4.42&4.47&0.44&0.29&-22.51&-22.36&-22.62&25.06&23.90\\[-1.0pt]&&0.40&&0.18&&0.00&0.21&\\[-1.0pt]
J033212.20-274530.1&4.40&4.24&3.94&3.49&0.56&0.61&-22.08&-21.96&-22.18&22.86&22.17\\[-1.0pt]&&0.47&&0.28&&0.01&0.09&\\[-1.0pt]
J033212.31-274527.4&\nodata&\nodata&3.42&3.06&0.56&0.61&-20.97&-20.84&-21.06&24.00&23.31\\[-1.0pt]&&\nodata&&0.33&&0.01&0.18&\\[-1.0pt]
J033212.47-274224.2&---&---&3.30&3.30&1.12&0.65&-19.45&-19.32&-19.55&23.61&22.79\\[-1.0pt]&&---&&0.39&&0.00&0.06&\\[-1.0pt]
J033214.26-274254.2&\nodata&\nodata&3.03&3.07&0.14&0.46&-19.82&-19.71&-19.91&25.45&24.70\\[-1.0pt]&&\nodata&&0.18&&0.11&0.19&\\[-1.0pt]
J033214.45-274456.6&---&---&4.65&4.82&0.56&0.59&-20.03&-19.90&-20.13&25.29&24.72\\[-1.0pt]&&---&&0.79&&0.02&0.06&\\[-1.0pt]
J033214.65-274136.6&---&---&3.64&3.68&0.74&0.45&-22.48&-22.34&-22.59&26.56&25.41\\[-1.0pt]&&---&&0.15&&0.01&0.32&\\[-1.0pt]
J033214.68-274337.1&2.54&2.53&2.20&2.30&0.07&0.27&-20.53&-20.45&-20.60&25.01&24.06\\[-1.0pt]&&0.23&&0.05&&0.06&0.06&\\[-1.0pt]
J033214.73-274153.3&\nodata&\nodata&---&---&1.17&0.65&-19.58&-19.47&-19.67&23.97&23.15\\[-1.0pt]&&\nodata&&---&&0.00&0.07&\\[-1.0pt]
J033214.78-274433.1&---&---&4.65&4.82&0.69&0.72&-20.46&-20.31&-20.58&25.00&24.43\\[-1.0pt]&&---&&0.56&&0.02&0.05&\\[-1.0pt]
J033214.83-274157.1&---&---&---&---&0.50&0.55&-20.71&-20.59&-20.81&24.08&23.39\\[-1.0pt]&&---&&---&&0.01&0.09&\\[-1.0pt]
J033215.98-274422.9&4.51&4.46&3.19&3.21&0.50&0.53&-21.52&-21.42&-21.61&23.44&22.87\\[-1.0pt]&&0.74&&0.06&&0.01&0.06&\\\hline\\
J033216.19-274423.1&3.02&3.25&3.17&3.17&1.10&0.64&-19.21&-19.09&-19.30&23.82&23.00\\[-1.0pt]&&0.52&&0.68&&0.00&0.06&\\[-1.0pt]
J033217.11-274220.9&0.37&0.28&0.67&0.58&0.76&0.51&-18.85&-18.84&-18.84&26.34&25.00\\[-1.0pt]&&0.09&&0.03&&0.04&0.15&\\[-1.0pt]
J033217.12-274407.7&\nodata&\nodata&3.97&4.06&0.61&0.63&-20.29&-20.15&-20.39&25.07&24.50\\[-1.0pt]&&\nodata&&0.27&&0.01&0.25&\\[-1.0pt]
J033217.14-274303.3&3.62&3.47&2.84&2.67&0.44&0.52&-21.64&-21.53&-21.73&22.31&21.55\\[-1.0pt]&&0.12&&0.07&&0.00&0.02&\\[-1.0pt]
J033217.49-274436.7&3.85&3.79&3.45&3.48&0.56&0.59&-21.57&-21.45&-21.67&23.49&22.92\\[-1.0pt]&&0.44&&0.12&&0.01&0.11&\\[-1.0pt]
J033217.91-274122.7&4.71&4.65&4.92&4.91&0.49&0.32&-22.21&-22.07&-22.32&25.39&24.23\\[-1.0pt]&&1.07&&0.41&&0.00&0.48&\\[-1.0pt]
J033218.31-274233.5&3.96&3.82&4.72&4.30&0.49&0.60&-21.94&-21.80&-22.05&22.06&21.30\\[-1.0pt]&&0.19&&0.41&&0.00&0.04&\\[-1.0pt]
J033218.64-274144.4&\nodata&\nodata&3.91&4.00&0.73&0.45&-20.80&-20.67&-20.91&28.43&27.28\\[-1.0pt]&&\nodata&&0.44&&0.01&0.37&\\[-1.0pt]
J033218.74-274415.8&4.28&4.14&4.03&3.69&0.44&0.55&-21.00&-20.87&-21.11&22.90&22.14\\[-1.0pt]&&0.40&&0.35&&0.01&0.06&\\[-1.0pt]
J033219.02-274242.7&---&---&3.55&3.67&0.43&0.28&-21.62&-21.47&-21.73&25.66&24.50\\[-1.0pt]&&---&&0.17&&0.01&0.20&\\[-1.0pt]
J033219.48-274216.8&3.88&4.16&3.59&3.58&1.12&0.70&-20.93&-20.79&-21.03&21.68&21.08\\[-1.0pt]&&0.28&&0.30&&0.00&0.03&\\[-1.0pt]
J033219.59-274303.8&4.11&4.06&3.76&3.82&0.56&0.58&-21.94&-21.83&-22.03&23.27&22.70\\[-1.0pt]&&0.32&&0.09&&0.00&0.09&\\[-1.0pt]
J033219.77-274204.0&\nodata&\nodata&3.68&3.79&0.43&0.28&-20.85&-20.73&-20.95&26.61&25.45\\[-1.0pt]&&\nodata&&0.38&&0.01&0.43&\\[-1.0pt]
J033220.02-274104.2&3.90&3.76&3.82&3.40&0.66&0.71&-21.37&-21.22&-21.49&23.95&23.26\\[-1.0pt]&&0.48&&0.34&&0.01&0.19&\\[-1.0pt]
J033220.09-274106.7&4.60&4.09&4.22&4.39&0.76&0.47&-23.20&-23.05&-23.31&26.33&25.18\\[-1.0pt]&&0.61&&0.14&&0.00&1.67&\\[-1.0pt]
J033220.67-274446.4&3.55&3.50&3.94&4.03&0.69&0.71&-21.56&-21.41&-21.68&23.76&23.19\\[-1.0pt]&&0.30&&0.15&&0.01&0.14&\\[-1.0pt]
J033221.28-274435.6&5.73&5.54&4.93&4.51&0.63&0.73&-22.61&-22.46&-22.72&22.09&21.40\\[-1.0pt]&&0.54&&0.26&&0.00&0.03&\\[-1.0pt]
J033222.33-274226.5&\nodata&\nodata&\nodata&\nodata&0.51&0.35&-21.56&-21.41&-21.68&26.22&25.06\\[-1.0pt]&&\nodata&&\nodata&&0.01&0.09&\\[-1.0pt]
J033222.58-274141.2&---&---&4.23&3.86&0.47&0.58&-20.90&-20.77&-21.01&22.98&22.22\\[-1.0pt]&&---&&0.42&&0.00&0.06&\\[-1.0pt]
J033222.58-274152.1&---&---&---&---&0.07&0.08&-17.31&-17.29&-17.32&25.90&25.14\\[-1.0pt]&&---&&---&&0.08&0.18&\\[-1.0pt]
J033223.01-274331.5&---&---&4.70&4.88&0.67&0.69&-21.04&-20.91&-21.15&24.37&23.80\\[-1.0pt]&&---&&0.39&&0.01&0.35&\\[-1.0pt]
J033224.36-274315.2&1.55&1.31&0.37&0.33&0.38&0.25&-19.37&-19.38&-19.35&25.66&24.32\\[-1.0pt]&&0.19&&0.02&&0.02&0.09&\\[-1.0pt]
J033224.98-274101.5&3.80&3.64&2.56&2.42&0.52&0.63&-21.24&-21.11&-21.33&23.05&22.29\\[-1.0pt]&&0.28&&0.10&&0.01&0.03&\\[-1.0pt]
J033225.11-274425.6&1.35&1.14&1.25&1.06&0.20&0.13&-19.79&-19.74&-19.82&26.39&25.05\\[-1.0pt]&&0.25&&0.05&&0.02&0.12&\\[-1.0pt]
J033225.29-274224.2&---&---&3.84&3.59&0.27&0.31&-20.16&-20.13&-20.20&23.63&22.94\\[-1.0pt]&&---&&0.37&&0.01&0.03&\\[-1.0pt]
J033225.47-274327.6&\nodata&\nodata&4.00&3.54&0.64&0.70&-22.77&-22.62&-22.89&22.52&21.83\\[-1.0pt]&&\nodata&&0.21&&0.00&0.11&\\[-1.0pt]
J033225.85-274246.1&---&---&3.71&3.00&0.62&0.42&-21.43&-21.33&-21.52&26.27&25.00\\[-1.0pt]&&---&&0.24&&0.02&0.32&\\[-1.0pt]
J033225.97-274312.5&---&---&---&---&0.16&0.54&-19.98&-19.84&-20.09&27.16&26.21\\[-1.0pt]&&---&&---&0.12&0.19&\\[-1.0pt]
J033225.98-274318.9&\nodata&\nodata&4.62&3.71&0.50&0.33&-21.92&-21.80&-22.02&26.61&25.27\\[-1.0pt]&&\nodata&&0.24&&0.00&0.33&\\[-1.0pt]
J033226.05-274236.5&---&---&---&---&1.14&0.82&-21.47&-21.29&-21.60&28.16&26.89\\[-1.0pt]&&---&&---&&0.02&0.42&\\[-1.0pt]
J033226.71-274340.2&4.53&4.36&---&---&0.50&0.59&-20.58&-20.45&-20.69&23.72&22.96\\[-1.0pt]&&0.69&&---&&0.01&0.10&\\[-1.0pt]
J033227.18-274416.5&5.20&5.02&4.47&4.14&0.47&0.55&-23.18&-23.06&-23.27&21.11&20.42\\[-1.0pt]&&0.39&&0.21&&0.00&0.02&\\[-1.0pt]
J033227.62-274144.9&3.11&2.99&3.64&3.25&0.38&0.43&-21.40&-21.28&-21.49&23.28&22.59\\[-1.0pt]&&0.13&&0.21&&0.00&0.03&\\[-1.0pt]
J033227.70-274043.7&\nodata&\nodata&4.62&4.63&0.23&0.76&-22.34&-22.20&-22.44&24.60&23.65\\[-1.0pt]&&\nodata&&0.19&&0.03&0.21&\\[-1.0pt]
J033227.84-274136.8&3.36&3.32&3.92&4.03&0.47&0.31&-22.40&-22.25&-22.50&24.97&23.81\\[-1.0pt]&&0.44&&0.21&&0.01&0.24&\\[-1.0pt]
J033227.86-274313.6&---&---&2.60&2.46&0.88&0.54&-21.05&-20.91&-21.16&26.87&25.72\\[-1.0pt]&&---&&0.08&&0.01&0.16&\\\hline\\ \\
J033228.88-274129.3&5.57&5.51&3.94&4.02&0.69&0.71&-22.41&-22.26&-22.53&23.09&22.52\\[-1.0pt]&&1.07&&0.09&&0.00&0.09&\\[-1.0pt]
J033229.04-274432.2&---&---&---&---&1.07&0.71&-21.03&-20.85&-21.16&28.55&27.21\\[-1.0pt]&&---&&---&&0.02&0.10&\\[-1.0pt]
J033229.30-274244.8&2.85&2.84&2.68&2.75&0.25&0.82&-20.39&-20.26&-20.49&25.49&24.74\\[-1.0pt]&&0.39&&0.11&&0.08&0.12&\\[-1.0pt]
J033229.64-274030.3&2.54&2.19&2.92&2.37&0.38&0.25&-20.95&-20.86&-21.02&26.09&24.82\\[-1.0pt]&&0.28&&0.09&&0.01&0.14&\\[-1.0pt]
J033230.56-274145.7&2.51&2.48&2.07&2.09&0.10&0.33&-20.97&-20.89&-21.04&24.04&23.29\\[-1.0pt]&&0.15&&0.04&&0.04&0.05&\\[-1.0pt]
J033231.84-274329.4&---&---&---&---&0.49&0.32&-20.97&-20.82&-21.08&26.26&25.10\\[-1.0pt]&&---&&---&&0.02&0.12&\\[-1.0pt]
J033232.34-274345.8&2.27&2.24&1.76&1.91&0.28&0.17&-19.88&-19.82&-19.94&26.01&24.85\\[-1.0pt]&&0.35&&0.08&&0.02&0.10&\\[-1.0pt]
J033232.57-274133.8&---&---&0.94&0.87&0.02&0.03&-17.42&-17.42&-17.42&26.43&25.86\\[-1.0pt]&&---&&0.14&&0.13&0.21&\\[-1.0pt]
J033232.96-274106.8&1.97&2.14&1.72&1.79&0.73&0.38&-20.03&-19.97&-20.08&22.97&22.15\\[-1.0pt]&&0.07&&0.07&&0.00&0.02&\\[-1.0pt]
J033233.28-274236.0&---&---&4.19&3.38&0.69&0.46&-20.39&-20.23&-20.51&28.79&27.45\\[-1.0pt]&&---&&0.55&&0.03&0.85&\\[-1.0pt]
J033233.40-274138.9&3.53&3.49&3.09&3.24&0.41&0.26&-22.45&-22.34&-22.53&24.50&23.34\\[-1.0pt]&&0.26&&0.05&&0.00&0.06&\\[-1.0pt]
J033233.87-274357.6&3.44&3.42&4.01&4.06&0.23&0.73&-21.31&-21.17&-21.42&25.70&24.75\\[-1.0pt]&&0.40&&0.23&&0.03&0.25&\\[-1.0pt]
J033234.34-274350.1&3.31&3.19&2.66&2.42&0.52&0.58&-21.75&-21.62&-21.85&23.01&22.32\\[-1.0pt]&&0.28&&0.14&&0.01&0.09&\\[-1.0pt]
J033235.10-274410.7&2.40&2.38&1.57&1.59&0.17&0.55&-20.69&-20.57&-20.79&24.40&23.65\\[-1.0pt]&&0.32&&0.06&&0.13&0.08&\\[-1.0pt]
J033235.63-274310.2&4.90&4.37&4.30&3.47&0.56&0.39&-22.88&-22.73&-22.99&25.61&24.34\\[-1.0pt]&&1.10&&0.14&&0.00&0.19&\\[-1.0pt]
J033236.72-274406.4&2.97&2.86&3.30&2.97&0.56&0.61&-21.09&-20.96&-21.19&23.66&22.97\\[-1.0pt]&&0.23&&0.29&&0.01&0.07&\\[-1.0pt]
J033237.32-274334.3&4.42&4.26&2.96&2.68&0.62&0.67&-21.29&-21.15&-21.40&23.63&22.94\\[-1.0pt]&&0.96&&0.21&&0.01&0.11&\\[-1.0pt]
J033237.38-274126.2&6.15&5.98&4.96&4.32&0.63&0.69&-23.26&-23.10&-23.37&21.89&21.20\\[-1.0pt]&&0.81&&0.23&&0.00&0.05&\\[-1.0pt]
J033238.06-274128.4&5.11&4.95&3.73&3.33&0.67&0.72&-21.82&-21.66&-21.94&23.38&22.69\\[-1.0pt]&&1.04&&0.26&&0.01&0.21&\\[-1.0pt]
J033238.36-274128.4&4.36&4.33&3.85&3.88&0.17&0.58&-21.33&-21.22&-21.42&24.34&23.59\\[-1.0pt]&&0.79&&0.17&&0.04&0.18&\\[-1.0pt]
J033238.44-274019.6&---&---&4.26&4.33&0.51&0.35&-22.17&-22.02&-22.28&25.36&24.20\\[-1.0pt]&&---&&0.32&&0.01&0.36&\\[-1.0pt]
J033238.48-274313.8&3.42&3.67&2.48&2.49&0.55&0.29&-19.88&-19.83&-19.91&22.77&21.95\\[-1.0pt]&&0.62&&0.30&&0.00&0.03&\\[-1.0pt]
J033239.17-274026.5&3.68&3.63&3.47&3.53&0.12&0.35&-21.94&-21.83&-22.03&23.42&22.85\\[-1.0pt]&&0.25&&0.08&&0.04&0.08&\\[-1.0pt]
J033239.17-274257.7&5.12&5.41&4.18&4.14&1.19&0.70&-22.19&-22.05&-22.30&20.92&20.10\\[-1.0pt]&&0.49&&0.24&&0.00&0.02&\\[-1.0pt]
J033239.18-274329.0&3.50&3.06&---&---&0.60&0.41&-21.36&-21.22&-21.46&26.82&25.55\\[-1.0pt]&&0.76&&---&&0.02&0.17&\\[-1.0pt]
J033239.52-274117.4&---&---&4.78&4.79&0.47&0.32&-22.11&-21.97&-22.21&25.44&24.28\\[-1.0pt]&&---&&0.33&&0.00&0.38&\\[-1.0pt]
J033240.38-274338.3&3.47&3.03&3.26&2.64&0.47&0.32&-22.39&-22.27&-22.48&25.36&24.09\\[-1.0pt]&&0.47&&0.09&&0.01&0.15&\\[-1.0pt]
J033241.63-274151.5&---&---&3.10&2.94&0.73&0.41&-23.05&-22.95&-23.14&25.37&24.48\\[-1.0pt]&&---&&0.07&&0.00&0.16&\\[-1.0pt]
J033242.36-274238.0&5.32&5.14&5.31&4.83&0.52&0.63&-22.24&-22.10&-22.36&22.18&21.42\\[-1.0pt]&&0.62&&0.66&&0.00&0.06&\\[-1.0pt]
J033243.93-274232.4&4.62&4.11&4.88&3.94&0.91&0.64&-22.77&-22.59&-22.90&26.21&24.94\\[-1.0pt]&&0.80&&0.25&&0.00&0.34&\\[-1.0pt]
J033244.97-274309.1&---&---&---&---&0.86&0.49&-18.31&-18.15&-18.43&25.44&24.62\\[-1.0pt]&&---&&---&&0.04&0.35&\\[-1.0pt]

\label{tab:plotcolors}
\end{longtable}
\end{landscape}
\endgroup
 % Table
\begin{figure}[htbc]
\centering
\epsfig{file=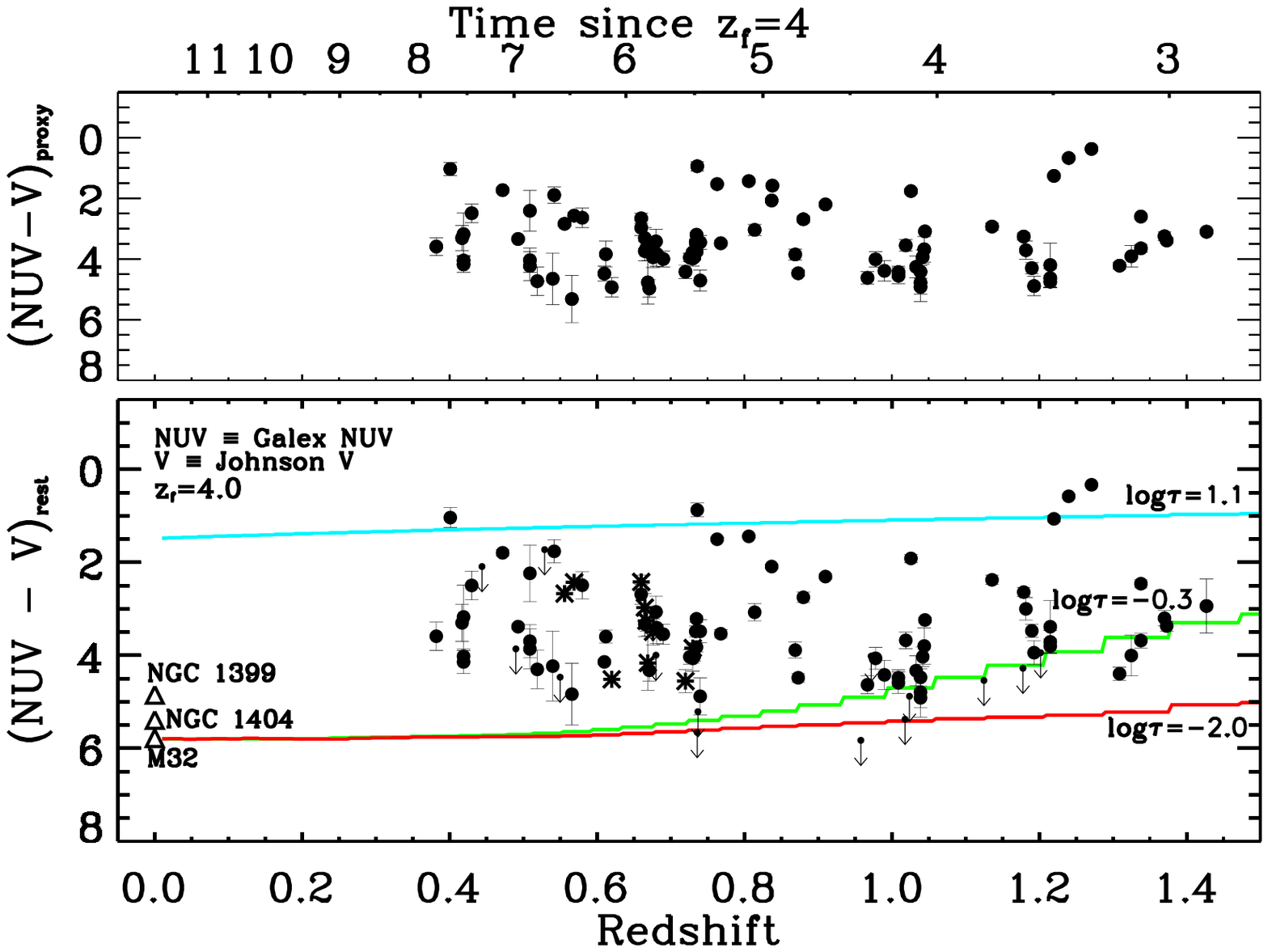, width=1.0\txw,clip=} 
\caption{\small {\bf Upper Panel} $\colon$ The observed (NUV--V)
  colors of the catalog of ETGs in the ERS field.  We calculate the
  observed colors by differencing the observed photometry for the
  combination of WFC/ACS filters that most closely matches that region
  of spectrum assessed by the NUV and Johnson V filters, respectively
  (see Table \ref{tab:proxy}). On the upper abscissa, we provide the
  time (Gyr) since $z_f$=4.0 for reference see text.  {\bf Bottom
    Panel} $\colon$ The \nuvv~colors of the final catalog of ETGs.
  We plot photometric and systematic (associated with the
  transformation function, see \S \ref{sec:analysis}) uncertainties
  for all detected ETGs. We plot ETGs detected in Radio and/or X-ray
  surveys of the GOODS-S field with an ``asterisk''
  (\textasteriskcentered).  Photometric upper limits, defined by the
  recovery limits discussed in \S \ref{sec:catalogueprod}, are
  overplotted as downward-pointing arrows.  We plot the colors of
  three, maximally old simple stellar evolution models derived from
  BC03, assuming a fixed redshift of formation ($z_f=4.0$), and a
  star-formation history defined by Equation \ref{eqn:burstmod} with
  log($\tau$[Gyr])$ \simeq$ 1.1 (Blue), $-$0.3 (Green) and $-$2.0 (Red).
  Note that the low redshift evolution of the \nuvv~color is an
  empirical fit to the UVX in quiescent ETGs at this redshift, and is
  not motivated by a physical theory of the stellar sources of the
  UVX.}
\label{fig:nuvcolors}
\end{figure}
 % Figure

\begin{figure}[htbc]
\centering
\epsfig{file=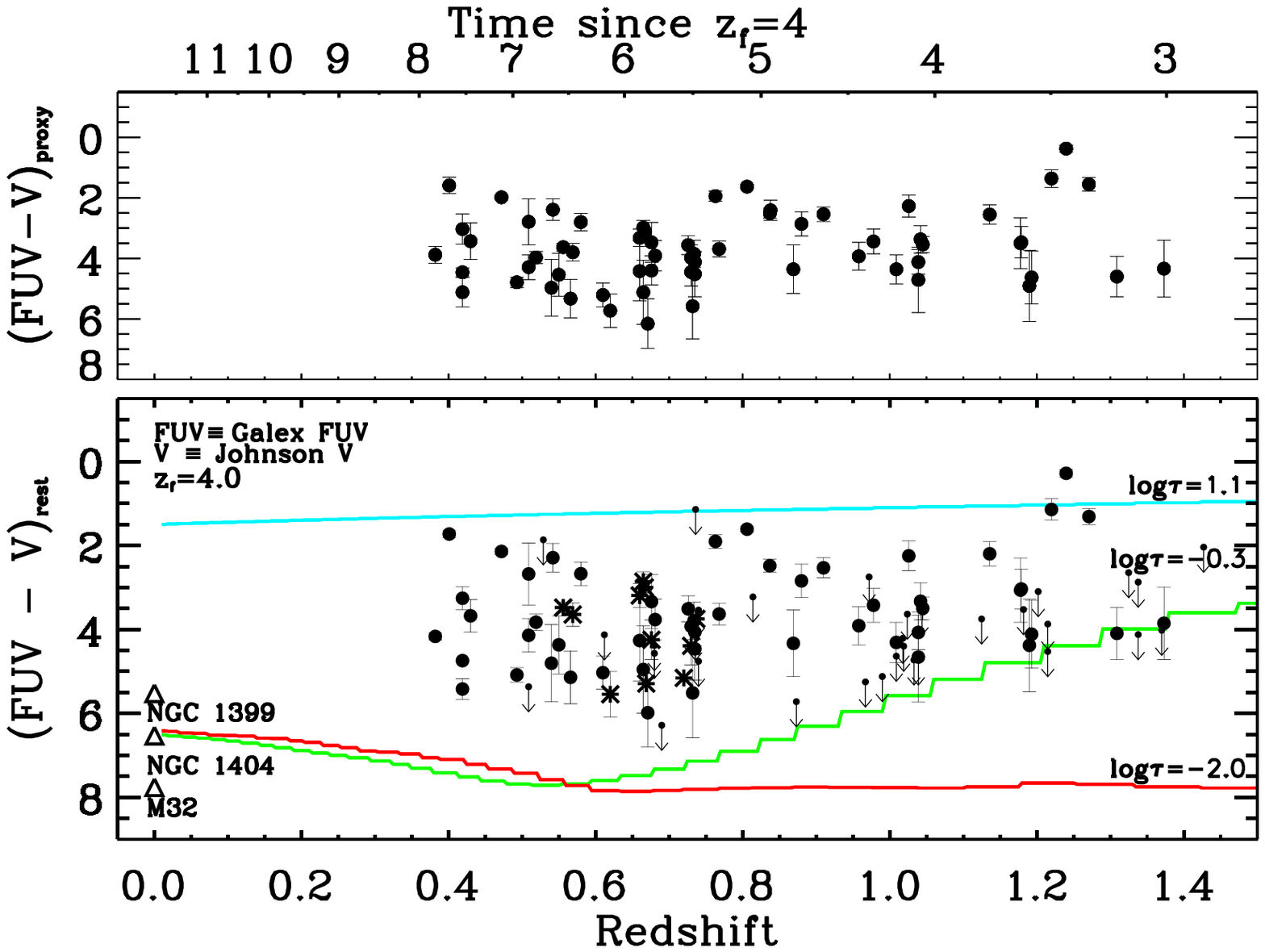, width=1.0\txw,clip=} 
\caption{\small Proxy ({\bf Upper Panel}) and \fuvv~({\bf Bottom
    Panel}) colors, and model tracks as in Figure
  \ref{fig:nuvcolors}.}
\label{fig:fuvcolors}
\end{figure}
 % Figure
\begin{figure}[htbc]
\centering
\epsfig{file=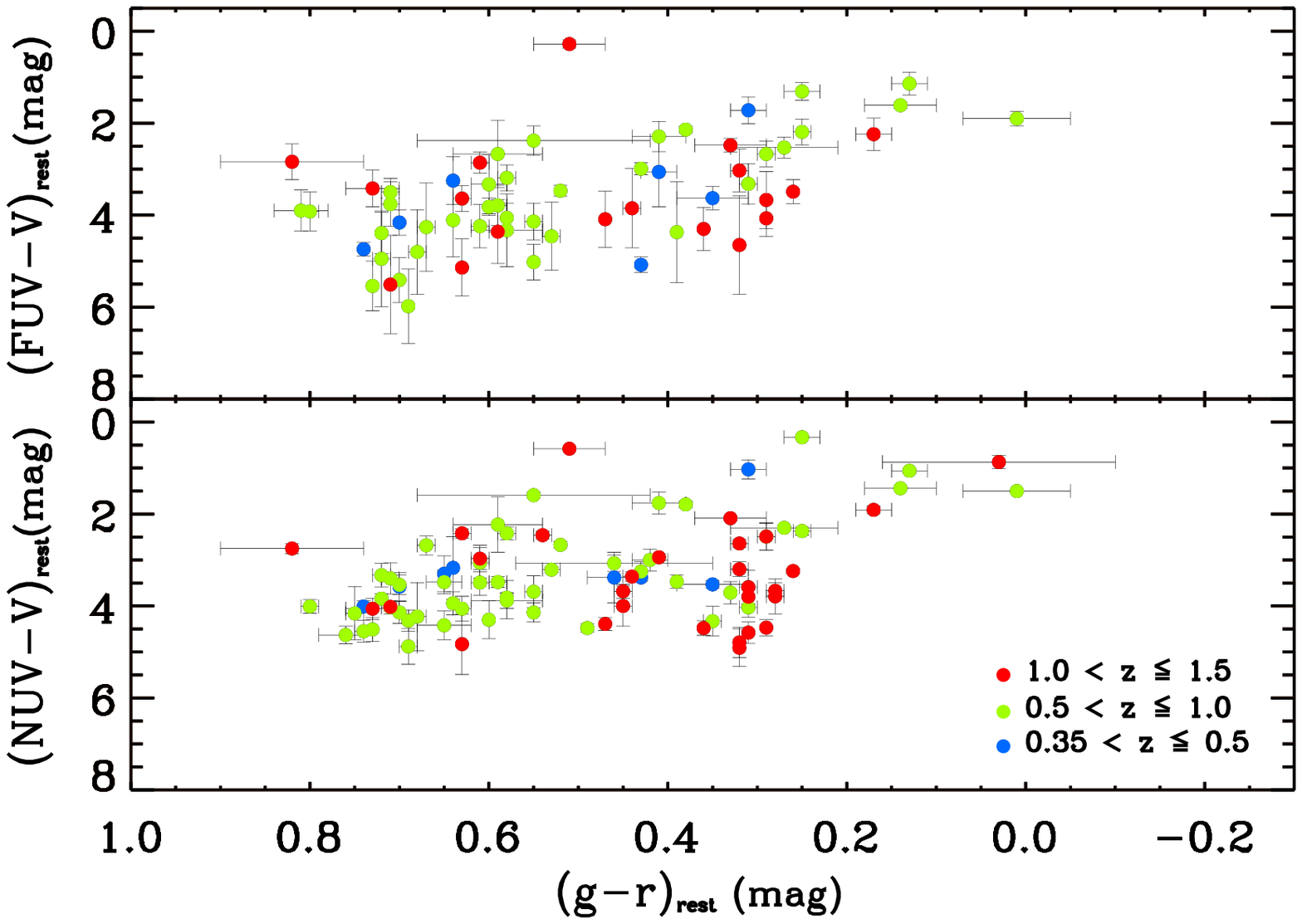,width=1.0\txw,clip=} 
\caption{\small {\bf Upper Panel} $\colon$ The \nuvv~and \gmr~colors
  of the ETGs are plotted.  {\bf Bottom Panel} $\colon$
  The \fuvv~and \gmr~colors of the catalog ETGs are plotted. The
  conversion between the observed and rest-frame colors of the ETGs is
  outlined in \S\ref{sec:analysis}.  All data are color-coded
  according to the redshift-color scheme defined in the bottom panel.
  The span of rest-frame colors in these panels likely indicates
  recent star-formation in many ETGs \citep[cf.][]{Ka07}. Paper II
  will model the star-formation histories of the ETGs in more
  detail. }
\label{fig:UVOPTcolors}
\end{figure}
 % Figure
%this file was produced in directory GMR_Dump on the laptop
\begin{figure}[htbc]
\centering
\epsfig{file=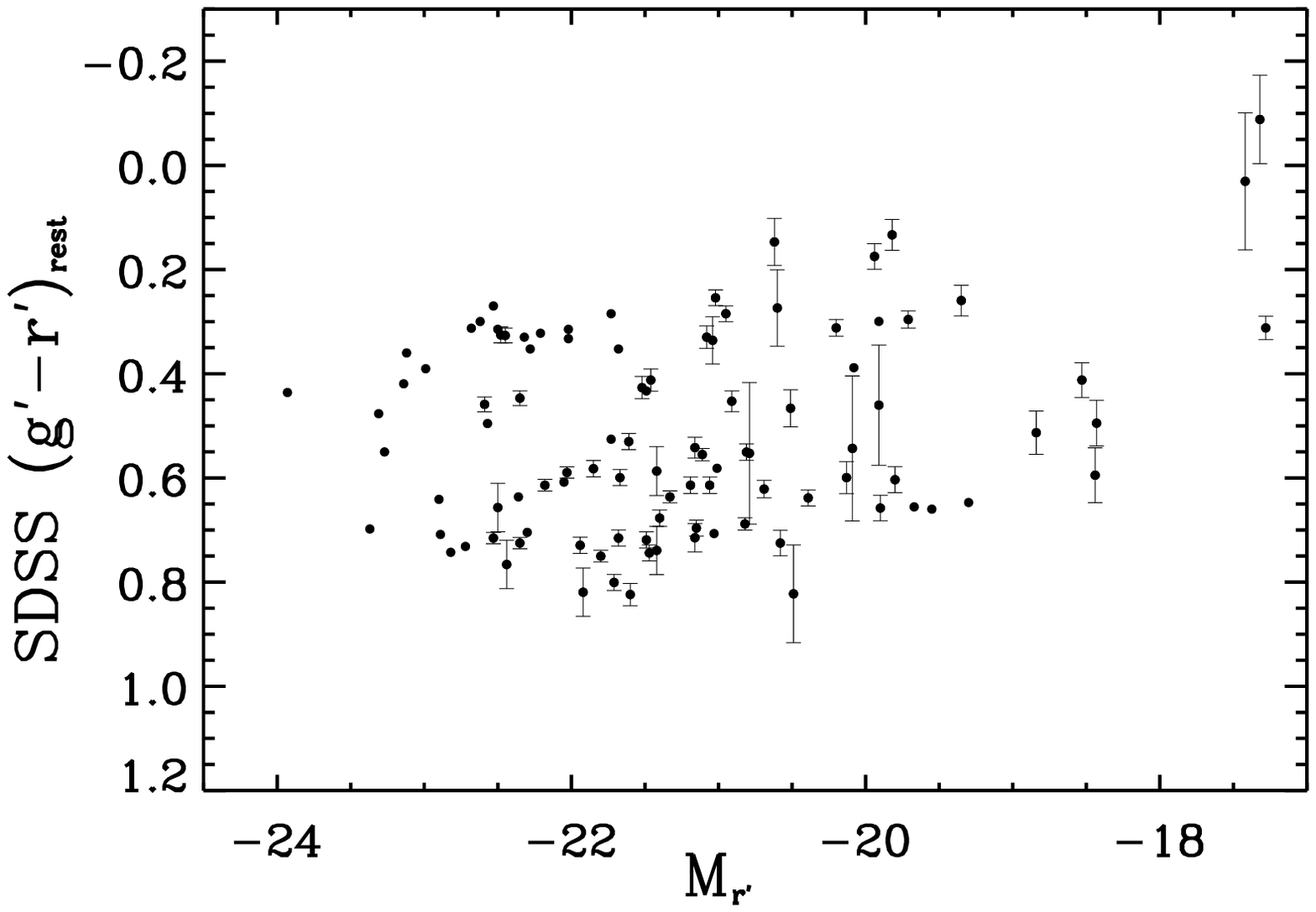, width=0.8\txw,clip=} 
\caption{\small The \gmr~colors of the ETGs.  For clarity, error bars
  are overplotted only for ETGs with measured (photometric and
  systematic) uncertainties greater than 0.01 mag. The broadband
  SED-fitting method for determining the absolute magnitudes is
  outlined in \S\ref{sec:absmag}.  See \S \ref{sec:analysis} for full
  details of the color transformation that we use to calculate the
  colors and photometric completeness limits plotted.}
\label{fig:optcmd}
\end{figure}
 % Figure
\begin{deluxetable}{ccc}
%\label{tab:proxy}
\tabletypesize{\scriptsize} \tablecaption{WFC3 UVIS estimated red-leak\tablenotemark{\dagger} for model ETGs}
\tablewidth{0pt} 
\tablehead{\colhead{Filter} & \colhead{BC03} & \colhead{CWW}}
\startdata 
F336W & 0.2\% & 2.9$\times 10^{-2}$\% \\
F275W & 1.2\% & 0.15\% \\
F225W & 3.5\% & 0.26\% \\
\enddata
\label{tab:redleaktab}
\tablenotetext{\dagger}{The red-leak is defined in \S \ref{sec:appendix}.}
%\comment{}
%\comment{The red-leaks were calculated for a grid of models selected to represent the reddest catalog galaxies, the quiescent, homogenously old ETGs. The grid of models were derived from the \ref{CWW80} Elliptical template and BC03 exponentially-declining star formation defined with $\tau$0.01 and an absolute age of 12Gyr.  For more details, see Section \ref{sec:appendix}.}
\end{deluxetable}
    
 % Supplemental
\begin{figure}[htbc]
\centering
\epsfig{file=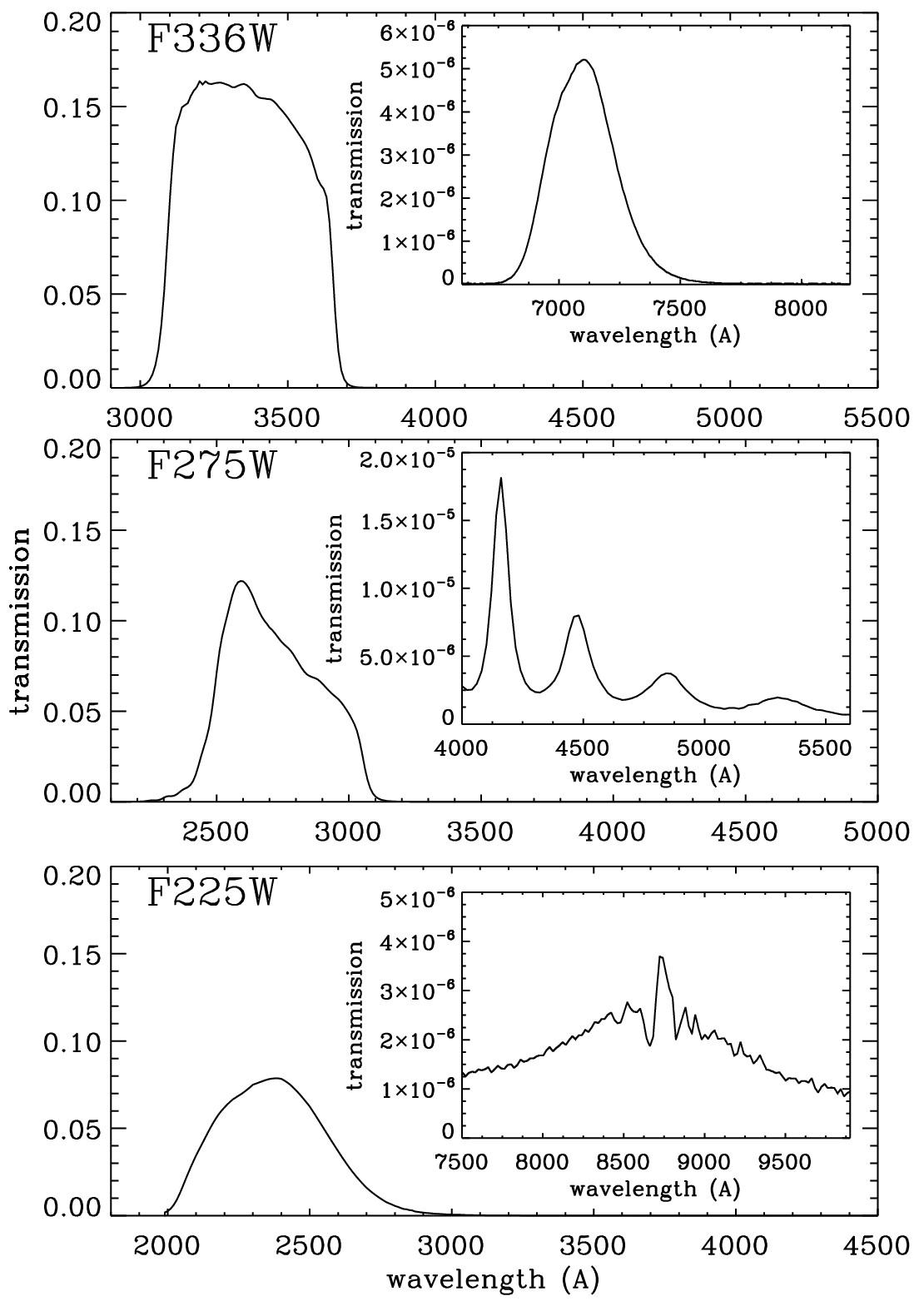, width=0.6\txw,clip=} 
\caption{WFC3 UVIS filter response curves.  The total throughput for
  the F225W, F275W, and F336W filters are shown here.  The inset in
  each panel illustrates the transmission of each filter at the
  wavelengths where the red-leak is most severe.  Note$\colon$ the
  range differs between each panel.  Using the BC03 and CWW template
  spectra, we estimate that for a typical ETG at
  $0.35$\,\,\lsim\,\,$z$\,\,\lsim\,\,$1.5$ the red-leak is {\cal R}
  \lsim\,3\%.  For more details, see Appendix \ref{sec:appendix} and
  Table \ref{tab:redleaktab}. }
\label{fig:redleakcurve}
\end{figure}
 % Figure


\thebibliography{109}
\linespread{0.3}%
\selectfont
\bibitem[Abraham et al.(2003)]{A03} Abraham, R.~G., van den Bergh, S., \& Nair, P.\ 2003, \apj, 588, 218%\vspace*{+0.1mm}
\bibitem[Atlee et al.(2009)]{A09} Atlee, D., Assef, R.~J., \& Kochanek, C.\ 2009, \apj, 694, 1539%\vspace*{+0.1mm}
\bibitem[Barkana and Loeb\,(2001)]{B01} Barkana, R., \& Loeb, A.\ 2001, Phys. Rep., 349, 125%\vspace*{+0.1mm}
\bibitem[Bell et al.(2004)]{B04} Bell, E., et al.\ 2004, \apj, 608, 752%\vspace*{+0.1mm}
\bibitem[Bertin \& Arnouts\,(1996)]{B96} Bertin, E., \& Arnouts, S.\ 1996, A\&A, 117, 393%\vspace*{+0.1mm}
\bibitem[Bezanson et al.(2009)]{B09} Bezanson, R., et al.\ 2009, \apj, 697, 1290%\vspace*{+0.1mm}
\bibitem[Brown et al.(2000)]{B00} Brown, T.~M., Bowers, C.~W., Kimble, R.~A., \& Ferguson, H. C.\ 2000, \apj, 529, L89%\vspace*{+0.1mm}
\bibitem[Brown et al.(2003)]{B03} Brown, T.~M., Ferguson, H.~C., Smith, E., Bowers, C.~W., Kimble, R.~A., Renzini, \& Rich, R.~M. \ 2003, \apj, 584, L69 %\vspace*{+0.1mm}
\bibitem[Bruzual and Charlot\,(2003)]{BC03} Bruzual, A.~G., \& Charlot, S.\ 2003, \mnras, 344, 1000%\vspace*{+0.1mm}
\bibitem[Burstein et al.(1988)]{B88} Burstein, D.,Bertola, F., Buson, L.~M., Faber, S.~M., Lauer, T.~R. \ 1988, \apj, 328, 440%\vspace*{+0.1mm}
\bibitem[Clemens et al.(2009)]{C09} Clemens, M.~S., Bressan, A., Nikolic, B., \& Rampazzo, R.\ 2009, MNRAS 392, 35%\vspace*{+0.1mm}
\bibitem[Colbert et al.(2001)]{C01} Colbert, J.~W., Mulchaey, J.~S., \& Zabludoff, A.~I.\ 2001, AJ, 121,808%\vspace*{+0.1mm}
\bibitem[Coleman, Wu and Weedman\,(1980)]{CWW80} Coleman, G.~D., Wu, C.~C., Weedman, D.~W. \ 1980, ApJS, 43, 393%\vspace*{+0.1mm}
\bibitem[Conselice et al.(2003)]{C03} Conselice, C.~J.\ 2003, ApJS, 147, 1 %\vspace*{+0.1mm}
\bibitem[Conselice et al.(2008)]{C08} Conselice, C.~J., Rajgor, S., \& Myers, R.\ 2008, MNRAS, 386, 909%\vspace*{+0.1mm}
\bibitem[Crockett et al.(2011)]{Cr10} Crockett, R.~M., et al. 2011, ApJ, 727, 115%\vspace*{+0.1mm}
\bibitem[Daddi et al.(2005)]{D05} Daddi, E., et al. \ 2005, \apj, 626, 680%\vspace*{+0.1mm}
\bibitem[Davis et al.(2007)]{D07} Davis, M., et al. \ 2007, \apj, 660, L1%\vspace*{+0.1mm}
\bibitem[Dickinson et al.(2003)]{D03} Dickinson, M., Giavalisco, M., \& the GOODS Team,\ 2003 {\it The Great Observatories Origins Deep Survey} in "The Mass of Galaxies at Low and High Redshift," eds. R. Bender \& A. Renzini, 324%\vspace*{+0.1mm}   
\bibitem[Donas et al.(2007)]{Do07} Donas, J., et al.\ 2007, \apjs, 173, 597%\vspace{+0.1mm}
\bibitem[Dorman et al.(1993)]{D93} Dorman, B., Rood, R., \& O'Connell, R.~W.\ 1993, \apj, 419, 596 %\vspace*{+0.1mm}
\bibitem[Dorman et al.(1995)]{D95} Dorman, B., O'Connell, R.~W., \& Rood, R.~T.\ 1995, \apj, 442, 105 %\vspace*{+0.1mm}
\bibitem[Eggen, Lynden-Bell, and Sandage\,(1962)]{ELS62} Eggen, O., Lynden-Bell, D., \& Sandage, A.\ 1962, \apj, 136, 748%\vspace*{+0.1mm}
\bibitem[Ferreras \& Silk\,(2000)]{FS00} Ferreras, I, \& Silk, J.\  2000, ApJ, 541, 37%\vspace{+0.1mm}
\bibitem[Ferreras et al.(2009)]{F09} Ferreras, I., et al.\ 2009, \apj, 706, 158%\vspace*{+0.1mm}
\bibitem[Fukugita et al.(1996)]{F96} Fukugita, M.\ 1996, AJ, 111, 174%\vspace*{+0.1mm}
\bibitem[Giacconi et al.(2002)]{G02} Giacconi, R., et al.\ 2002, ApJS, 139, 369%\vspace*{+0.1mm}
\bibitem[Giavalisco et al.(2004)]{G04} Giavalisco, M., et al.\ 2004, \apj, 600, L93%\vspace*{+0.1mm}
\bibitem[Gil de Paz et al.(2007)]{G07} Gil de Paz, A., et al.\ 2007, ApJS, 173, 185%\vspace*{+0.1mm}
\bibitem[Gilli et al.(2003)]{G03} Gilli, R., et al.\ 2003, \apj, 592, 791%\vspace*{+0.1mm}
\bibitem[Graves et al.(2009)]{G09} Graves, G.~J., Faber, S.~M., \& Schiavon, R.~P.\ 2009, \apj, 698, 1590%\vspace*{+0.1mm}
\bibitem[Greggio \& Renzini(1990)]{GR90} Greggio, L., \& Renzini, A.\ 1990, \apj, 364, 35%\vspace*{+0.1mm}
\bibitem[Han et al.(2007)]{H07} Han, Z., Podsiadlowski, P., \& Lynas-Gray, A.~E.\ 2007, MNRAS, 380, 1098%\vspace*{+0.1mm}
\bibitem[Jeong et al.(2009)]{Je09} Jeong, H., et al.\ 2009, \mnras, 398, 2028%\vspace{+0.1mm}
\bibitem[Kalirai et al.(2009a)]{K09a} Kalirai, J.~S., et al.\ 2009a, Instrument Science Report WFC3, 2009-21%\vspace*{+0.1mm}
\bibitem[Kalirai et al.(2009b)]{K09b} Kalirai, J.~S., MacKenty, J., Bohlin, R., Brown, T., Deustua, S., Kimble, R.~A., \& Riess, A. \ 2009b, Instrument Science Report WFC3, 2009-30%\vspace*{+0.1mm}
\bibitem[Kaviraj et al.(2007a)]{K07} Kaviraj, S., et al.\ 2007a, ApJS, 173, 619%\vspace*{+0.1mm}
\bibitem[Kaviraj et al.(2007b)]{Ka07} Kaviraj, S., Rey, S.-C., Rich, R.~M., Yoon, S.~J., \& Yi, S.~K., \ 2007b, MNRAS, 381, L74%\vspace*{+0.1mm}
\bibitem[Kaviraj et al.(2008)]{K08} Kaviraj, S., et al.\ 2008, MNRAS, 388, 67%\vspace*{+0.1mm}
\bibitem[Kaviraj et al.(2009)]{Ka09} Kaviraj, S., Peirani, S., Khochfar, S., Silk, J., \& Kay, S., \ 2009, MNRAS, 394, 1713%\vspace*{+0.1mm}
\bibitem[Kaviraj\,(2010)]{K10} Kaviraj, S. \ 2010, MNRAS, 406, 382%\vspace*{+0.1mm}
\bibitem[Kaviraj\,(2011)]{Ka11} Kaviraj, S., Tan, K.-M., Ellis, R.~S., Silk, J., \ 2005, MNRAS, 411, 2148%\vspace*{+0.1mm}
\bibitem[Komatsu et al.(2011)]{Ko11} Komatsu, E., et al.\ 2011, ApJS, 192, 18%\vspace*{+0.1mm}
\bibitem[Kondo\,(1987)]{K87} Kondo, Y. \ 1987, {\it Exploring the universe with the IUE satellite}. Dordrecht $\colon$ Reidel. 787 pp.%\vspace*{+0.1mm} 
\bibitem[Kron, Koo, \& Windhorst (1985)]{K85} Kron, R.~G., Koo, D.~C., Windhorst, R.~A. 1985 A\&A,  146, 38%\vspace*{+0.1mm}
\bibitem[Lee et al.(2005)]{Le05} Lee, Y.-W., et al.\ 2005, \apjl, 619, L103%\vspace*{+0.1mm}
\bibitem[Le F\`{e}vre et al.(2004)]{LeF04} Le F\`{e}vre, O., et al.\ 2004, A\&A, 428, 1043%\vspace*{+0.1mm}
\bibitem[Lisker\,(2008)]{L08} Lisker, T., \ ApJS, 179, 319%\vspace*{+0.1mm}
\bibitem[Lotz et al.(2004)]{L04} Lotz, J.~M., Primack, J., \& Madau, P. 2004, AJ, 128, 163%\vspace*{+0.1mm}
\bibitem[Lucero and Young\,(2007)]{L07} Lucero, D.~M., \& Young, L.~M. 2007, AJ, 134, 2148%\vspace*{+0.1mm}
\bibitem[Luo et al.(2010)]{L10} Luo, B., et al.\ 2010, ApJS, 187, 560 %\vspace{+0.1mm}
\bibitem[Madau et al.(1998)]{M98} Madau, P., Pozzetti, L. \& Dickinson, M. 1998, ApJ, 498, 106%\vspace*{0.1mm}
\bibitem[Martin et al.(1997)]{M97} Martin, C., et al.\ 1997, {\it Bull. Amer. Astron. Soc.} 191$\colon$\#63.04%\vspace*{+0.1mm}
\bibitem[Marchesini et al.(2009)]{M09} Marchesini, D., et al.\ 2009, \apj,701, 1765%\vspace*{+0.1mm}
\bibitem[Mignoli et al.(2005)]{Mi05} Mignoli, M. et al.\ 2005, A\&A, 437, 883%\vspace*{+0.1mm}
\bibitem[Miller et al.(2008)]{M08} Miller, N., et al.\ 2008, ApJS, 179, 114%\vspace*{+0.1mm}
\bibitem[Morganti et al.(2006)]{M06} Morganti, R., et al.\ 2006, MNRAS, 371, 157%\vspace*{+0.1mm}
\bibitem[Morrissey et al.(2005)]{M05} Morrissey, P., et al.\ 2005, ApJS, 619, L7%\vspace*{+0.1mm}
\bibitem[O'Connell\,(1980)]{OC80} O'Connell, R.~W. 1980, \apj, 236, 430%\vspace*{+0.1mm}
\bibitem[O'Connell\,(1999)]{OC99} O'Connell, R.~W. 1999, ARA\&A, 37, 603%\vspace*{+0.1mm}
\bibitem[Ohl et al.(1998)]{O98} Ohl, R.~G., et al.\ 1998, \apj, 505, L11%\vspace*{+0.1mm}
\bibitem[Oke and Gunn\,(1983)]{O83} Oke, J.~B., \& Gunn, J.~E. 1983, \apj, 266, 713%\vspace*{+0.1mm}
\bibitem[Park \& Lee(1997)]{PL97} Park, J.-H., \& Lee, Y.-W.\ 1997, \apj, 476, 28%\vspace*{+0.1mm}
\bibitem[Papovich et al.(2001)]{P01} Papovich, C., Dickinson, M., \& Ferguson, H.~C.\ 2001, \apj, 559, 620%\vspace*{+0.1mm}
\bibitem[Pasquali et al.(2006)]{P06} Pasquali, A., et al.\ 2006, \apj, 636, 115%\vspace*{+0.1mm} 
\bibitem[Patton et al.(2002)]{P02} Patton, D.~R., et al.\ 2002, \apj, 565, 208%\vspace*{+0.1mm}
\bibitem[Peirani et al.(2010)]{P10} Peirani, S., et al.\ 2010, MNRAS, 405, 2327%\vspace*{+0.1mm}
\bibitem[Popesso et al.(2009)]{P09} Popesso, P., et al.\ 2009, A\&A, 494, 443%\vspace*{+0.1mm} 
\bibitem[Postman et al.(2005)]{P05} Postman, M., et al.\ 2005, \apj, 623, 721%\vspace*{0.1mm}
\bibitem[Ravikumar et al.(2007)]{R207} Ravikumar, C.~D., et al.\ 2007, A\&A, 465, 1099%\vspace*{+0.1mm}
\bibitem[Ree et al.(2007)]{R07} Ree, C., et al.\ 2007, ApJS, 173, 607 %\vspace*{+0.1mm}
\bibitem[Rose\,(1985)]{R85} Rose, J.~A. 1985, AJ, 90, 1927%\vspace*{+0.1mm}
\bibitem[Ryan et al.(2010)]{R10} Ryan, R.~E., et al., 2010, ApJ, arXiv$\colon$1007.1460%\vspace*{+0.1mm} 
\bibitem[Salim \& Rich\,(2010)]{Sa10} Salim, S., \& Rich, R.~M. 2010, \apj, 714, 290%\vspace*{+0.1mm}
\bibitem[Schawinski et al.(2007)]{S07} Schawinski, K., et al.\ 2007, ApJS, 173, 512%\vspace*{+0.1mm}
\bibitem[Schweizer et al.(1990)]{S90} Schweizer, F., Seitzer, P., Faber, S.~M., Burstein, D., Dalle Ore, C.~M., \& Gonz\'{a}lez, J.~J. \ 1990, ApJ, 364, L33%\vspace*{+0.1mm} 
\bibitem[Scott et al.(2009)]{Sc09} Scott, N., et al.\ 2009, MNRAS, 398, 1835%\vspace*{+0.1mm}
\bibitem[Strolger et al.(2004)]{St04} Strolger, L.-G. et al.\ 2004, \apj, 613, 200%\vspace{+0.1mm}
\bibitem[Suh et al.(2010)]{S10} Suh, H., Jeong, H., Oh, K., Yi, S.~K., Ferreras, I., \& Schawinski, K. \ 2010, \apj, 187, 374%\vspace*{+0.1mm}
\bibitem[Szokoly et al.(2004)]{S04} Szokoly, G., et al.\ 2004, ApJS, 155, 271%\vspace*{+0.1mm}  
\bibitem[Tantalo et al.(1996)]{T96} Tantalo, R., Chiosi, C., Bressan, A., \& Fagotto, F. 1996, A\&A, 311, 361%\vspace*{+0.1mm}
\bibitem[Taylor-Mager et al.(2007)]{V07} Taylor-Mager, V.~A., Conselice, C.~J., Windhorst, R.~A., \& Jansen, R.~A.\ 2007, \apj, 659, 162%\vspace*{+0.1mm}
\bibitem[Tinsley\,(1980)]{T80} Tinsley, B.~M.\ 1980 {\it Fund. of Cosmic Physics}, 5, 287%\vspace*{+0.1mm}
\bibitem[Toomre and Toomre\,(1972)]{T72} Toomre, A., \& Toomre, J.\ 1972, \apj, 178, 623%\vspace*{+0.1mm}
\bibitem[Trager et al.(2000)]{T00} Trager, S.~C., Faber, S.~M., Worthey, G., \& Gonz\'{a}lez, J.~J. \ 2000, AJ, 119,1645%\vspace*{+0.1mm}
\bibitem[Vanden Berk et al.(2001)]{vdB01} Vanden Berk, D.~E., et al.\ 2001, AJ, 122, 549%\vspace*{+0.1mm}
\bibitem[van Dokkum\,(2005)]{VD05} van Dokkum, P.~G.\ 2005, AJ, 130, 2647%\vspace*{+0.1mm}
\bibitem[van Dokkum et al.(2010)]{VD10} van Dokkum, P.~G., et al.\ 2010, \apj, 709, 1018%\vspace*{+0.1mm}
\bibitem[Vanzella et al.(2008)]{V08} Vanzella, E., et al.\ 2008, A\&A, 478, 83%\vspace*{+0.1mm}
\bibitem[von den Linden et al.(2007)]{vdL07} von den Linden, A., Best, P.~N., Kauffmann, G., White, S.~D.~M. \ 2007, MNRAS, 379, 867%\vspace*{+0.1mm}
\bibitem[Weiner et al.(2005)]{Wie05} Weiner, B.~J., et al.\ 2005,\apj, 620, 595%\vspace*{+0.1mm}
\bibitem[Willmer et al.(2006)]{Wi06} Willmer, C.~N.~A., et al. \ 2006, \apj, 647, 85%\vspace*{+0.1mm} 
\bibitem[Windhorst et al.(2011)]{W11} Windhorst, R.~A., et al.\ 2011, ApJS, 193, 27%\vspace*{+0.1mm}
\bibitem[Worthey et al.\ (1994)]{W94} Worthey, G., Faber, S.~M., Gonz\'{a}lez, J.~J., \& Burstein, D.\ 1994, ApJS, 94, 687%\vspace*{+0.1mm}
\bibitem[Yi et al.(1995)]{Y95} Yi, S.~K., Afshari, E., Demarque, P., \& Oemler, A., Jr.\ 1995, ApJL, 453, L69 %\vspace*{+0.1mm}
\bibitem[Yi et al.(1998)]{Y98} Yi, S.~K, Demarque, P., \& Oemler, A., Jr.\ 1998, \apj, 492, 480%\vspace*{+0.1mm}
\bibitem[Yi et al.(1999)]{Y99} Yi, S.~K, Lee, Y.-W., Woo, J.-H., Park, J.-H., Demarque, P., \& Oemler, A. \ 1999, \apj, 513, 128%\vspace*{+0.1mm}
\bibitem[Yi et al.(2003)]{Y03} Yi, S.~K., Kim, Y, Demarque, P.\ 2003, ApJS, 144, 259%\vspace*{+0.1mm}
\bibitem[Yi \& Yoon\,(2004)]{Y04} Yi, S.~K. \& Yoon,S-J.\ 2004, Ap\&SS, 291, 205%\vspace*{+0.1mm}
\bibitem[Yi et al.(2005)]{Yi05} Yi, S.~K., et al.\ 2005, ApJS, 619, L111%\vspace*{+0.1mm}
\linespread{1.0}%
\selectfont
\end{document}